\documentclass[amsmath, amsfonts, nofootinbib, reprint, superscriptaddress]{revtex4-2}
\usepackage{amsmath}
\usepackage[english]{babel}
\usepackage[utf8]{inputenc}
\usepackage{amsthm}
\usepackage{mathtools}
\usepackage{physics}
\usepackage{xcolor}
\usepackage{graphicx}
\usepackage{adjustbox}
\usepackage{placeins}
\usepackage[T1]{fontenc}
\usepackage{lipsum}
\usepackage{csquotes} \usepackage[pdftex, pdftitle={Article}, pdfauthor={Stephen Fairhurst}]{hyperref} 

\usepackage{soul} 

\usepackage[acronym]{glossaries}

\newacronym{gw}{GW}{gravitational wave}
\newacronym{snr}{SNR}{signal-to-noise ratio}
\newacronym{psd}{PSD}{power spectral density}
\newacronym{pdf}{pdf}{probability density function}

\newcommand\numberthis{\addtocounter{equation}{1}\tag{\theequation}}

\begin{document}
\title{Simple parameter estimation using observable features of gravitational-wave signals}

\author{Stephen Fairhurst}
    \email[Correspondence email address: ]{fairhursts@cardiff.ac.uk}\affiliation{Gravity Exploration Institute, School of Physics and Astronomy, Cardiff University, Cardiff, CF24 3AA, United Kingdom}
\author{Charlie Hoy}
    \affiliation{Gravity Exploration Institute, School of Physics and Astronomy, Cardiff University, Cardiff, CF24 3AA, United Kingdom}
    \affiliation{University of Portsmouth, Portsmouth, PO1 3FX, United Kingdom}
\author{Rhys Green}
    \affiliation{Gravity Exploration Institute, School of Physics and Astronomy, Cardiff University, Cardiff, CF24 3AA, United Kingdom}
\author{Cameron Mills}
    \affiliation{Gravity Exploration Institute, School of Physics and Astronomy, Cardiff University, Cardiff, CF24 3AA, United Kingdom}
    \affiliation{Albert-Einstein-Institut, Max-Planck-Institut for Gravitationsphysik, D-30167 Hannover, Germany}
    \affiliation{Leibniz Universitat Hannover, D-30167 Hannover, Germany}

\author{Samantha A. Usman}
    \affiliation{Gravity Exploration Institute, School of Physics and Astronomy, Cardiff University, Cardiff, CF24 3AA, United Kingdom}

\date{\today} 

\begin{abstract}
Using simple, intuitive arguments, we discuss the expected accuracy with which astrophysical parameters can be extracted from an observed gravitational wave signal.  The observation of a chirp like signal in the data allows for measurement of the component masses and aligned spins, while measurement in three or more detectors enables good localization.  The ability to measure additional features in the observed signal --- the existence or absence of power in i) the second gravitational wave polarization, ii) higher gravitational wave multipoles or iii) spin-induced orbital precession --- provide new information which can be used to significantly improve the accuracy of parameter measurement.  We introduce the \texttt{simple-pe} algorithm which uses these methods to generate rapid parameter estimation results for binary mergers. We present results from a set of simulations, to illustrate the method, and compare results from \texttt{simple-pe} with measurements from full parameter estimation routines.  The \texttt{simple-pe} routine is able to provide initial parameter estimates in a matter of CPU minutes, which could be used in real-time alerts and also as input to significantly accelerate detailed parameter estimation routines.
\end{abstract}


\maketitle

\section{Introduction} 
\label{sec:introduction}

Gravitational wave astronomy has quickly evolved from the first observation in 2015~\cite{Abbott:2016blz} to now regular observations of black hole binary mergers~\cite{abbott2019gwtc, LIGOScientific:2021usb, LIGOScientific:2021djp,venumadhav2020new,Olsen:2022pin,Nitz:2021zwj}. Future improvements to the existing ground-based gravitational-wave detectors~\cite{TheLIGOScientific:2014jea,acernese2014advanced} are expected to increase the frequency of observations further, with mergers being observed daily, or even more frequently, in the coming years~\cite{KAGRA:2013rdx}. With a large number of observed signals, we can expect that many of them are not unique, and mostly serve to improve the sampling of the underlying astrophysical populations.  However, there will be a small number of signals which probe new areas of the parameter space, for example due to having particularly large or small masses and spins (see e.g.~\cite{Abbott:2020khf}); significant mass ratios (see e.g.~\cite{LIGOScientific:2020stg,Abbott:2020khf}); clear evidence of neutron star structure; eccentricity (see e.g.~\cite{Romero-Shaw:2020thy}) and, most tantalizingly, evidence for physics beyond Einstein's relativity.  There is, then, a desire to be able to, quickly and easily, determine which signals are likely to provide interesting results so that energy can be focused on them.

Detailed parameter estimation routines have already been developed~\cite{Veitch:2014wba,Biwer:2018osg,Ashton:2018jfp,Lange:2018pyp,Romero-Shaw:2020owr,Ashton:2021anp,Smith:2019ucc,Gabbard:2019rde,Green:2020hst,Dax:2021tsq,Dax:2022pxd,Shen:2019vep,Green:2020dnx,Williams:2021qyt,Zackay:2018qdy,Pathak:2022ktt, Islam:2022afg}, and are routinely used to recover the parameters of observed signals.  Furthermore, increasingly accurate gravitational waveforms have been developed, which incorporate ever more physical effects -- higher multipoles, accurate treatment of black hole spins, inclusion of accurate neutron star equation of state, use of numerical relativity results, eccentricity, beyond GR, etc.~\cite{Bohe:2016gbl,Cotesta:2018fcv,Cotesta:2020qhw, Ossokine:2020kjp,Babak:2016tgq,Pan:2013rra, Husa:2015iqa,Khan:2015jqa,London:2017bcn, Hannam:2013oca,Khan:2018fmp,Khan:2019kot, Varma:2019csw,Varma:2018mmi,Pratten:2020fqn, Garcia-Quiros:2020qpx,Pratten:2020ceb, Estelles:2020osj,Estelles:2020twz,Estelles:2021gvs,Hamilton:2021pkf}.  Thus, we can infer the parameters of the system using ever more sophisticated methods.  However, there are two issues.  First, as more physical effects are added to the waveforms, the time taken to generate these waveforms, and to sample the expanding parameter space, increases; although there has been recent effort to reduce this computational cost through bespoke optimisations~\cite{Devine:2016ovp,Knowles:2018hqq,gadre2022fully}, and by harnessing machine learning techniques~\cite{Khan:2020fso,Thomas:2022rmc}. Second, the parameter estimation routines provide estimates and uncertainties, but typically do not identify the features in the waveform that enable the measurement of given parameters with the stated accuracy.

There is a long history of work aimed at understanding, at a more basic level, how parameters can be extracted from the observed gravitational waveform and providing some idea of the expected accuracy of measurements.  For example, Refs.~\cite{cutler1994gravitational,Poisson:1995ef,Baird:2012cu,Farr:2015lna,Ng:2018neg} give early examples of investigations into measurements of masses and the degeneracy of mass and spin.  With increased interest in multi-messenger astronomy, various methods to understand gravitational wave localization have also been developed~\cite{Fairhurst:2010is,Wen:2010cr,Rodriguez:2013mla,Zhao:2017cbb,Vallisneri:2007ev, Grover:2013sha}.  The impact of higher gravitational wave multipoles were also examined in Refs.~\cite{Kalaghatgi:2019log,Mills:2020thr} and the impact of spin-induced orbital precession in detail in Refs.~\cite{Fairhurst:2019_2harm,fairhurst2019will,Green:2020ptm,Pratten:2020igi,Krishnendu:2021cyi}.  All along, warnings of using approximation techniques to investigate the full, high-dimensional parameter space in a single analysis have been given \cite{Vallisneri:2007ev, Rodriguez:2013mla}.

In this paper, we synthesize the physical insights mentioned above to provide a hierarchical understanding of parameter recovery from gravitational wave observations.  To do so, we begin with the basic information --- the observation of a gravitational wave chirp in one detector.  In this case, the shape of the waveform can be used to infer some details of the masses and spins.  For lower mass binaries, where the merger doesn't contribute too significantly to the signal power, the chirp mass is measured with good accuracy, while for higher masses the total mass determines the waveform during merger and ringdown.  Additional information about the phasing of the system allows for inference of the mass-ratio and the components of black hole spin aligned with the orbital angular momentum.  A single detector provides essentially no information about sky location, other than what can be inferred probabilistically (the signal is more likely to come from a sky location where the detector is more sensitive).  We can infer a maximum distance for the source.  However, for most cases, an accurate distance measurement is not possible as it is degenerate with the orientation and sky location.

As more features are observed, it is possible to extract additional astrophysical parameters from the signal.  Extra measurements arise from either the observation of the signal in additional detectors, or from the observation of additional waveform features in a detector or network of detectors.  If the signal is observed in more than one detector, this enables localization of the source and measurement, in principle, of both gravitational-wave polarizations.  The relative amplitude and phase of the second polarization provide additional constraints on the distance to and orientation of the binary.  Additional waveform features include higher gravitational-wave multipoles and spin-induced orbital precession.  In both cases, these features can be considered as adding additional components to the gravitational wave signal which are, to a good approximation, orthogonal to the dominant chirp waveform.  The significance of higher multipoles is typically more pronounced for systems with more unequal masses.  The relative amplitudes of the higher multipoles also depend upon the orientation, with many higher multipoles vanishing for face-on systems.  Thus observation of higher multipoles can allow for improved measurement of mass ratio and orientation of the binary.  The observation of precession requires non-zero in-plane spin components and allows inferences about the in-plane spins as well as the orientation of the binary.  In this paper, we consider the impact of adding each of the above features, and how it can improve the parameter recovery.  

The paper is laid out as follows. In Section \ref{sec:waveform} we describe the observable features of the waveform, in Section \ref{sec:approx_pe} we describe how the observable features of the waveform can be used to infer the system parameters.  In Section \ref{sec:simple_pe} we provide details of an implementation with results in Section, including a comparison between our results and those obtained with Bilby \cite{Ashton:2018jfp,Romero-Shaw:2020owr,Smith:2019ucc}, in Section \ref{sec:pe_comp}.  In Section \ref{sec:discussion} we provide a summary and discussion of future work.  In Appendix \ref{app:waveform} we provide additional details of the waveform decomposition and in \ref{app:equal_mass}
provide results for a low \gls*{snr} signal. \section{The observable features in a gravitational waveform}
\label{sec:waveform}

The gravitational waveform observed at a detector is given by
\begin{equation}\label{eq:h_det}
h_{X}(t) =
	\mathrm{Re} \left[ F^{X}  h \right] 
\end{equation}
where
\begin{equation}\label{eq:f}
    F^{X} = F^{X}_{+} + i F^{X}_{\times}
\end{equation}
and $F^{X}_{+, \times}$ are the detector response functions for the detector $X$ which depend upon the location of the source relative to the detector, and 
\begin{equation}\label{eq:complex_h}
h := h_{+} - i h_{\times}
\end{equation}
where $h_{+, \times}$ are the two polarizations of the gravitational wave, which depend upon the details of the source.  In this paper, we restrict attention to black hole binary mergers for which the signal is relatively short-lived, so that we can treat $F_{+, \times}$ as constant over the duration of the signal.\footnote{This approximation is appropriate for binary mergers in the advanced detector network, but breaks down for low-mass mergers in next-generation detectors \cite{Chan:2018csa}.
}

The gravitational waveform emitted during a binary merger can naturally be decomposed into a set of spin-weighted spherical harmonics as \cite{Thorne:1980ru, Kidder:2007rt}
\begin{equation}\label{eq:waveform_lm}
h(t) =
\sum_{\ell \ge 2} \sum_{m = -\ell}^{\ell} h_{\ell, m}(t, \vec{\lambda}) {}^{-2}Y_{\ell, m}(\theta, \phi)
\end{equation}
where ${}^{-2}Y_{\ell, m}$ is the spin-weighted spherical harmonic of weight $-2$, $\theta$ and $\phi$ give the orientation of the observer relative to a co-ordinate system used to identify
the spherical harmonics, $\vec{\lambda}$ encodes the physical parameters of the system (masses, spins, etc) and $t$ is the time.  Here, and throughout this section, we follow the notation used in \cite{Khan:2019kot, Brown:2007jx}.
The frequency evolution of the harmonics depends upon the orbital frequency of the binary, with the dominant harmonic being the $(\ell, m) = (2, 2)$ having a frequency which is double the orbital frequency during the inspiral phase.  Various models for the gravitational waveform emitted during the merger of quasi-circular black hole binary mergers have been developed in recent years, see e.g. \cite{London:2017bcn, Pratten:2020ceb, Cotesta:2018fcv, Varma:2019csw}.

For binaries where the spins are misaligned with the orbital angular momentum, neither the orientation of the spins nor the magnitude and orientation of orbital angular momentum remain fixed and both precess around the direction of the total angular momentum, which does remain approximately constant \cite{Apostolatos:1994mx}.  This orbital precession leads to amplitude and phase modulations in the observed gravitational wave signal, on the precession timescale which is typically slower than the orbital period.  These modulations can be interpreted as the beating of different harmonics whose frequencies differ by multiples of the precession frequency \cite{Fairhurst:2019_2harm, Lundgren:2013jla}. Thus, for a binary with misaligned spins, precession will cause a splitting of each $(\ell, m)$ multipole in Eq.~(\ref{eq:waveform_lm}) into multiple harmonics whose frequencies differ by the precession frequency.

In many cases, the multipoles for a precessing system are approximated by ``twisting up'' \cite{Hannam:2013oca, Boyle:2011gg}
the multipoles of the non-precessing counterpart based upon the evolution of the orientation of the orbital angular momentum.  The direction of the orbital angular momentum $\hat{\mathbf{L}}$, relative to the total angular momentum $\hat{\mathbf{J}}$ is given by two angles: the opening angle $\beta$ ($\cos \beta = \hat{\mathbf{L}} \cdot \hat{\mathbf{J}}$) and the precession phase, $\alpha$ relative to a fixed orientation, $\alpha_{o}$ (also denoted $\phi_{JL}$).  Then, the precession frequency is given by 
\begin{equation}\label{eq:omega_p}
    \Omega_{p} = \dot{\alpha}
\end{equation}
To fully describe a co-precessing co-ordinate system, we require a third Euler angle $\epsilon$, defined via
\begin{equation}\label{eq:epsilon}
\dot{\epsilon} = \dot{\alpha} \cos \beta \, ,
\end{equation}
which determines the rate of rotation of the co-precessing frame.

The multipoles for a precessing system are given by
\begin{equation}\label{eq:prec_decomp}
h_{\ell, m}(t) = \sum_{n = -\ell}^{\ell} h^{\mathrm{NP}}_{\ell, n}(t) D^{\ell}_{n, m}(\alpha(t), \beta(t), \epsilon(t))
\end{equation}
where $h^{\mathrm{NP}}_{\ell, n}$ denotes the waveform for the equivalent non-precessing system or, equivalently, the waveform observed in a frame that is co-precessing with the binary,  $D^{\ell}_{n, m}$ denotes the Wigner D-matrix,
\begin{equation}
D^{\ell}_{n, m}(\alpha, \beta, \epsilon) = e^{i m \alpha} d^{\ell}_{n, m}(-\beta) e^{-i n \epsilon} \, ,
\end{equation}
and $d^{\ell}_{n, m}$ the Wigner d-matrix given, for example, in \cite{Brown:2007jx}.
It is straightforward to insert the expression for the precessing multipoles, Eq.~(\ref{eq:prec_decomp}), into the multipole expansion of the waveform, Eq.~(\ref{eq:waveform_lm}), to obtain the waveform for a precessing binary.  To do so, we first note that the co-ordinate system $(\theta, \phi)$ is naturally aligned with the total orbital angular momentum $\hat{\mathbf{J}}$. Therefore, if the system is viewed in a direction $\hat{\mathbf{N}}$, the angle $\theta = \theta_{JN}$ is the angle between $\hat{\mathbf{J}}$ and $\hat{\mathbf{N}}$. In addition, the orientation of the $x$-axis
is specified relative to the (initial) precession phase so that $\phi = - \alpha_{o}$. Therefore,
\begin{equation}\label{eq:prec_modes}
h(t) = \sum_{\ell, m, n} {}^{-2}Y_{\ell, m}(\theta_{JN}, -\alpha_{o}) D^{\ell}_{n, m}(\alpha, \beta, \epsilon) h^{\mathrm{NP}}_{\ell, n}(t, \vec{\lambda}) \, .
\end{equation}

Waveform models have been developed to generate accurate representations of the leading multipoles in the gravitational waveform \cite{London:2017bcn, Pratten:2020ceb, Cotesta:2018fcv, Varma:2019csw}. For example, numerous models provide the (2,2), (3,3), (4,4) multipoles and in addition the (2,1) and (3,2) multipoles. 

\subsection{Waveform components}
\label{ssec:wf_modes}

The gravitational waveform, $h(t)$, as given in Eq.~(\ref{eq:prec_modes}), is expressed as an infinite sum of waveform components.  However, only a small number of these make a significant contribution to the waveform.  By identifying the most significant waveform components, and restricting attention to them, we can simplify the waveform with little loss of accuracy.
In previous works \cite{Mills:2020thr}, we have shown that, for non-precessing waveforms, it is the (2, 2) and (3, 3) multipoles which are most significant across the majority of the parameter space, with the (4, 4) also contributing significantly for systems with high, comparable masses.  Similarly in \cite{Fairhurst:2019_2harm, Green:2020ptm}, we have shown that the two leading precession harmonics of the (2, 2) mode provide the dominant contribution, although see \cite{McIsaac:2023ijd} for examples of highly precessing systems where the third precession harmonic also contributes significantly.  Here, we present the expansion in terms of precession harmonics and higher multipoles simultaneously.

In \cite{Fairhurst:2019_2harm}, we demonstrated that the leading (2,2) waveform could be decomposed into five precession harmonics and, furthermore, that these precession harmonics formed a natural hierarchy, with each subsequent mode suppressed by an additional power of
\begin{equation}\label{eq:b}
    b = \tan(\beta/2) \, .
\end{equation}
For the majority of systems, the opening angle is significantly smaller than $45^{\circ}$, so that $b \lesssim 0.4$ (see Figure 3 of \cite{Fairhurst:2019_2harm} for details).  The opening angle only approaches $45^{\circ}$ for systems with unequal masses and a large spin on the more massive black hole --- in this case, the spin can be comparable to the orbital angular momentum. Consequently, for the majority of the binary black hole parameter space, $b$ can be used as a small expansion parameter, and terms of higher order in $b$ can be neglected.
In Appendix \ref{app:waveform}, we perform a decomposition for a generic waveform, comprising several multipoles and demonstrate that each multipole can be written in terms of precession harmonics with increasing powers of $b$. 

Let us restrict attention to the most significant waveform component and the leading sub-dominant contributions.  The leading waveform component, which we denote $h_{22, 0}$ arises as the dominant precession contribution to the (2, 2) multipole.  The two most significant sub-leading contributions are $h_{22, 1}$, the leading precession correction to the (2, 2) multipole, and $h_{33, 0}$ and the leading contribution the (3, 3) multipole.  The waveform can be written as:
\begin{align}\label{eq:leading_prec_hm}
    h &\approx
    \frac{d_{o}}{d_{L}} 
    \frac{(
    e^{2i\phi_o} h_{22,0} + 
     \tau^{4} e^{-2i\phi_o} h_{22,0}^{\ast})}
    {(1 + \tau^{2})^{2}} 
    \nonumber \\
    &
    + 
    \frac{d_{o}}{d_{L}} 
    \frac{4\tau (
    e^{2i\phi_o} h_{22,1} -
    \tau^{2} e^{-2i\phi_o}  h_{22,1}^{\ast})}{(1 + \tau^{2})^{2}} 
    \nonumber \\
    &
    + 
    \frac{d_{o}}{d_{L}} 
    \frac{\tau (
    e^{3i\phi_o} h_{33,0} + 
    \tau^{4} e^{-3i\phi_o}
     h_{33,0}^{\ast})}{(1 + \tau^{2})^{3}}  
    \, ,
\end{align}
where $d_{L}$ is the luminosity distance, $d_{o}$ is a fiducial distance, $\phi_{o}$ is the reference phase and
\begin{equation}
    \tau = \tan(\theta_{JN}/2)
\end{equation}
describes the orientation of the binary.\footnote{In Appendix \ref{app:waveform}, we have not explicitly extracted the $d_{L}$ or $\phi_{o}$ factors from the waveform.  It is straightforward to do this by re-defining the waveform components to be those associated with a binary at distance $d_{o}$ and coalescence phase $\phi_{o} = 0$.}

The detailed expression for the waveform in terms of the co-precessing harmonics are given in Equations (\ref{ap_eq:h22}) and (\ref{ap_eq:h_33k}).  Most notably, the sub-dominant precession harmonic is reduced in amplitude by a factor $b$ and offset in frequency from the dominant harmonic by the precession frequency $\Omega_{p}$, so that 
\begin{equation}\label{eq:h221}
    h_{22, 1} = b e^{-i (\alpha - \alpha_{o})} h_{22,0}
\end{equation}
In addition, the (3,3) multipole has a frequency which is 1.5 times the (2,2) multipole.  

We are interested in the relative amplitudes of the different waveform components.  This is most usefully given in terms of the expected \gls*{snr} for each component in a gravitational wave detector. Given a detector with a sensitivity given by the \gls*{psd} $S(f)$, the expected \gls*{snr} of a waveform $h$ is
\begin{equation}\label{eq:rho_h}
    \rho_{h} := | h | = \sqrt{ (h | h)}
\end{equation}
where the inner product, $(a | b)$, between two time-series $a(t)$ and $b(t)$ is defined as
\begin{equation}\label{eq:inner_product}
    (a | b) = 4 \, \mathrm{Re} \int df 
    \frac{\tilde{a}(f) \tilde{b}^{\ast}(f)}{S(f)}
\end{equation}
and $\tilde{a}(f)$, $\tilde{b}(f)$ are the Fourier transforms of $a(t)$ and $b(t)$, respectively.  

We denote the amplitude of the $h_{22,0}$ component as $\sigma$ (consistent with e.g. \cite{Allen:2005fk}), and define the amplitudes of the other components \textit{relative} to this.  Therefore, we have
\begin{align}\label{eq:wf_amps}
| h_{k}| &=: \sigma \alpha_{k}
\end{align}
where, by definition $\alpha_{22,0} = 1$, 
\begin{align}
    \alpha_{22,1} = \bar{b} \quad \mathrm{and} \quad
\alpha_{33,0} = \alpha_{33} \, .
\end{align}
The quantity $\bar{b}$ is the average value of the $b = \tan(\beta/2)$ over the observed waveform, and this gives the relative amplitude of the precession harmonic.  The amplitude of the higher multipole is used to define $\alpha_{33}$.  As discussed in detail in \cite{Mills:2020thr, Fairhurst:2019_2harm}, both of these quantities are generally significantly less than unity.  In obtaining the waveform as given in Equation (\ref{eq:leading_prec_hm}), we have neglected terms which are of order $b^{2}$ or $b \, \alpha_{33}$.  In addition, we have neglected terms of order $\alpha_{\ell m}$ for $(\ell, m) \neq (3, 3)$.  In some regions of parameter space, the (4, 4) multipole can be more significant than the (3, 3).  While we don't consider that case in this paper, the results in the following sections could easily be re-derived for an alternative higher-multipole.

\begin{figure*}
    \centering\includegraphics[width=\linewidth]{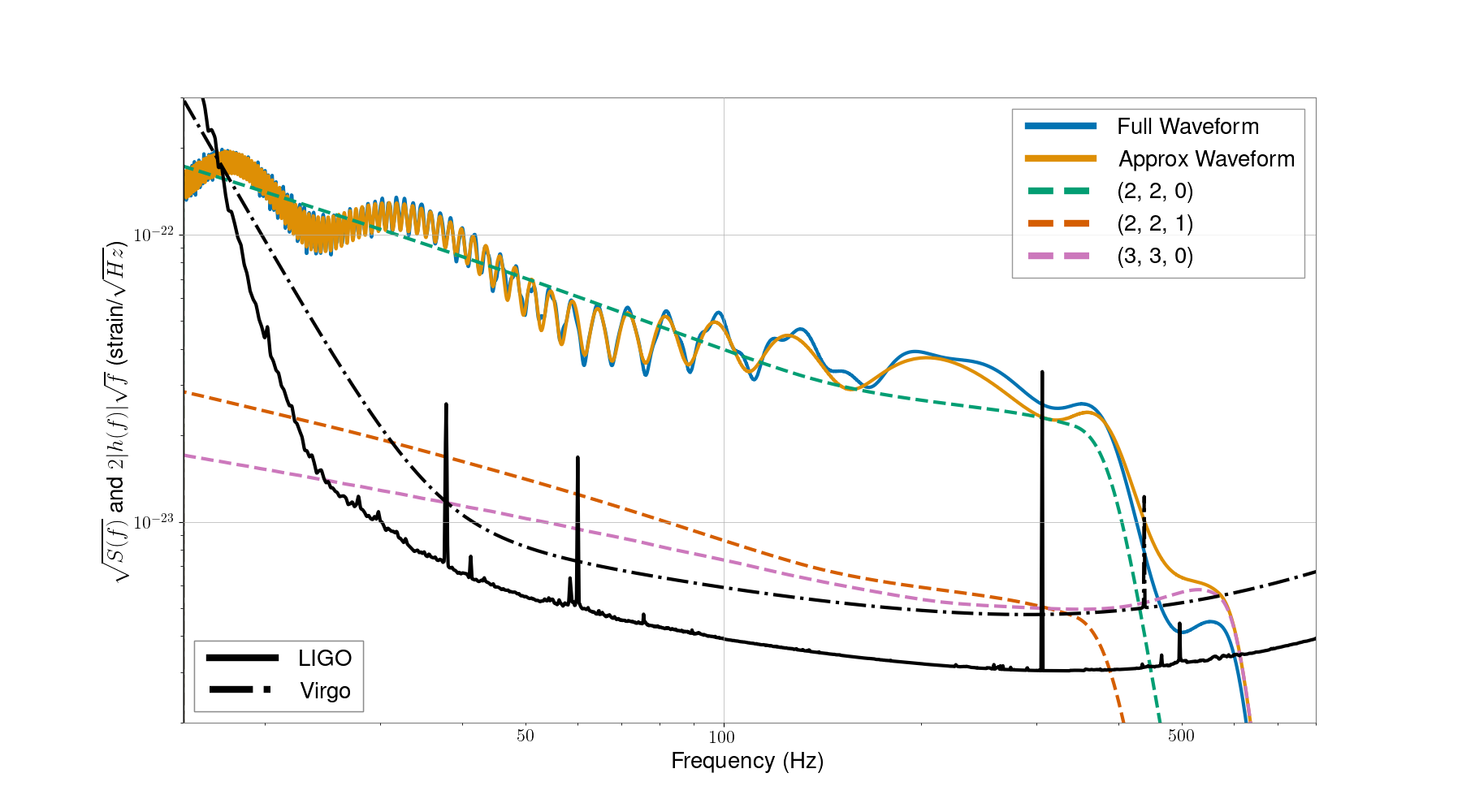}
    \caption{The gravitational waveform emitted by a $40-10 M_{\odot}$ black hole binary, with aligned spin of $0.5$ and in-plane spin of $0.4$ on the larger black hole and zero spin on the smaller, oriented at an angle $\theta_{JN} = 0.6$.  In the plot we show the full waveform and, in addition, the leading contributions to the waveform: the (2, 2, 0) component which is the leading contribution to the (2, 2) multipole, the (2, 2, 1) component which is the leading -order precession correction to the (2, 2) multipole and the (3, 3, 0) component which is the leading contribution to the (3, 3) multipole.  The (2, 2) multipole is the most significant and the (3, 3) is the next largest. In addition, we show the expected detector sensitivities for the advanced LIGO and Virgo observatories for the fourth gravitational-wave observing run (O4)~\cite{O4PSD}.
    }
    \label{fig:waveform_modes}
\end{figure*}

In Figure \ref{fig:waveform_modes} we show the contribution of the different waveform components to the full waveform.  It is clear that the (2, 2, 0) waveform is dominant, while the (2, 2, 1) and (3, 3, 0) waveforms have similar amplitudes which are significantly smaller than the (2, 2, 0).  Furthermore, these three components provide an excellent approximation to the full waveform.  The overall network \gls*{snr} of the signal is set to 25 which givens \glspl*{snr} of 14.5, 18.6 and 8.3 in H1, L1 and V1 respectively.  This signal has a network \gls*{snr} of 4.4 in both the (3, 3, 0) and (2, 2, 1) waveform components.  Finally, we can calculate the overlap between the full waveform and the approximate waveform,
\begin{equation}
    \mathrm{O} = 
    \frac{(h | h^{\prime})}{|h| |h^{\prime}|}
\end{equation}
[Note that in the above, we do not maximize the overlap over phase or time but require them to be identical in the full waveform and the approximate one].  The overlap between the full waveform and the (2, 2, 0) component is 0.965, while the overlap with the 3-component waveform is 0.997.  Thus, our approximate waveform is only distinguishable from the full waveform at a \gls*{snr} of over 30, which is larger than any observed binary black hole \gls*{snr} observed in O1-O3 \cite{LIGOScientific:2021djp} (see e.g. \cite{Lindblom:2008cm, Baird:2012cu} and Section \ref{sec:approx_pe} for details of the distinguishability criteria).

\subsection{Orthogonalization of Waveform Components}
\label{ssec:wf_orth}

In Section \ref{sec:approx_pe}, we argue that identifying power in the leading precession or higher multipole waveform components can play a crucial role in improving parameter estimates, as only a subset of the parameter space will be consistent with the observation of additional features.  In addition, non-observation allows us to exclude regions of parameter space which predict the presence of an observable feature.   So far, we have made the simplifying assumption that the different waveform components, specifically the precession harmonics and higher multipole waveforms, are orthogonal to the leading (2, 2, 0) waveform.  In many cases, the assumption of orthogonality between waveform components is reasonable, particularly between the (2, 2, 0) waveform and the higher multipoles \cite{Mills:2020thr}.  However, this does break down in certain regions of parameter space, most notably at higher masses where there are fewer waveform cycles in the observable band of the detector \cite{Fairhurst:2019_2harm}.  

To obtain \gls*{snr} in higher multipoles or precession which is orthogonal to the leading waveform component, we must project the waveform component for the mode $k$ onto the space orthogonal to the leading mode.  Since the relative phase between the waveform components depends upon the physical parameters of the system it is a free parameter.  Hence the projection of mode $k$ which is orthogonal to the leading component (with an arbitrary phase) is
\begin{equation}\label{eq:h_perp}
    h_{k, \perp} = h_{k} 
     - \frac{(h_{o} | h_{k})}
     {|h_{o}|^{2}} h_{o}
     - \frac{(e^{i\pi/2} h_{o} | h_{k})}
     {|h_{o}|^{2}}
      e^{i\pi/2}h_{o} \, ,
\end{equation}
where the waveform $e^{i\pi/2} h_{o}$ is simply $h_{o}$ rotated through $90^{\circ}$.
Let us define the complex overlap between the waveform mode $k$ and the dominant mode as
\begin{equation}
    o_k = 
     \frac{(h_{o} | h_{k})}
     {|h_{o}| |h_{k}|}
     + i \frac{(e^{i\pi/2} h_{o} | h_{k})}
     {|h_{o}| |h_{k}|}
\end{equation}
then the orthogonal \gls*{snr} in the mode $k$ is reduced by a factor $\sqrt{1 - |o(k)|^{2}}$,
\begin{equation}
    \rho_{k, \perp} = \rho_{k} \sqrt{1 - |o_k|^{2}} \, .
\end{equation}
In addition, the total \gls*{snr} in the signal is
\begin{equation}
    \rho^{2} = \rho_{o}^{2} + \sum_{k \neq o} \left\{ \rho_{k}^{2} + 2 \,\mathrm{Re}[o_k]\rho_{o} \rho_{k}  \right\} \,.   
\end{equation}
The overlap between the two modes will always lead to a reduction in the perpendicular \gls*{snr} of the mode $k$.  However, the total \gls*{snr} can be increased or decreased depending upon the phase of the overlap between the two signal components.  See \cite{Fairhurst:2019_2harm} for a discussion of this in the context of precession. 
 \subsection{The two gravitational-wave polarizations}
\label{ssec:wf_network}

We have, in Equation (\ref{eq:leading_prec_hm}), presented an expression for the gravitational waveform emitted by a precessing binary, restricting to the leading-order precession effects as well as the most significant signal multipoles. We observe a hierarchical decomposition of the signal in terms of the precession parameter $\bar{b}$ and higher-multipole amplitude $\alpha_{33}$.
It is tempting to identify the viewing angle, encoded in $\tau = \tan(\theta_{JN}/2)$, as an additional expansion parameter, and separate terms in Equation (\ref{eq:leading_prec_hm}) which appear with higher powers of $\tau$. However, for a single detector, it is not possible to distinguish the two gravitational-wave polarizations encoded in $h_{k}$ and $h_{k}^{\ast}$. When we generalize to a network of detectors, $\tau$ does become an appropriate expansion parameter.

In Equation (\ref{eq:h_det}) we gave the observed signal in a gravitational-wave detector, as a function of both the gravitational-wave signal $h$ and the detector's antenna response $F$. In many cases, including here, it is more natural to work with the left and right \textit{circular} polarizations of the gravitational wave. The detector response to the circular polarizations is $F$ and $F^{\ast}$, respectively, and the observed waveform can be expressed as
\begin{equation}
	h_{X}(t) = \mathrm{Re}
	\left[ \sum_{k \in \textrm{modes}} h_{k} 
	\left[ F^{X} \mathcal{A}^{R}_{k} 
    +  (F^{X})^{\ast} \mathcal{A}^{L}_{k} 
    \right] \right] \, .
\end{equation}
where, for our purposes, the sum over modes is restricted to (2, 2, 0), (2, 2, 1) and (3, 3, 0). The amplitudes of these modes can be read off from Equation (\ref{eq:leading_prec_hm}) as
\begin{align*}\label{eq:circular_amps}
\mathcal{A}_{22,0}^{R} &= \frac{d_{o}}{d_L} \frac{e^{2i\phi_{o}} }{(1 + \tau^{2})^2} , 
&
\mathcal{A}_{22,0}^{L} &= \frac{d_{o}}{d_L} \frac{\tau^{4} e^{2i\phi_{o}}}{(1 + \tau^{2})^2} , 
\\
\mathcal{A}_{22,1}^{R} &= \frac{d_{o}}{d_L} \frac{ 4 \tau e^{2i\phi_{o}} }{(1 + \tau^{2})^2} , 
&
\mathcal{A}_{22,1}^{L} &= \frac{d_{o}}{d_L} \frac{ -4 \tau^{3} e^{2i\phi_{o}}}{(1 + \tau^{2})^2} , \\
\mathcal{A}_{33,0}^{R} &= \frac{d_{o}}{d_L} \frac{4 \tau e^{3i\phi_{o}}}{(1 + \tau^{2})^3} ,
&
\mathcal{A}_{33,0}^{L} &= \frac{d_{o}}{d_L} \frac{4 \tau^{5} e^{3i\phi_{o}}}{(1 + \tau^{2})^3} . \numberthis \\
\end{align*}

When expressing the waveform amplitudes in terms of left and right circular polarizations, they separate in powers of $\tau = \tan(\theta_{JN}/2)$, so that the waveform is right circularly polarized at $\tau = 0$ (a face-on signal), and $\mathcal{A}^{L}_{k} = 0$. It is therefore tempting to introduce $\tau$ as an expansion parameter, in a similar way to $b$ and $\alpha_{33}$. However, the analogy is not exact.  The astrophysical population of binaries is expected to be randomly oriented, so that $\cos \theta_{JN}$ is uniformly distributed between $-1$ and $1$. Thus, there will be systems taking all possible values of $\tau$, including edge-on systems for which $\tau \approx 1$ and face-off systems for which $\tau \rightarrow \infty$. For a face-away signal, it is natural to change to a coordinate
\begin{equation}
  \gamma := \cot\left( \frac{\theta_{JN}}{2}\right) = \tan\left( \frac{\pi - \theta_{JN}}{2} \right)
\end{equation}
so that $\gamma = 1/\tau$, and $\gamma = 0$ corresponds to a left circularly polarized waveform. Therefore, we can always perform an expansion in the smaller of $\tau$ and $\gamma$.\footnote{In what follows, we will restrict to $\tau \le 1$, with the understanding that the calculation can be easily repeated for $\tau > 1$ by switching to the variable $\gamma$.}
Next, we note that the amplitude of gravitational wave emission in the (2, 2, 0) component is strongest for $\tau \approx 0$ (and $\gamma \approx 0$) . Consequently, there will be an observation bias towards sources which are close to face-on or face-away \cite{Schutz:2011tw}, and indeed the majority of signals are expected to be observed with $\theta_{JN} < 60^{\circ}$, for which $\tau = 1/\sqrt{3}$ --- in which case the amplitude of the second circular polarization is reduced by a factor of 9. Thus, it is reasonable to include the orientation angle as our third expansion parameter, and keep only leading terms in $\tau$. 

Since a single gravitational wave interferometer is only sensitive to one polarization of the signal, we require a network of detectors\footnote{Alternatively a single gravitational wave observatory comprising multiple interferometers, such as the Einstein Telescope, can be used to infer the polarization content}to measure the polarization content of the signal. Therefore, we extend our analysis to a gravitational wave detector network. To do so, we define the multi-detector inner product as the sum over individual detector contributions
\begin{equation}
  \mathbf{(a | b)} := \sum_{X \in \mathrm{dets}} (a_{X} | b_{X} )_{X} \, ,
\end{equation}
where the subscript $X$ on the inner product denotes the fact that the \gls*{psd} varies between detectors. The expected network \gls*{snr} is simply the quadrature sum over detectors of the individual detector \glspl*{snr}.  The expected \gls*{snr} of each polarization and each waveform component in the network of detectors is
\begin{align}\label{eq:snr_det}
  (\rho_{k}^{R, L})^{2} = 
  \boldsymbol{\sigma}^{2} |\boldsymbol{F}|^{2} \alpha_{k}^{2} |\mathcal{A}_{k}^{R, L}|^{2} \, ,
\end{align}
where $\boldsymbol{\sigma}^{2}$ and $|\mathbf{F}|^{2}$ denote vector dot products over the space of detectors. Thus, the network \gls*{snr} depends upon the detector sensitivities, $\sigma$, the network response $\mathbf{F}$, the relative significance of the waveform component, $\alpha_{k}$, and its overall amplitude, $\mathcal{A}_{k}^{R, L}$.

To measure the left circular polarization, we are interested in the power orthogonal to the right circular polarization. Therefore, we need to calculate the overlap between the two circular polarizations. 
To do so, it is convenient to introduce the concept of the \textit{dominant polarization} frame \cite{Klimenko:2011hz, Harry:2010fr}. The detector response functions $\mathbf{F}$ depend upon the unknown polarization angle of the source. In many cases, it is more convenient to fix a preferred polarization frame when considering the network response and then include the polarization angle $\psi$ in the description of the waveform. To this end, we introduce the weighted network response
\begin{equation}\label{eq:w}
  \mathbf{w} = \boldsymbol{\sigma} \mathbf{F} e^{- 2i \psi} \, ,
\end{equation}
where $\psi$ is the polarization angle. Then $\mathbf{w}$ is simply a vector describing the sensitivity of each detector to the gravitational wave signal, as encoded by the product of the detector's sensitivity, $\boldsymbol{\sigma}$ and the response to the gravitational wave, $\mathbf{F}$. 
We fix the polarization angle by working in the dominant polarization frame for which the network is maximally sensitive to the $+$ polarization and, consequently, minimally sensitive to the $\times$ polarization \cite{Klimenko:2011hz, Harry:2010fr}. In the dominant polarization frame, $\mathbf{w}$ satisfies
\begin{equation}
  \mathbf{w_{+} \cdot w_{\times}} = 0 
  \quad \Leftrightarrow \quad
   \mathbf{w \cdot w} = \mathbf{w^{\ast} \cdot w^{\ast}}
\end{equation}

We characterize the network by its overall sensitivity to the dominant polarization, which is given by $|\mathbf{w}_{+}|$. The sensitivity to the second polarization is $|\mathbf{w}_{\times}|$. Following \cite{Usman:2018imj}, we define the relative sensitivity between to $+$ and $\times$ linear polarizations which is given by
\begin{equation}\label{eq:alpha_net}
  \alpha_{\mathrm{net}} = 
  \frac{|\mathbf{w}_{\times}|}{|\mathbf{w}_{+}|} \, .
\end{equation}

In many cases, the detector network is significantly more sensitive to a single polarization of gravitational waves. For example \cite{Klimenko:2011hz, Usman:2018imj}, the typical sensitivity to the second polarization for the advanced LIGO-Virgo network is 0.3. As additional detectors are added to the network, both the overall sky coverage and sensitivity to the second polarization improves.

In general, the two circular polarizations are not orthogonal. The complex overlap between the left and right circular polarizations is
\begin{align}
  o_{L} &= \frac{\mathbf{ ( F h_{k} | 
  F^{\ast} h_{k} ) }}{{|\mathbf{F h_{k}|^{2}}}} + 
  \frac{i \mathbf{(F h_{k} e^{i\pi/2}| 
  F^{\ast} h_{k} )}}{{|\mathbf{F h_{k}|^{2}}}}
  \nonumber \\
  & = 
  \left[\frac{1 - \alpha_{\mathrm{net}}^{2}}{1 + \alpha_{\mathrm{net}}^{2}} \right]
  (\cos 4\psi + i \sin 4 \psi) \, ,
\end{align}
where, to obtain the result, we have used the form of $\mathbf{w}$ given in Equation (\ref{eq:w}) as well as the definition of $\alpha_{\mathrm{net}}$ from Equation (\ref{eq:alpha_net}). 
For a single detector or network sensitive to only one polarization ($\alpha_{\mathrm{net}} = 0$), the two polarizations are completely degenerate, as expected. For a network with equal sensitivity to the two polarizations ($\alpha_{\mathrm{net}} = 1$), the two circular polarizations are orthogonal. For a typical signal in the advanced LIGO-Virgo network, $\alpha_{\mathrm{net}} \approx 0.3$ and the overlap between the two circular polarizations is $o_{L} \approx 0.9$. This places us in a very different situation than for precession and higher multipoles where the overlap is typically small. 

When attempting to identify the presence of the second circular polarization, we must identify the power that is orthogonal to the leading polarization. This is obtained by projecting a left-circular signal onto the space orthogonal to that spanned by right-circular signals, in the same way as we did in Equation (\ref{eq:h_perp}) to obtain $h_{k, \perp}$.  The fact that there is significant overlap between the polarizations has two major impacts. First, the observable power in the second polarization is significantly reduced, 
\begin{align}
  \rho_{L, \perp}& =
  \rho_{L} 
  \sqrt{1 - |o_L|^{2}} 
  \nonumber \\
  & = \left[\frac{2\alpha_{\mathrm{net}}}{1 + \alpha_{\mathrm{net}}^{2}} \right] 
  \rho_{L} \, .
\end{align}  
Second, the fact that the overlap is large means that the overall power in the signal can vary considerably as the polarization angle changes. Specifically, the network \gls*{snr} is
\begin{align}\label{eq:rho_net}
  \rho_{\mathrm{net}}^{2} 
  = \mathbf{(h | h)} 
	 = \boldsymbol{\sigma}^2 |\mathbf{F}|^{2} 
 & \sum_{k} 
	 \alpha_{k}^{2}
  \Biggl\{
	 (\mathcal{A}^{L}_{k})^{2} + 
	 (\mathcal{A}^{R}_{k})^{2} 
	 \\
	 & 
	 +2 \mathcal{A}^{L}_{k} 
	 \mathcal{A}^{R}_{k} 
 \left[\frac{1 - \alpha_{\mathrm{net}}^{2}}{1 + \alpha_{\mathrm{net}}^{2}}\right] 
 \cos 4\psi
	 \Biggr\} \, .
 \nonumber
\end{align}
As before, we have made the approximation that $\alpha_{k}$ is the same for each detector in the network. This is a reasonable approximation --- the overall sensitivity of the detectors is captured by $\sigma$. Provided the \textit{shape} of the \gls*{psd} is similar between detectors, then the relative importance of the different waveform components will also be similar.

\subsection{The observed waveform}
\label{ssec:final_waveform}

We have now identified three expansion parameters which enable us to identify the dominant contribution to the waveform, and the leading sub-dominant contributions. In particular, we parametrize the precession contribution through $b = \tan(\beta/2)$, the higher multipoles through $\alpha_{\ell m}$, their amplitude relative to the $(2, 2)$ mode, and the second polarization through the binary orientation $\tau = \tan(\theta_{JN/2})$.

Keeping only the leading terms and the most significant sub-leading terms in each of these parameters, we obtain the waveform as
\begin{align}\label{eq:final_waveform}
  h &\approx
  \frac{d_{o}}{d_{L}} \left[
  \frac{e^{2i(\phi_o + \psi)}}{(1 + \tau^{2})^{2}} h_{22,0}
  + \frac{\tau^{4} e^{2i(\phi_o - \psi)}}{(1 + \tau^{2})^{2}} h_{22,0}^{\ast}
  \right.
  \nonumber \\
  &
  \quad \left.+ 
  \frac{4\tau e^{2i(\phi_o + \psi)}}{(1 + \tau^{2})^{2}} h_{22,1} 
  + 
  \frac{4 \tau
  e^{3i\phi_o + 2i\psi }}{(1 + \tau^{2})^{3}} 
  h_{33,0}
  \right]
  \, .
\end{align}
Here, we have chosen to absorb the (unknown) polarization angle into the expression for the waveform, and subsequently work in a fixed polarization frame.
The waveform is comprised of four terms. The first is the dominant contribution --- it is the right circularly polarized waveform for the leading contribution of the (2,2) mode. The other contributions are all sub-dominant in different ways. The second term is the left-circularly polarized contribution. This is down-weighted by the fourth power of $\tau = \tan (\theta/2) \le 1$ and, additionally, the observable power is reduced due to the significant overlap between the left and right circular polarized signals. The third term is the leading precession correction which is down-weighted by the precession amplitude $b = \tan(\beta/2)$ as well as $\tau$. The final term is the most-significant higher multipole contribution to the waveform, which is down-weighted by the reduced amplitude of the higher multipole, encoded in $\alpha_{33}$. We note that the formalism is equally applicable to predominantly left-circularly polarized signals, for which we simply convert to $\gamma = \cot(\theta_{JN}/2)$ and cases where a different multipole, for example (4,4), is the second most significant.

The expected \gls*{snr} in the leading mode is given by
\begin{equation}\label{eq:rho_o}
   \rho_{o} = \frac{d_{o}}{d_{L}}
  \frac{ |\boldsymbol{\sigma} \mathbf{F}| }
   {(1 + \tau^{2})^{2}} \, .
\end{equation}
The expected \gls*{snr} in the different sub-dominant waveform components are
\begin{align}
   \rho_{L} &= \rho_{o} 
   \left[\frac{2 \alpha_{\mathrm{net}} \tau^{4}}{1 + \alpha_{\mathrm{net}}^{2}}\right]
   \, \label{eq:rho_2pol}
   \\
   \rho_{33} &= \rho_{o}
   \, \left[\frac{2 \alpha_{33} \tau}{1 + \tau^{2} }\right]
   \label{eq:rho_33}
   \\
   \rho_{\mathrm{prec}} &= \rho_{o}
   \left[ 4 \bar{b} \tau  \right]\, \label{eq:rho_prec}
\end{align}
The values of $b$, $\alpha_{33}$ and $\alpha_{\mathrm{net}}$ are given, for different binary parameters and network configurations in \cite{Fairhurst:2019_2harm}, \cite{Mills:2020thr} and \cite{Usman:2018imj} respectively.
In the expressions above, we have explicitly included the overlap between the two polarizations, which gives rise to the $2 \alpha_{\mathrm{net}}/(1 + \alpha_{\mathrm{net}}^{2})$ factor, but neglected the overlaps with the precession and higher mode signals as these are typically smaller. Additionally, we have not included the impact of the additional waveform components on the recovered \gls*{snr} consistent with the leading mode. \section{Extracting astrophysical parameters from an observed signal}
\label{sec:approx_pe}

Once a gravitational wave signal has been observed, the challenge is to extract the astrophysical parameters of the source.  Over the years, numerous methods have been developed to obtain parameter estimates, typically using Bayesian methods and densely sampling the parameter space~\cite{Veitch:2014wba,Biwer:2018osg,Ashton:2018jfp,Lange:2018pyp,Romero-Shaw:2020owr,Ashton:2021anp,Smith:2019ucc,Gabbard:2019rde,Green:2020hst,Dax:2021tsq,Dax:2022pxd,Shen:2019vep,Green:2020dnx,Williams:2021qyt,Tiwari:2023mzf}.  Here, we take a different approach and attempt to identify the  \textit{primary} feature that enables the measurement of one, or a combination of, the astrophysical parameters of a source.  By doing so, we build up an intuitive understanding of how the parameters can be extracted from the observed waveform.  We consider a quasi-circular (non-eccentric) binary described by fifteen parameters: the masses, $m_{1}$ and $m_{2}$, of the black holes, their spins, denoted by the vectors $\mathbf{S}_{1}$ and $\mathbf{S}_{2}$, the orientation of the binary given by the phase $\phi_{o}$, inclination angle $\theta_{JN}$ and source polarization $\psi$, and the location of the system relative to the earth, given by the sky location $(\alpha, \delta)$, distance $d_{L}$ and arrival time $t_{o}$ of the signal.

In Section \ref{sec:waveform}, we demonstrated that the waveform can be expressed as a dominant component, with three sub-dominant contributions which are the leading order corrections.  
The signal can be decomposed in terms of three expansion parameters: 
$\tau = \tan\theta_{JN}/2$, where $\theta_{JN}$ is the angle between the binary's total angular momentum and the line of sight; $b = \tan{\beta/2}$, where $\beta$ is the opening angle between the orbital and total angular momentum and governs the observability of precession effects; $\alpha_{\ell, m}$, the sensitivity of the network to the leading subdominant multipole, $(\ell, m)$, relative to the (2, 2) multipole.
By expressing the waveform in this way, we are able to identify the impact that the observability (or otherwise) of these three features has upon the parameter recovery.  In some cases, the next-order corrections will be observable but, as we argue later, they are unlikely to dramatically impact the parameter recovery.  Each of the features above enables us to break a degeneracy between parameters.  Identification of the next-order terms merely allows us to refine the measurements, but doesn't lead to an ability to measure entirely new features.

Throughout this section, we provide examples using signal shown in Figure \ref{fig:waveform_modes}: the gravitational waveform emitted during the merger of a $40 M_{\odot}$ -- $10 M_{\odot}$ black hole binary with aligned spins of $0.5$ on the larger black hole and $0$ on the smaller black hole.  The system is placed at a distance to provide a total network \gls*{snr} of 25 in the LIGO-Virgo at projected O4 sensitivity~\cite{O4PSD}.  We vary the distance, viewing angle and in-plane spins of the system to investigate the impact of observability of precession, higher multipoles and the second circular polarization on parameter recovery.

\subsection{Parameter measurement accuracy}
\label{ssec:match}

Given an observed signal $s$ in a gravitational wave detector, the likelihood ratio for the data to contain a signal $h$, rather than just Gaussian noise $n$, is given by
\begin{equation}\label{eq:like}
    \Lambda(\vec{\lambda}, s) = \frac{\exp
    [-\tfrac{1}{2} (s - h(\vec{\lambda}) | s - h(\vec{\lambda}))]}{\exp[-\tfrac{1}{2} (s | s )]} \, ,
\end{equation}
where the inner product $(a|b)$ was previously introduced in Equation (\ref{eq:inner_product}).  For a network of detectors, the likelihood is the product of likelihoods for individual detectors.  
In order to calculate the posterior distribution for the parameters $\vec{\lambda}$, we use Bayes formula
\begin{equation}
    p(\vec{\lambda} | s ) \propto \Lambda(\vec{\lambda}, s) \pi(\vec{\lambda}) \, ,
\end{equation}
where $\pi(\vec{\lambda})$ is the prior distribution on the parameters $\vec{\lambda}$. 

Let us return to Eq.~(\ref{eq:like}) and consider the case where the data $s$ is well approximated by a signal with parameters $\hat{\lambda}$, in the sense that $(s | h(\vec{\lambda}) ) \approx (h(\hat{\lambda}) | h(\vec{\lambda}) )$.  When considering simulated signals, it is natural to identify $h(\hat{\lambda})$ with the known signal.  More generally, as discussed in Section \ref{ssec:max_snr}, it is straightforward to identify the peak likelihood and identify this as $\hat{\lambda}$.  Then, the peak likelihood is related to the expected \gls*{snr}, defined in Equation (\ref{eq:rho_h}), through
\begin{equation}
    \Lambda(\hat{\lambda}, s) \approx \exp[ \rho_{h}^{2}/2] \, .
\end{equation}
We can also explore the features of the likelihood as the parameters $\vec{\lambda}$ are varied.  Substituting into Equation (\ref{eq:like}), we see that the likelihood depends upon the difference between the waveforms
\begin{equation}
    \Lambda(\vec{\lambda}| \hat{\lambda}) \propto 
    \exp \left[-\tfrac{1}{2} | h(\hat{\lambda}) - h(\vec{\lambda})|^{2} \right]
\end{equation}
Next, following \cite{Baird:2012cu}, we relate this to the similarity between waveforms, as characterized by the match, $M$, defined as
\begin{equation}
    M(h_{1}, h_{2}) = \max_{dt, d\phi} \frac{(h_{1} | h_{2})}{|h_{1}| |h_{2}|} \, ,
\end{equation}
where $dt$ and $d\phi$ denote the time and phase offset between the two waveforms, respectively.
In particular, we can re-express the likelihood as
\begin{equation}\label{eq:like_match}
    \Lambda(\vec{\lambda}| \hat{\lambda} ) \propto 
    \exp \left[ - \frac{\rho_{h}^{2}}{2} \left(1 -  M^{2}\right) \right] \, .
\end{equation}

As a final approximation, we assume that the match varies quadratically with the difference in parameters $\delta \vec{\lambda}$.  This is true at leading order, but at low \gls*{snr} and for large dimensional parameter spaces, this approximation breaks down \cite{Vallisneri:2007ev}.  Nonetheless, the quadratic approximation can be useful in investigating the properties of the signal.  To use it, we construct the waveform metric, $g_{ab}$, defined through
\begin{equation}
    M(\delta\vec{\lambda}) \approx 1 - g_{ab} \delta\lambda^{a}\delta\lambda^{b} 
    \quad \mathrm{where} \quad
    \delta \vec{\lambda} = \hat{\lambda} - \vec{\lambda} \, .
\end{equation}
Then, the likelihood is approximated as
\begin{equation}\label{eq:like_metric}
    \Lambda(\vec{\lambda}| \hat{\lambda} ) \propto 
    \exp \left[ - \rho_{h}^{2} ( g_{ab} \delta \lambda^{a} \delta \lambda^{b} ) \right] \, .
\end{equation}
The eigenvectors and eigenvalues of the matrix $g_{ab}$ provide, respectively, the principal directions and the measurement accuracy in these directions.  Specifically, the approximate contour containing a fraction $p$ of the posterior distribution, for a signal with \gls*{snr} $\rho_{h}$ is given by
\begin{equation}
    g_{ab} \delta \lambda^{a} \delta \lambda^{b} = \frac{\chi^{2}_{k}(1-p)}{2 \rho_{h}^{2}}
\end{equation}
where $k$ is the dimension of the parameter space under consideration and $\chi^{2}_{k}(1-p)$ is the chi-square value with $k$ degrees of freedom for which there is a $(1-p)$ probability of obtaining that value or larger.  We will typically be interested in generating 90\% contours, and will be working in 2, 3 or 4 dimensions, in which case the thresholds are $g_{ab} \delta \lambda^{a} \delta \lambda^{b} \le  2.3,\, 3.1,\, 3.9 \, \rho_{h}^{-2}$ for $k = 2, 3, 4$ respectively.

\subsection{The chirp waveform}
\label{ssec:chirp}

We begin by restricting attention to the dominant component of the waveform, arising from the right circularly polarized,\footnote{As before we assume $\tau < 1$ so that the waveform is preferentially right-circular polarized.  The calculation is equally applicable to left-circular waveforms under the replacement $\tau \rightarrow \gamma$.}
leading contribution to the (2, 2) harmonic and neglect sub-dominant contributions from the second polarization, higher multipoles and precession.  As is well known, the amplitude and phase evolution of the waveform can be used to extract the masses and (aligned-)spins of the black holes \cite{Cutler:1994ys, Hanna:2019ezx}.  

The amplitude and phase evolution of the binary merger waveform is given, at large separations, by the post-Newtonian expansion \cite{Blanchet:2006zz} of Einstein's equations and, close to and at merger, from numerical simulations of binary systems which are combined to provide full models of the gravitational wave signal ~\cite{Bohe:2016gbl,Cotesta:2018fcv,Cotesta:2020qhw, Ossokine:2020kjp,Babak:2016tgq,Pan:2013rra, Husa:2015iqa,Khan:2015jqa,London:2017bcn, Hannam:2013oca,Khan:2018fmp,Khan:2019kot, Varma:2019csw,Varma:2018mmi,Pratten:2020fqn, Garcia-Quiros:2020qpx,Pratten:2020ceb, Estelles:2020osj,Estelles:2020twz,Estelles:2021gvs,Hamilton:2021pkf}.  In the inspiral stage, the leading order evolution of the waveform is given by the chirp mass
\begin{equation}
    \mathcal{M} = \frac{(m_{1} m_{2})^{3/5}}{(m_{1} + m_{2})^{1/5}} = M \eta^{3/5} \, ,
\end{equation}
where $m_{1}$ and $m_{2}$ are the masses of the two components, $M$ is the total mass and $\eta$ the symmetric mass ratio given by
\begin{equation}
    M = m_{1} + m_{2}; \quad
    \eta = \frac{m_{1} m_{2}}{M^{2}} \, .
\end{equation}
Corrections to the phasing arise at subsequent post-Newtonian orders (powers of $v/c$), with the first correction at 1PN depending upon the mass ratio, $\eta$, and the coefficient at 1.5PN order depending also upon the components of the black hole spins which are aligned with the orbital angular momentum.  Consequently, the chirp mass is the best-measured mass parameter, while the mass ratio and binary spins are typically less well constrained.

\begin{figure}
    \centering
    \includegraphics[width=\linewidth]{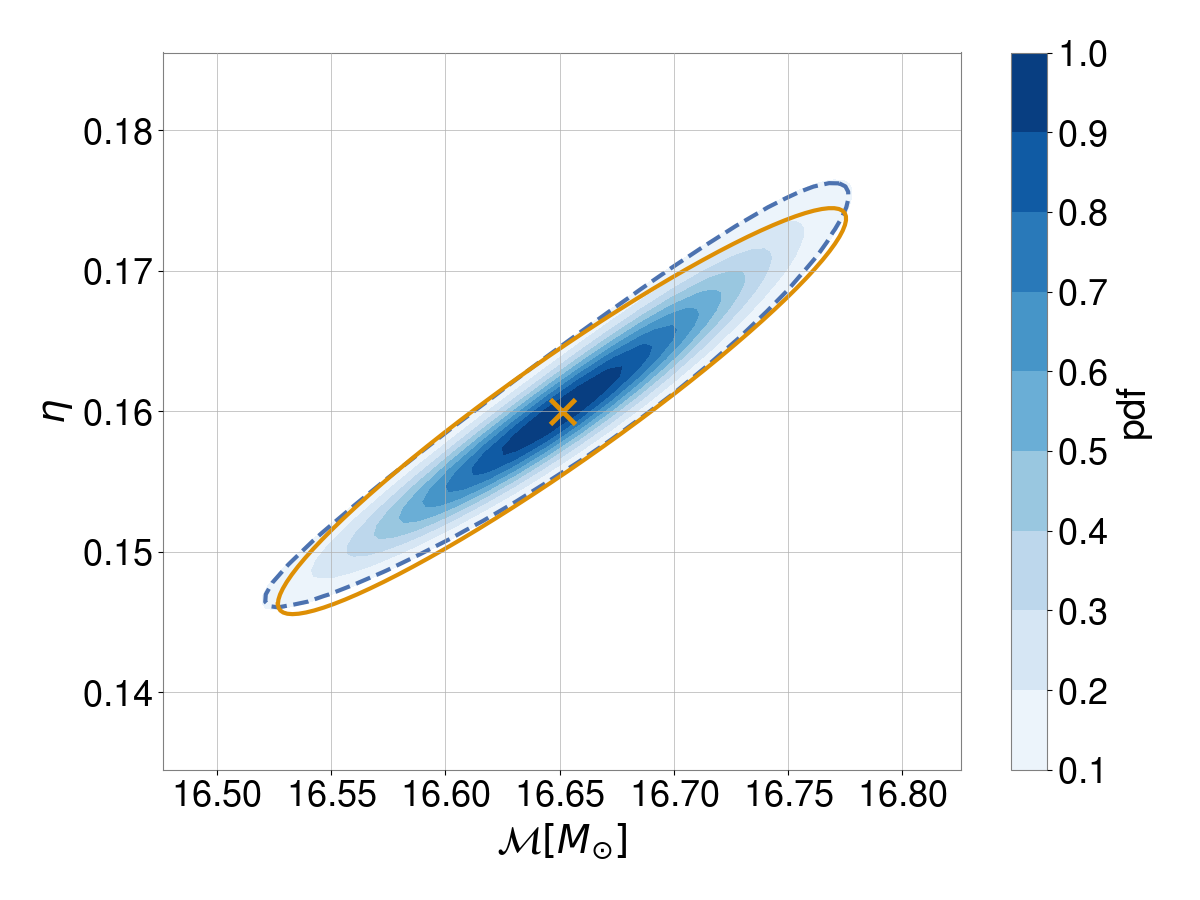}
    \caption{The posterior distribution for chirp mass and mass ratio.  The contours show the posterior \gls*{pdf}, the dashed blue contour shows the 90\% region obtained from the pdf and the solid orange ellipse shows the approximate 90\% region.  In this, and all following plots, the \gls*{pdf} is plotted so that the value at the peak is unity. 
    }
    \label{fig:mchirp_eta}
\end{figure}

In Figure \ref{fig:mchirp_eta}, we show the posterior probability distribution for our fiducial source ($m_{1} = 40 M_{\odot}$ and $m_{2} = 10 M_{\odot}$ with \gls*{snr} 25) when varying only the masses.  The posterior distribution is generated  by calculating the match across the mass space and substituting into Equation (\ref{eq:like_metric}) to obtain the likelihood.  The ellipse represents the approximate 90\% confidence interval obtained from calculating the metric over the two-dimensional mass space.  The metric provides a good approximation to the likelihood, although the fact that the lower probability contours are curved (rather than elliptical) shows that the simple quadratic approximation is beginning to break down.  

\begin{figure}
    \centering
    \includegraphics[width=\linewidth]{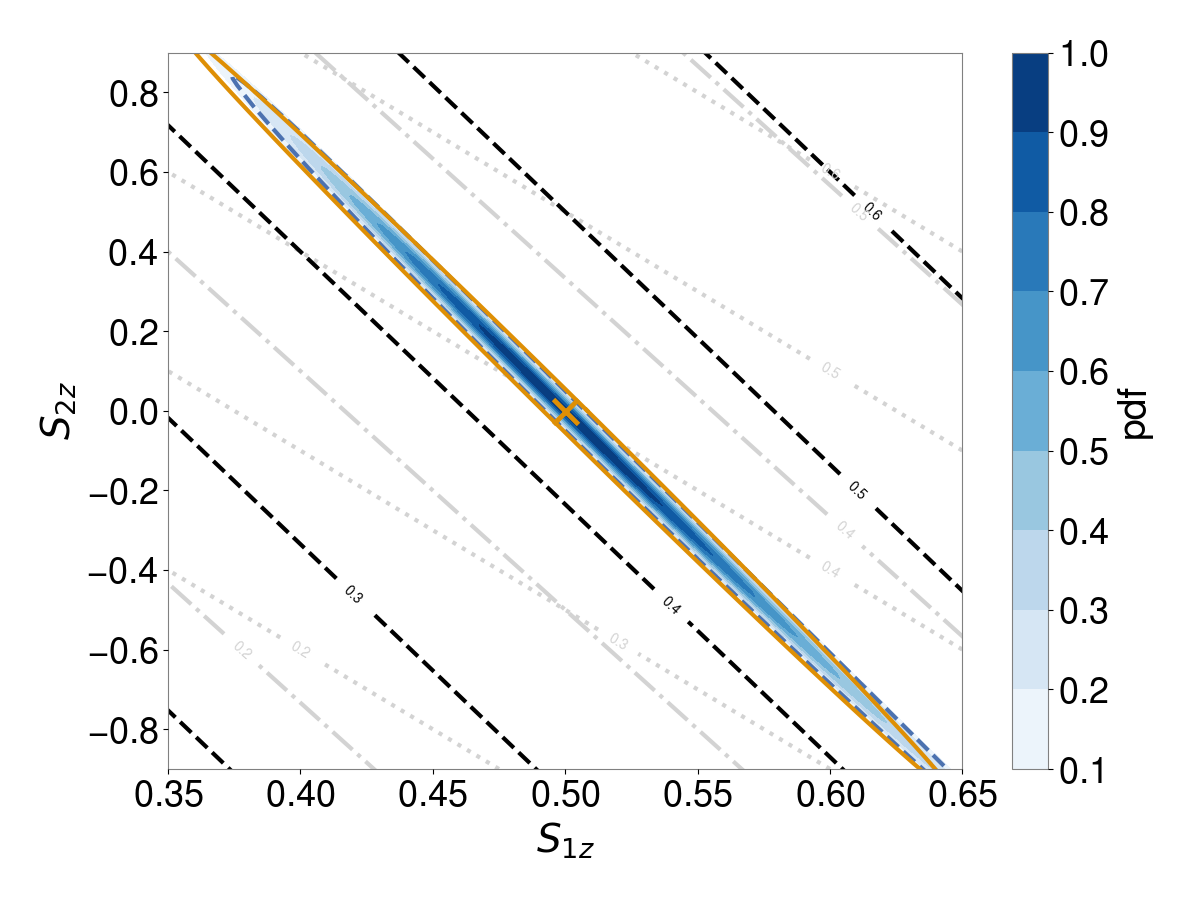}
    \caption{The posterior distribution for the aligned spins, with other parameters kept fixed.  The contours show the posterior \gls*{pdf}, the dashed blue contour shows the 90\% region obtained from the pdf and the solid orange ellipse shows the approximate 90\% region. Lines of constant $\chi_{\mathrm{eff}}$, $\chi_{\mathrm{hu}}$ and $\chi_{\mathrm{align}}$ are plotted as grey dotted lines, grey dash-dotted lines and black dashed lines respectively.}
    \label{fig:aligned_spins}
\end{figure}

In Figure \ref{fig:aligned_spins}, we show the posterior probability distribution for the components of the black hole spins aligned with the orbital angular momentum.  We denote the aligned spin components as
\begin{equation}
    \chi_{1z} = \frac{\mathbf{S_{1} \cdot \hat{L}}}{m_{1}^{2}} 
    \quad \mathrm{and} \quad
    \chi_{2z} = \frac{\mathbf{S_{2} \cdot \hat{L}}}{m_{2}^{2}} \, .  
\end{equation}    
We keep the other parameters of the signal fixed while allowing the aligned spins to vary.  As expected, there is a clear degeneracy between the inferred spins.  And, as with the mass space, the metric approximation provides a good description of the likelihood distribution.  On the figure, we have also plotted lines of constant effective spin, 
\begin{equation}
    \chi_{\mathrm{eff}} = \frac{m_{1} \chi_{1z} + m_{2} \chi_{2z}}{m_{1} + m_{2}} \, ,
\end{equation}
which is typically used to describe a binary's in-plane spin. Since our fiducial system has aligned spin only on the larger black hole, $\chi_{\mathrm{eff}}$ doesn't accurately describe the spin degeneracy, as can be seen in Figure \ref{fig:aligned_spins}.

We therefore use an alternative effective spin parameter throughout the remainder of this paper, which accurately describes the spin-degeneracy shown in Figure \ref{fig:aligned_spins}, and attribute this spin to both black holes equally (i.e. $\chi_{1z} = \chi_{2z} = \chi_{\mathrm{align}}$).\footnote{We do not make this restriction on the simulated signals, only on our parameter recovery.}
We use,
\begin{equation}
    \chi_{\mathrm{align}} = \frac{m_{1}^{\alpha} \chi_{1z} + m_{2}^{\alpha} \chi_{2z}}{m_{1}^{\alpha} + m_{2}^{\alpha}} \, ,
\end{equation}
where $\alpha = \tfrac{4}{3}$. This was chosen since it accurately describes the spin-degeneracy for low mass systems (including the one considered here), it is similar to $\chi_{\mathrm{hu}}$ which describes the number of orbits before merger~\cite{Healy:2018swt}, as can be seen in Figure \ref{fig:aligned_spins}, and it has the nice property that it is equal to $\chi_{1z}$ when $\chi_{1z} = \chi_{2z}$.

\begin{figure*}
    \centering
    \includegraphics[width=\linewidth]{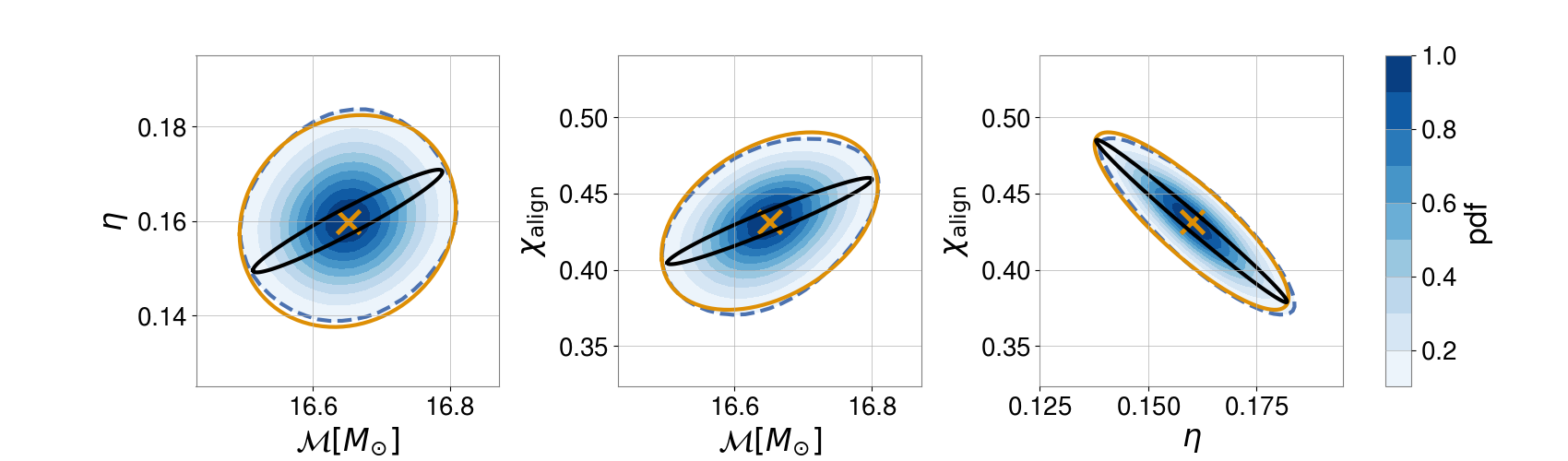}
    \caption{The posterior distribution for the chirp mass, symmetric mass ratio and effective spin.  Each plot shows a two-dimensional slice through the parameter space, where the value of the third parameter is chosen to maximize the value of the likelihood.  The blue dashed contour shows the 90\% region and the orange ellipse shows the approximate 90\% region based on calculating the parameter space metric, as discussed in detail in Section \ref{sec:approx_pe}. The black ellipse shows the region based on fixing the third parameter to equal the simulated value.
    }
    \label{fig:mass_spin_measurement}
\end{figure*}

Having restricted to a single aligned spin parameter, we can calculate the posterior distribution across the remaining three-dimensional parameter space of masses and (aligned-)spins.  Figure \ref{fig:mass_spin_measurement} shows the likelihood on two-dimensional slices through the parameter space.  The likelihood distribution at each point in the $\mathcal{M}$--$\eta$ space is obtained both by maximizing the match over the aligned spin values and evaluating the likelihood using the maximized match.  The distribution for other pairs of parameters is calculated similarly.  In addition, we plot ellipses corresponding to the 90\% contours using either the two-dimensional metric (fixing the third parameter) or a three-dimensional metric projected into the two-dimensional space under consideration.  The three-dimensional metric accurately reproduces the likelihood distribution. The two-dimensional metric significantly underestimates the parameter uncertainties due to correlations between the parameters, particularly mass ratio and spin \cite{Hannam:2013oca}.  For example, in Figure \ref{fig:mchirp_eta}, the symmetric mass ratio is bounded between $\eta \in [0.145, 0.175]$ (mass ratio between 3.5 and 4.5 to 1) while allowing the spin to vary increases the range to $\eta \in [0.125, 0.195]$ (a mass ratio between 3:1 and 6:1). 

In all cases, we observe that the quadratic approximation given by the metric provides a good fit to the posterior distribution.  In particular, the principal directions of the metric match those of the distribution and the surfaces of constant probability are reasonably well-described by ellipses.  However, there are some discrepancies, most notably the shape of contours in the $\mathcal{M}-\chi$ space and the asymmetry of contours in the $\eta-\chi$ space.  

The overall amplitude, phase and time of arrival of the signal also carry physical information.  The amplitude of the observed gravitational wave scales with the mass of the system and is inversely proportional to the distance.  Furthermore, the signal amplitude varies with the orientation of the binary relative to the line of sight, and the binary's location relative to the detector network.  These facts are encoded in our expression for the \gls*{snr} in the leading mode, Eq.~(\ref{eq:rho_o}), which we restate here:
\begin{equation*}
     \rho_{o} =  \frac{d_{o}}{d_{L}}
     \frac{|\mathbf{F} \boldsymbol{\sigma}| }
     {(1 + \tau^{2})^{2}} \, , 
\end{equation*}
where, as before $|\mathbf{F} \boldsymbol{\sigma}|$ denotes a sum over the detectors in the network. The variation of masses and spins impacts $\sigma$, the location of the source impacts $F$, while the orientation is encoded in $\tau$.  All of these combine to limit the accuracy with which the distance to the source can be measured.  The phase of the \gls*{snr} also provides information about the signal.  In particular, looking at Equation (\ref{eq:final_waveform}), we see that the phase of the waveform is given by $2(\phi + \psi)$.  Thus, the measurement of the \gls*{snr}  phase enables us to determine a combination of coalescence phase and polarization angle.

Finally, we note that the observed gravitational wave signal is redshifted as it travels to the detector.  This reduces the frequency of the waveform by an overall factor of $(1 + z)$.  For black hole binary systems, the frequency content of the observed waveform also scales with the total mass of the binary.  Consequently, when the distance/redshift to the system is not known, we are only able to infer the redshifted mass $\mathcal{M} = \mathcal{M}^{\mathrm{source}} (1 + z)$, and not the source mass.  In the remainder of the paper, all results are shown in terms of the redshifted chirp mass $\mathcal{M}$.

\subsection{Observation in a Network of Detectors}
\label{ssec:network}

In a network of detectors, we independently measure \gls*{snr}, signal phase and time of arrival in each of the detectors. In addition to improving the accuracy of mass and spin measurements by increasing the observed \gls*{snr}, this also enables measurement of the sky location of the source and the second gravitational-wave polarization.

\subsubsection{Source localization}
\label{ssec:loc}

Localization of a transient gravitational-wave source in a network of detectors is primarily achieved through timing: the relative time delays between the observed signal at the different detectors can be inverted to provide a sky region from which the source originates \cite{Fairhurst:2009tc, Fairhurst:2010is}.   The timing accuracy in a single detector is given by $\sigma_{t} = (2\pi \rho \sigma_{f})^{-1}$ where $\sigma_{f}$ is the frequency bandwidth of the system and $\rho$ is the \gls*{snr}. With two detectors, timing alone can restrict the source to a ring on the sky, although it is often possible to identify a most likely region on the ring based upon the relative amplitude and phase of the signal in the two detectors \cite{Singer:2015ema, Nitz:2017svb}.  With three detectors, the source can be localized by timing to two regions of the sky, one above and the other below the plane formed by the three detectors.  Observation in three detectors also affords three measurements of the signal amplitude and phase.  In many cases, consistency with a gravitational-wave signal comprised of two polarizations enables the identification of a preferred sky region \cite{Singer:2015ema, Fairhurst:2017mvj}.

\begin{figure}
    \centering
    \includegraphics[width=0.98\linewidth]{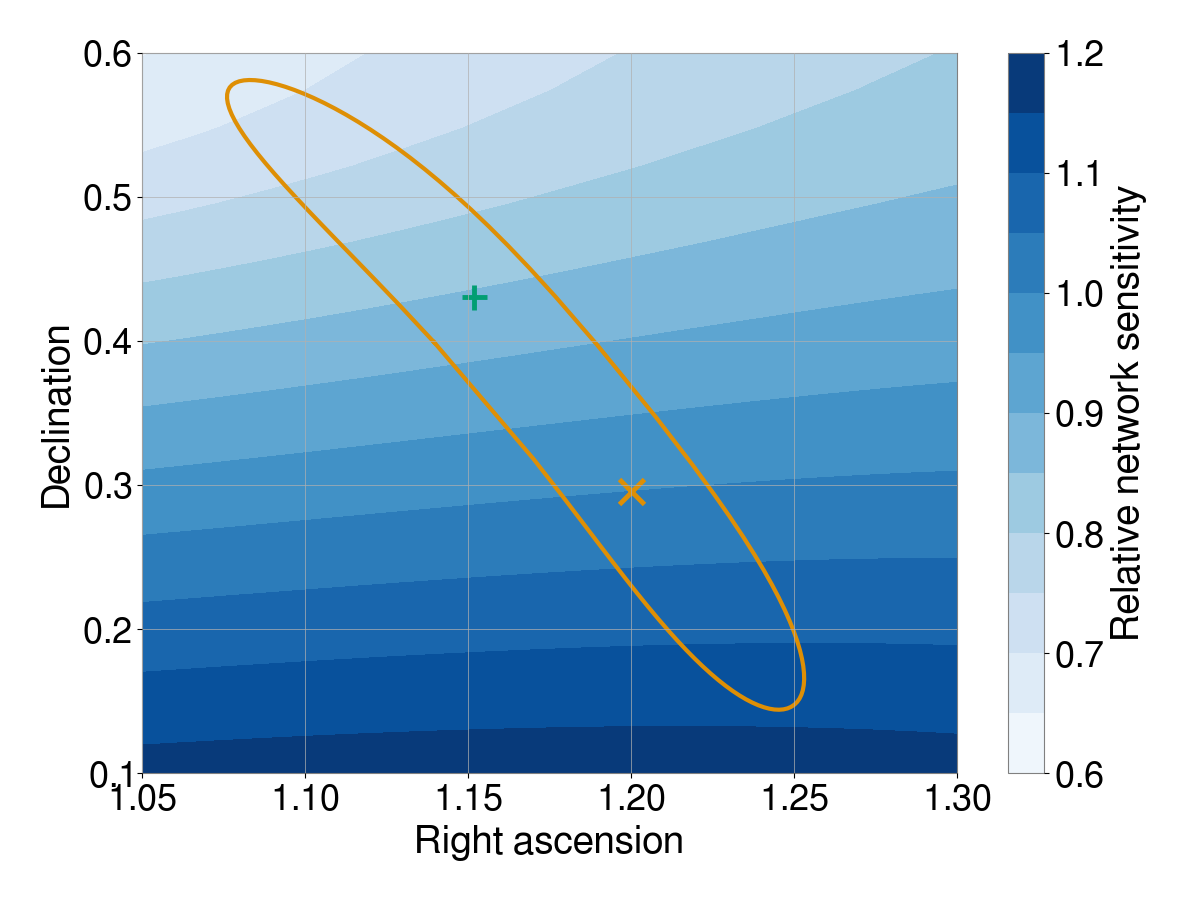}
    \includegraphics[width=0.98\linewidth]{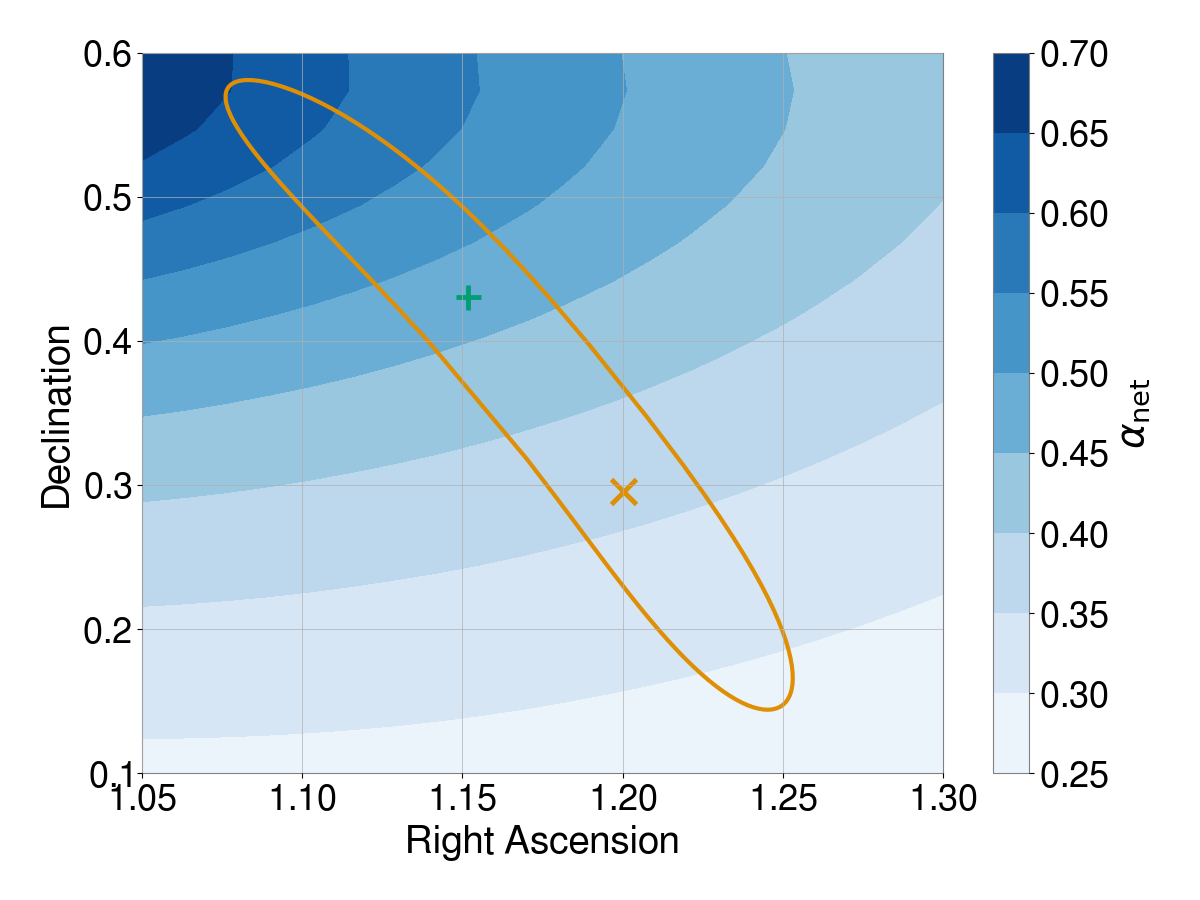}
    \caption{The 90\% source localization ellipse for our simulated signal.  The localization is obtained using time of arrival and consistency in amplitude and phase of the gravitational wave signal in the detectors \cite{Fairhurst:2017mvj}.  The orange $\times$ shows location of the simulated signal, the green + shows the ``mirror'' location (reflection in the plane of the detectors).  The top panel shows the localization ellipse overlaid on the network sensitivity while the bottom shows the sensitivity to the second polarization, $\alpha_{\mathrm{net}}$.
    }
    \label{fig:localization}
\end{figure}

For the purposes of this paper, we are not interested in a detailed discussion of source localization.  Nonetheless, uncertainty in the sky location of the source will impact the inference of other parameters.  Most notably, an accurate estimate of the detector response $\boldsymbol{F}$ in Equation (\ref{eq:rho_o}) enables an accurate measurement of the distance to the source $d_{L}$.  Similarly, the sensitivity of the network to the second polarization, encoded in $\alpha_{\mathrm{net}}$ determines the expected \gls*{snr} in the left circular polarization or, inverting the problem, measurement of $\alpha_{\mathrm{net}}$ and the \gls*{snr} in the left circular polarization enables inference of the binary orientation from Eq.~(\ref{eq:rho_2pol}), as we discuss in detail below.

Figure \ref{fig:localization} shows the inferred localization for a simulated signal. The  signal has a total \gls*{snr} of 25 in the LIGO-Virgo network, using the expected sensitivity of the fourth observing run \cite{KAGRA:2013rdx}.  For the given sky location and detector sensitivity, that translates to a \gls*{snr} of 14.7 in H1, 18.5 in L1 and 8.3 in V1. Using only timing information, the event is localizes to an area of $135 \deg^{2}$.  By requiring a consistent amplitude and phase across the detectors, the localization area improved to $80 \deg^{2}$.  This localization is poorer than achieved for some high \gls*{snr} events, such as GW170817 \cite{abbott2017gw170817} and GW190814 \cite{Abbott:2020khf} due to the selected sky location, close to the plane of the detectors. For this event, large changes in sky position led to relatively small impact on the time of arrival of the source.

\subsubsection{Binary orientation and distance}
\label{ssec:2nd_pol}
 
\begin{figure*}
    \centering
    \includegraphics[width=0.49\linewidth]{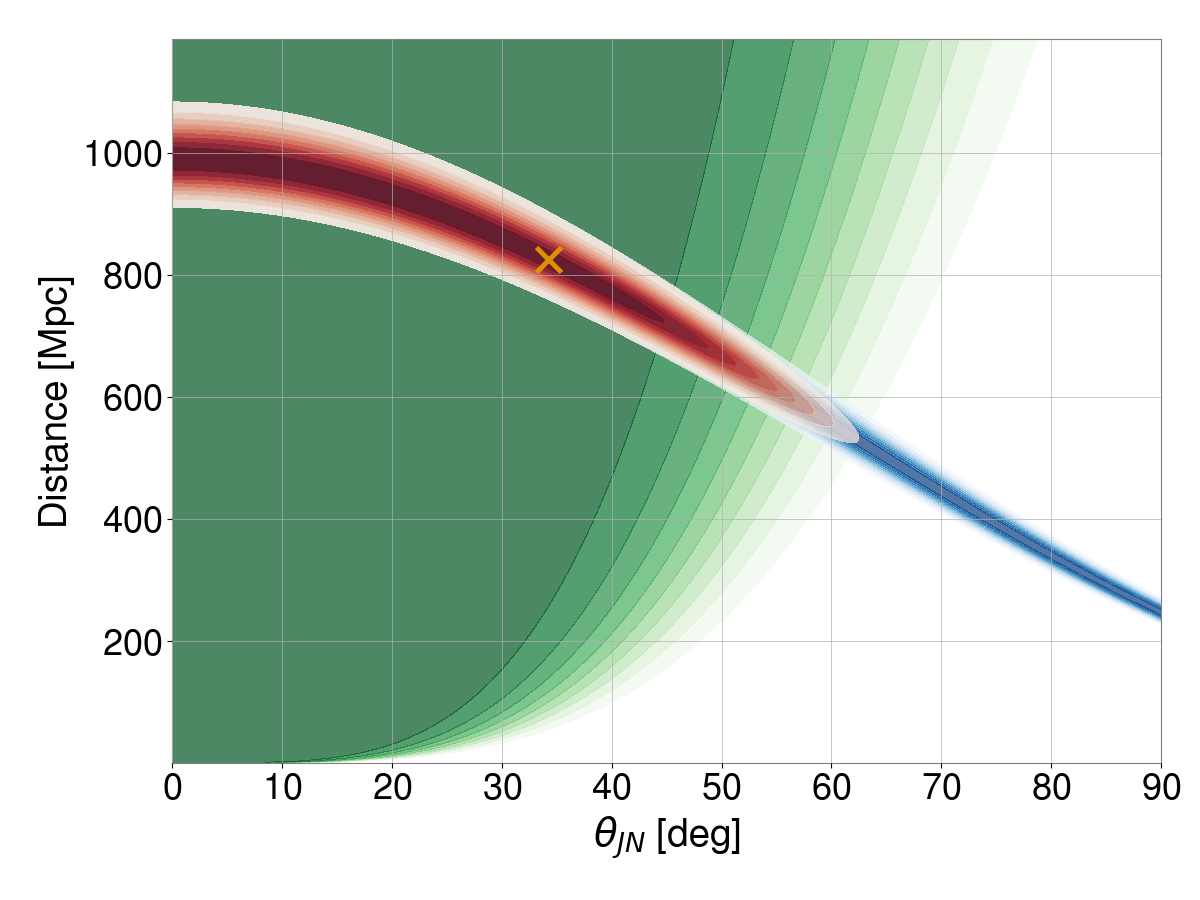}
    \includegraphics[width=0.49\linewidth]{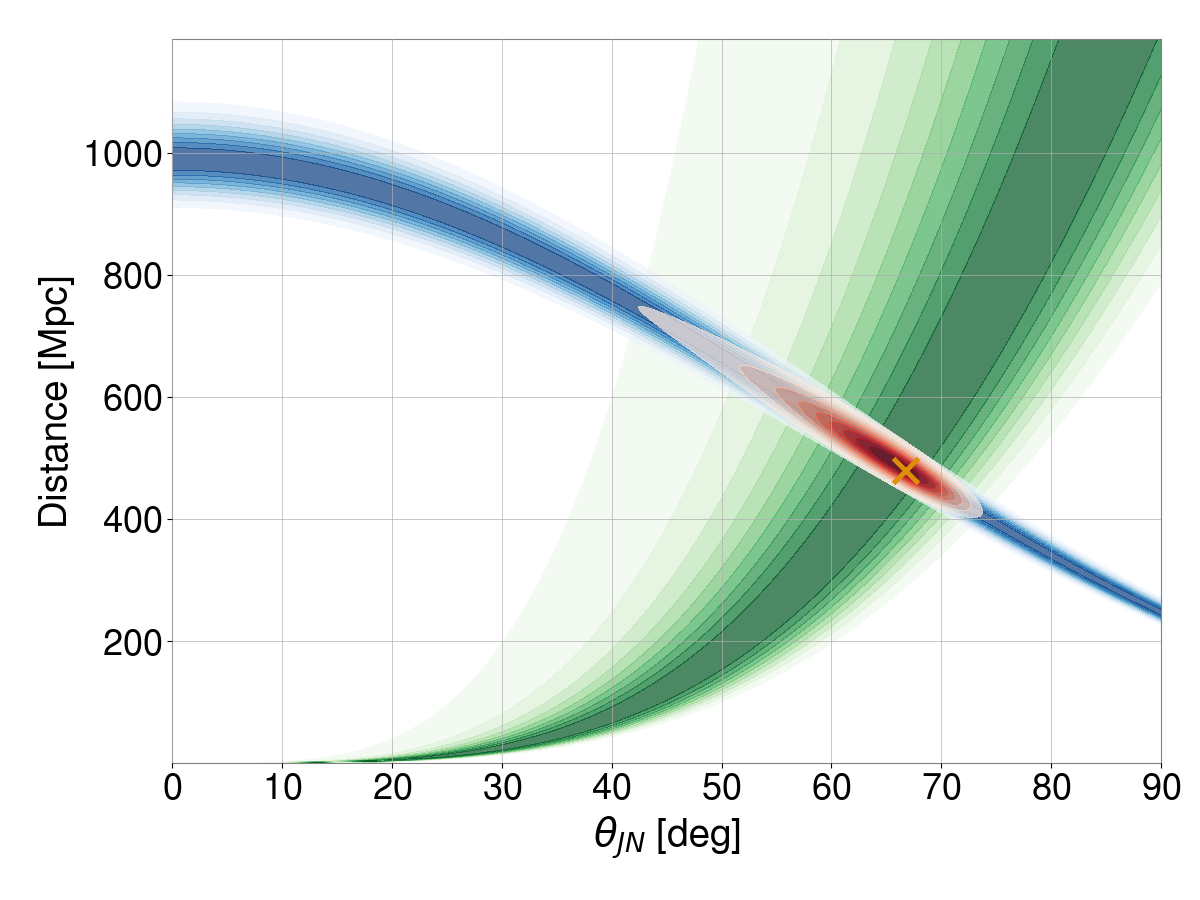}
    \caption{Inference of the distance to and orientation of the binary based upon the observed \gls*{snr} in the two circular polarizations.  The blue contours show the probability density from the observed \gls*{snr} in the right polarization, the green contours show the \gls*{pdf} from \gls*{snr} in the left polarization and the red regions show the combined \gls*{pdf}.  For both examples, the \gls*{snr} of the source is 25.  For the left figure, the binary is inclined at $35^{\circ}$, giving an \gls*{snr} of 0.15 in the left circular polarization.  For the right figure, the binary is inclined at $67^{\circ}$ giving an \gls*{snr} of 3 in the second polarization.
    }
    \label{fig:dist_inc}
\end{figure*}

Observation of a signal in a network of detectors enables the measurement of the second gravitational-wave polarization.  In Section \ref{ssec:wf_network} we obtained expressions for the expected \gls*{snr} in the left and right circular polarizations, with the ratio between them, which depends upon $\alpha_{\mathrm{net}}$ and $\tau$, given in Equation (\ref{eq:rho_2pol}), and repeated below:
\begin{equation*}
   \rho_{L} = \rho_{o} 
   \left[\frac{2 \alpha_{\mathrm{net}} \tau^{4}}{1 + \alpha_{\mathrm{net}}^{2}}\right] \, .
\end{equation*}
For our example signal, the sensitivity to the second polarization is $\alpha_{\mathrm{net}} = 0.35$, which varies between $0.3$ and $0.65$ over the localization region (as shown in Figure \ref{fig:localization}). Thus, sensitivity to the second polarization is reduced by a factor of $2 \alpha_{\mathrm{net}}/(1 + \alpha_{\mathrm{net}}^{2}) \approx 0.6$ relative to the leading polarization.  Then, measurement of $\rho_{o}$ and $\rho_{L}$, coupled with a knowledge of $\alpha_{\mathrm{net}}$ provide an estimate of the binary orientation, encoded in $\tau$.

In Figure \ref{fig:dist_inc}, we show how the binary orientation can be restricted based upon the measurement of the \gls*{snr} in the two polarizations.  As is clear from the equation above, the masses, spins and overall network sensitivity will not impact the estimation of $\tau$.  Nonetheless, they do impact the inferred distance.  Consequently, for simplicity of presentation, we consider the case where the masses, spins and sky location of the source are fixed.  Then the measured \gls*{snr} in the right circular polarization provides a measurement of $(1 + \cos\theta_{JN})^{2} d_{L}^{-1}$.  Similarly, measurement of the \gls*{snr} in the left circular polarization provides a measurement of $(1 - \cos\theta_{JN})^{2} d_{L}^{-1}$. In the figure, we show two example signals, with binaries inclined at  $35^{\circ}$ and $67^{\circ}$ respectively.  For each, we show the region in distance--$\theta_{JN}$ space consistent with the observed \glspl*{snr}.  For an expected \gls*{snr} $\hat{\rho}$ the measured squared \gls*{snr} will be non-centrally $\chi^{2}$ distributed with a non-centrality parameter $\hat{\rho}^{2}$ and two degrees of freedom\cite{Allen:2005fk, Mills:2020thr}.  Thus, for any measured \gls*{snr}, we can infer region in the distance-inclination space that would give an expected \gls*{snr} consistent with the observation.  The fractional uncertainty in \gls*{snr} is proportional to $\rho^{-1}$ and, consequently, the observation of the (lower \gls*{snr}) left-circular polarization provides a significantly weaker constraint than the right-circular polarization.

For the binary at $35^{\circ}$, there is negligible power observable in the second polarization and the binary is consistent with being face-on.  However, the binary orientation cannot be accurately measured and can only be restricted to lie in the range $\theta_{JN} \lesssim 50^{\circ}$.  The system inclined at $67^{\circ}$, has an \gls*{snr} of 3 in the left-circular polarization so that the system is no longer consistent with a circular polarized gravitational wave.  The  binary orientation can now be restricted to lie in the range $40^{\circ} \lesssim \theta_{JN} \lesssim 70^{\circ}$.  In these examples, the likelihood is shown as a function of distance and binary orientation.  As discussed in \cite{Usman:2018imj}, it is more appropriate to use a distance prior which is uniform in (comoving-)volume and an orientation distribution flat in $\cos\theta_{JN}$.  These distributions will add further weight to large distances and face-on systems making it more difficult to identify inclined sources.

Finally, we note that the phase of the second polarization has a different dependence on the polarization angle $\psi$, as can be seen in Eq.~(\ref{eq:final_waveform}).  Thus, observation of both polarizations enables measurement of both the coalescence phase $\phi$ and the polarization $\psi$.

\subsection{Higher order multipoles}
\label{ssec:higher_harm}

As discussed in detail in Section \ref{sec:waveform}, all gravitational wave signals will contain contributions from multipole moments other than the (2, 2).  For the majority of signals, we do not expect to observe these multipoles, as their amplitude will be too small.  However, if it is possible to observe additional harmonics or place limits on the power contained in them, then we can further constrain the range of parameters consistent with an observed signal.  The \gls*{snr} in the (3, 3, 0) waveform component, relative to the (2, 2, 0) component is given in Eq.~(\ref{eq:rho_33}).  It scales linearly with $2\tau(1 + \tau^2)^{-1} = \sin \theta_{JN}$ and $\alpha_{33}$, the relative significance of the (3, 3, 0) harmonic.  The value of $\alpha_{33}$ increases with the mass of the binary and is higher for binaries with more unequal components (see figure 2 in \cite{Mills:2020thr} for details).  Therefore, the (3, 3, 0) component is most significant in unequal mass binaries viewed away from face-on (or face-off).

\begin{figure}
    \centering
    \includegraphics[width=\linewidth]{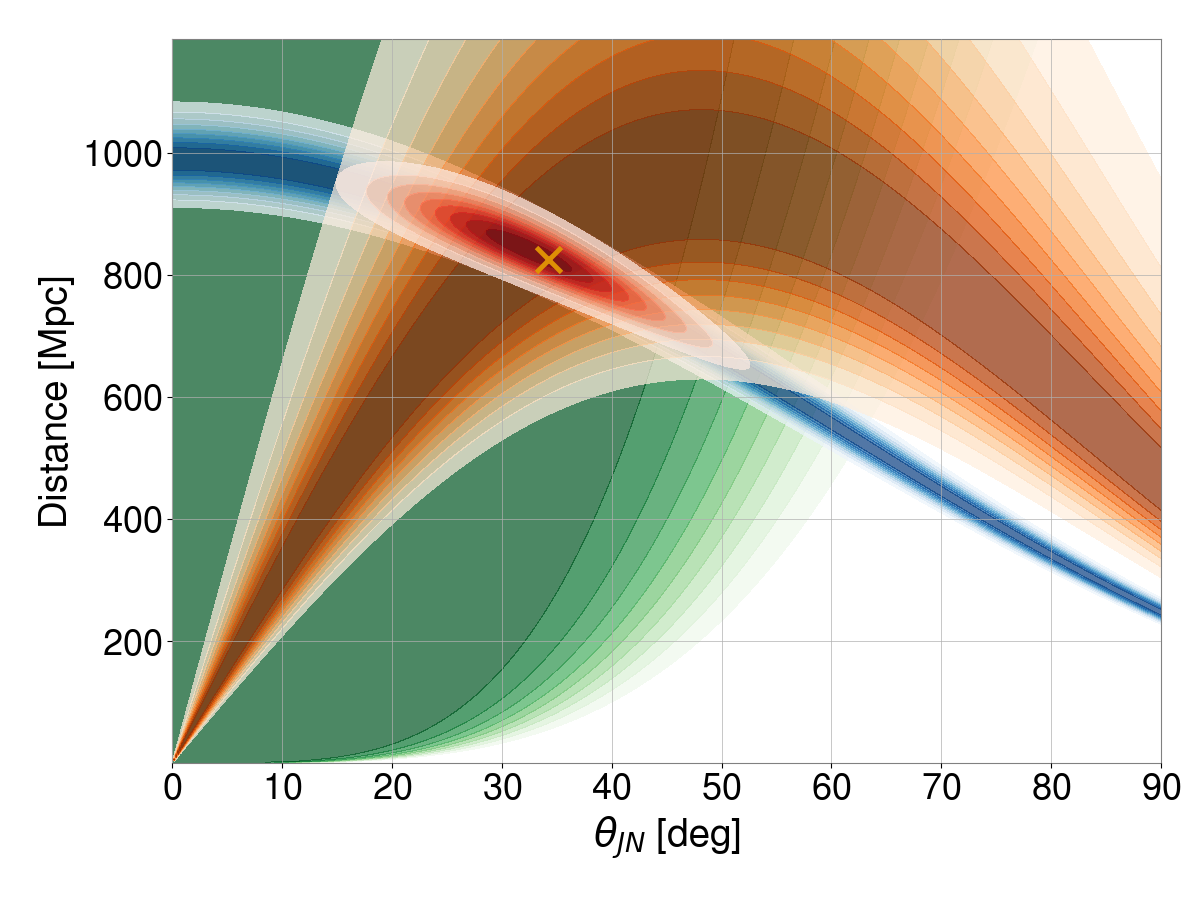}
    \caption{The restriction of the distance to and inclination of the binary based upon the observed \gls*{snr} in the two circular polarizations and the (3, 3, 0) harmonic.  The blue contours show the probability density from the observed \gls*{snr} in the right polarization, the green from power orthogonal to the right polarization, the orange from the (3, 3, 0) harmonic and the red show the combined distribution.  In calculating the (3, 3, 0) harmonic contours, we have kept the mass ratio fixed to the true value, allowing it to vary will broaden this distribution.  The \gls*{snr} of the source is 25, and it is inclined at an angle of $35^{\circ}$, giving an \gls*{snr} of 0.15 in the second polarization and $4.3$ in the (3, 3, 0) harmonic.
    }
    \label{fig:dist_inc_33}
\end{figure}

Given an observed \gls*{snr} in the (3, 3, 0) waveform, we can obtain a region in the distance-inclination plane which is consistent with the observed signal, overlaid on the constraints from the two polarizations of the (2, 2, 0) waveform.   For concreteness, we use the same system as before, a binary with masses of $40 M_{\odot}$ and $10 M_{\odot}$, which is inclined at $\theta_{JN} = 35^{\circ}$.  This gives a \gls*{snr} of 4.4 in the (3, 3, 0) waveform.  In Fig.~\ref{fig:dist_inc_33}, we show how measurement of the \gls*{snr} in the (3, 3, 0) waveform can be used to restrict the distance and orientation of the binary.  Since this is a binary with a significant mass ratio, the (3, 3, 0) waveform plays a much more significant role in determining the orientation of the binary than the second polarization, which has negligible \gls*{snr}.  The observation of the (3, 3, 0) waveform clearly shows that the binary is not face-on, with $\theta_{JN} \gtrsim 15^{\circ}$.  For the events GW190412 and GW190814, it was observation of power in the (3, 3, 0) waveform which enabled measurement of the binary orientation.

\begin{figure}
    \centering
    \includegraphics[width=\linewidth]{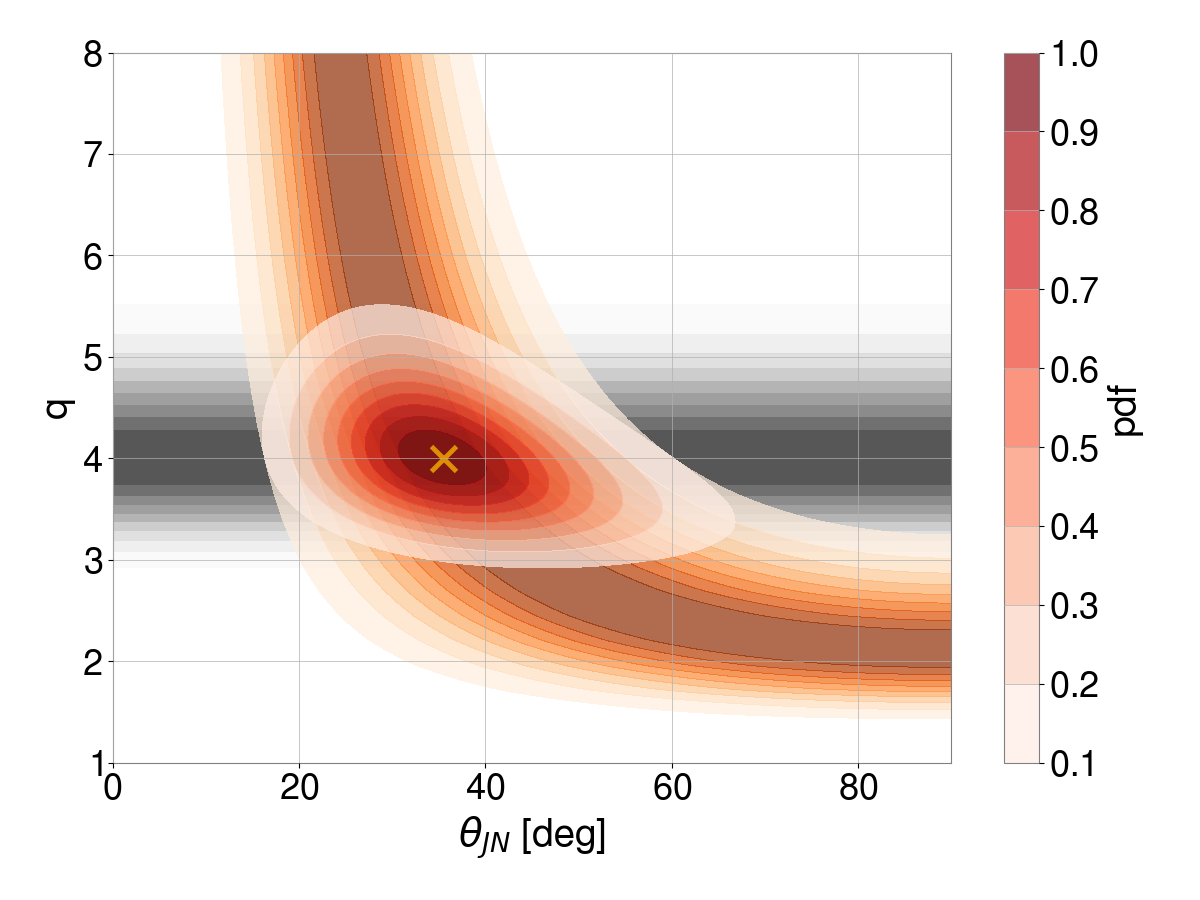}
    \caption{The restriction of the mass ratio to and inclination of the binary based upon the observed \gls*{snr} in the (3, 3, 0) multipole.  The orange band gives the posterior based only on measurement of the (3, 3, 0) multipole, the grey band is the region of mass-ratio space consistent with the (2, 2, 0) waveform and the red contours give the region consistent with both measurements. 
    }
    \label{fig:q_inc_33}
\end{figure}

In Fig.~\ref{fig:q_inc_33}, we show the region in mass-ratio and binary orientation that is consistent with a given observed value of $\rho_{33}$.  The range of $\theta_{JN}$ at $q=4$ corresponds to that in Figure \ref{fig:dist_inc_33}.  However, when we allow mass ratio to vary, the allowed region encompasses close-to-equal-mass systems which are significantly inclined or unequal mass systems which are close-to face on.  Since the mass-ratio is already restricted by the observed leading-order waveform, as discussed in Section \ref{ssec:chirp}, the measurement of the higher multipole \gls*{snr} can be used to restrict the binary orientation, as shown on the figure.  It is straightforward to add additional multipoles to this analysis, and the relative power will typically have a different dependence on $\theta_{JN}$.  However, additional multipoles will likely refine the measurements but probably not significantly improve them.

The measurement of the phase of the (3, 3, 0) waveform can be used to extract measurements of both the signal's polarization and phase angle.  Looking at Equation (\ref{eq:final_waveform}), we see that the phase of the (3, 3, 0) waveform differs from the (2, 2, 0) by the coalescence phase $\phi$.  Thus, observation of both waveform components allows for measurement of the phase and, consequently, also the polarization.

\subsection{Precession}
\label{ssec:prec}

\begin{figure}
    \centering
    \includegraphics[width=\linewidth]{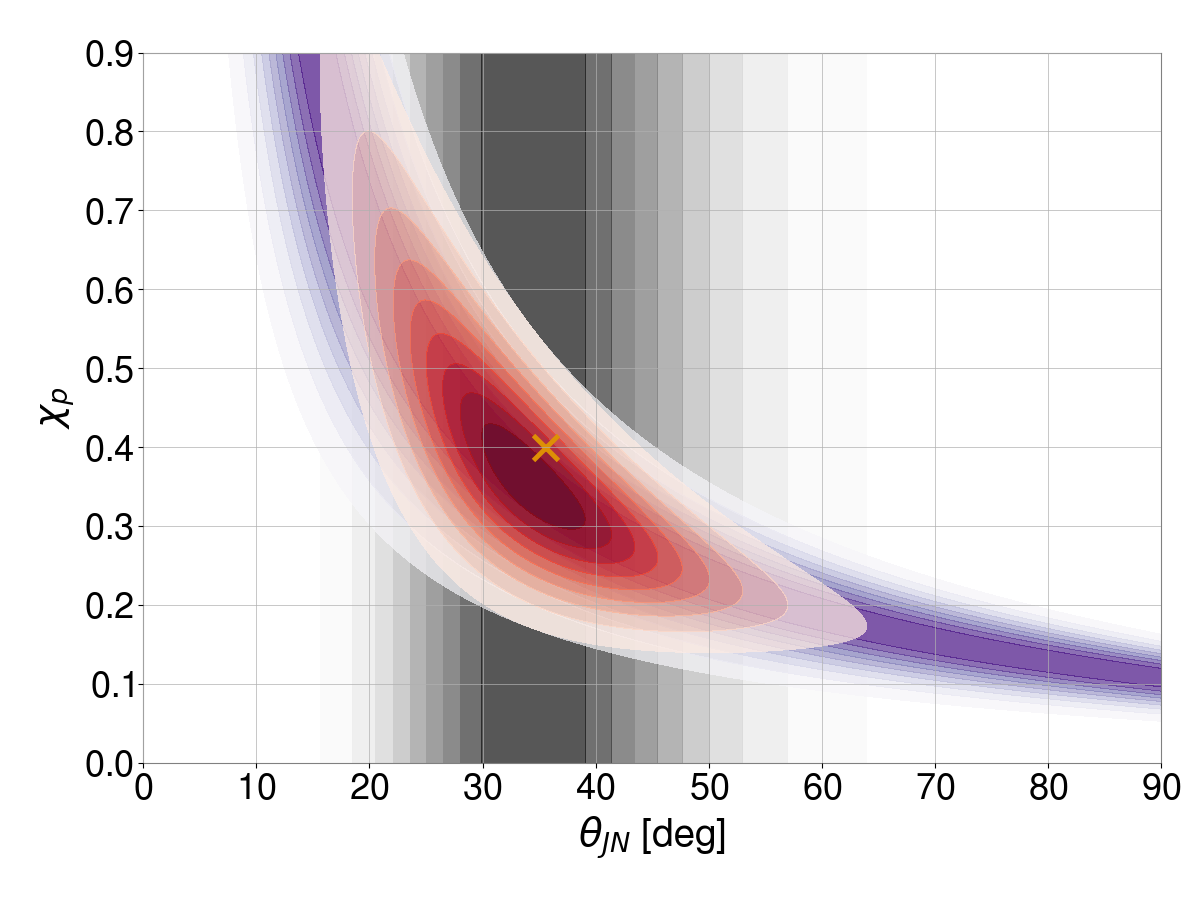}
    \caption{The restriction of the in-plane spin and the binary orientation besed upon the observed \gls*{snr} in precession, $\rho_{p}$.  The purple contours show the probability density from a system with $\chi_{\mathrm{p}} = 0.4$ inclined at $35^{\circ}$, giving an observed \gls*{snr} in precession of $4.0$.  Since the precessing spin is otherwise unconstrained, measurement of $\rho_{p}$ provides minimal restriction to the binary orientation -- the inclination is restricted to be above $15^{\circ}$.  However, if the orientation is already constrained, e.g. from the observation of higher modes, as indicated by the grey band, then the range of permitted values of $\chi_{\mathrm{p}}$ can be significantly reduced, as indicated by the red region.
    }
    \label{fig:chip_inc_prec}
\end{figure}

Black hole spins which are mis-aligned with the orbit leads to precession of the orbital plane \cite{Apostolatos:1994mx} which manifests as amplitude and phase modulations of the signal.  As discussed in Section \ref{ssec:wf_modes}, precession leads to a splitting of the gravitational-wave multipoles.  In particular, the (2, 2) multipole is split into five, where the (2, 2, 0) harmonic is the leading term and the (2, 2, 1) harmonic is the first-order precession correction.  The \gls*{snr} in the (2, 2, 1) precession harmonic, relative to the leading (2, 2, 0) harmonic, is given in Equation (\ref{eq:rho_prec}) as $4\tau \bar{b}$, where $\bar{b}$ is the average value of $b = \tan(\beta/2)$ and $\beta$ is the opening angle between the total and orbital angular momenta. To leading order during the inspiral phase,
\begin{equation}
    \tan{\beta} = \frac{S_{\perp}}{L + S_{\parallel}} \, .
\end{equation}
where $S_{\perp}$ and $S_{\parallel}$ are the perpendicular and parallel components of the spins and $L$ is the orbital angular momentum.  The effective precession spin parameter, $\chi_{p}$, is obtained by averaging the in-plane spins of the system over a precession cycle, so that $S_{\perp} \approx m_{1}^{2} \chi_{p}$.
Thus, measurement of precession \gls*{snr} allows us to infer a combination of the precession spin and binary orientation.  

In Figure \ref{fig:chip_inc_prec} we show the region of $\chi_{\mathrm{p}}$--$\theta_{JN}$ parameter space which is consistent with a precession \gls*{snr} of 4.\footnote{In the figure, the simulated value is slightly offset from the centre of the inferred region.  This is due to the fact that there is a small amount of power in the left-circular polarization in the (2, 2, 1) harmonic which we do not account for when inferring $\chi_{p}$ and $\theta_{JN}$ from $\rho_{p}$}
In this case, the signal can clearly be identified as precessing and, therefore, both the in-plane spin and binary orientation are bounded away from zero.  Nonetheless, there remains a broad range of parameter space consistent with the observation, ranging from maximal in-plane spins for binaries inclined at $15^{\circ}$ to edge-on binaries with $\chi_{\mathrm{p}} \approx 0.1$. 

In contrast to the left-circular polarization and higher multipoles, the observation of precession is unlikely to lead to a significant improvement in the measurement of the binary orientation, distance or phase.  The reason for this is that the amplitude and phase of the precession \gls*{snr} depend upon the in-plane spins, encoded in the precession spin $\chi_{\mathrm{p}}$ and the precession phase $\alpha_{o}$.  Thus, measurement of the precession \gls*{snr} enables a measurement of $\chi_{\mathrm{p}}$ while measurement of the phase of the precession \gls*{snr} enables us to extract the precession phase $\alpha_{o}$, as can be seen from Equation (\ref{eq:final_waveform}).\footnote{See \cite{Hoy:2021aaa} for a discussion of the measured precession \gls*{snr} in existing gravitational-wave events and \cite{Roulet:2022kot} for a discussion of the measurability of the precession phase.}
Of course, if the binary orientation has already been restricted through observation of a second gravitational wave polarization or higher multipoles, then this can lead to significant restrictions on $\chi_{\mathrm{p}}$, as is shown in Figure \ref{fig:chip_inc_prec}.  As an example, the event GW190814 \cite{Abbott:2020khf} was observed to \gls*{snr} $\approx 6$ in the (3, 3) multipole but minimal \gls*{snr} in precession, enabling the inference of a very low spin for the primary.

\begin{figure}
    \centering
    \includegraphics[width=\linewidth]{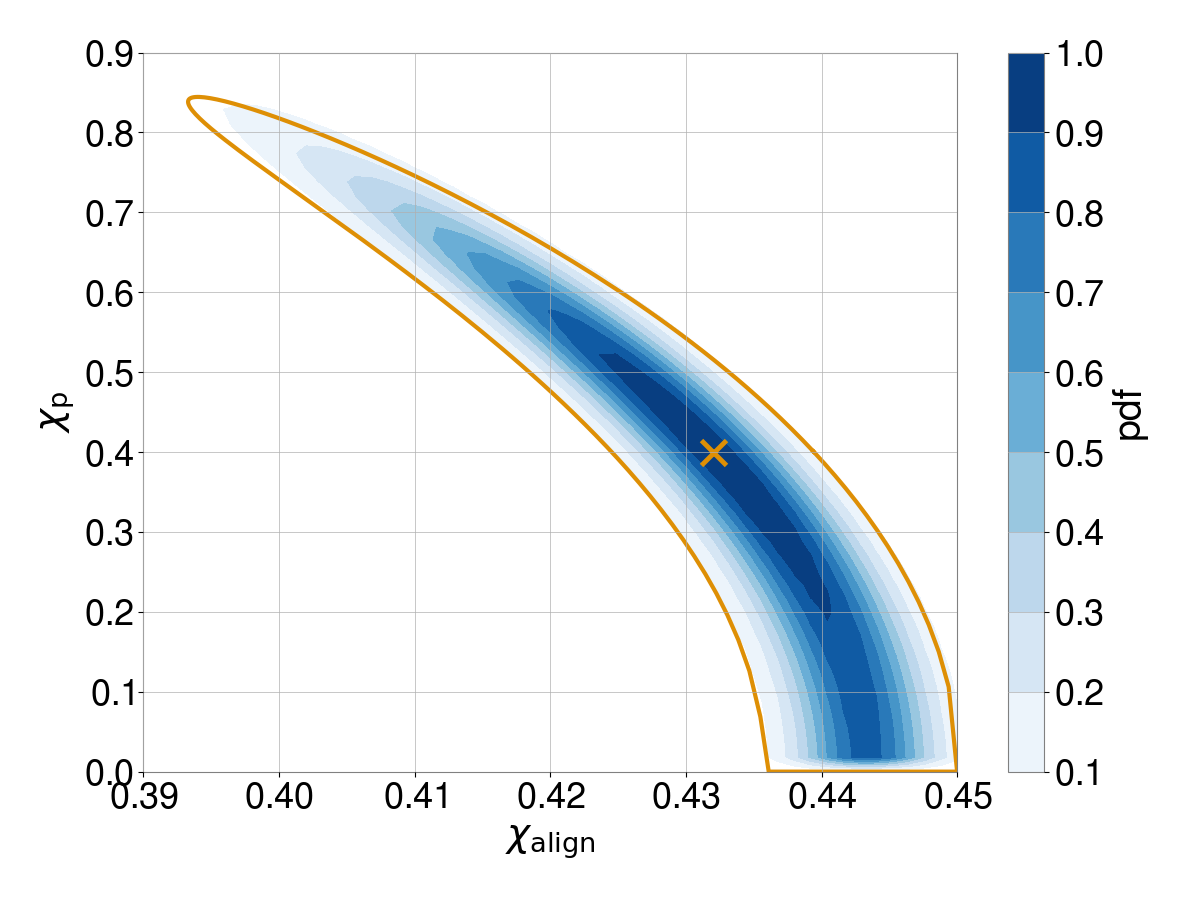}
    \caption{The posterior distribution for the precessing and aligned spins of the system, keeping other parameters fixed.  The contours are generated by calculating the match, the orange contour is generated in the $\chi_{\mathrm{p}}^{2}$--$\chi_{\mathrm{align}}$ space, but plotted against $\chi_{\mathrm{p}}$.  
    }
    \label{fig:chip_chi_eff}
\end{figure}

In-plane spins will impact the phase evolution of the (2, 2, 0) waveform component.  This can be seen in Eq.~(\ref{ap_eq:h22}), where the waveform acquires an additional phase of $\exp[2i(\alpha - \alpha_{o} - \epsilon)]$ relative to the non-precessing signal. Under the approximation that the opening angle $\beta$ is small and approximately constant, we can simplify Equation (\ref{eq:epsilon}) to obtain
\begin{equation}
    \epsilon \approx \alpha (1 -\beta^{2}/2) + \mathrm{const} \, .
\end{equation}
Therefore, the phase is approximately quadratic in $\beta$.  Furthermore, for small values, the opening angle is linearly dependent upon $\chi_{\mathrm{p}}$. 
Thus, the phasing due to precession will, to leading order, scale with $\chi_{\mathrm{p}}^{2}$.  
To investigate the impact of this, we should re-examine the accuracy with which the masses and aligned spins can be measured when we also allow for precession.  

In Figure \ref{fig:chip_chi_eff}, we show the posterior distributions for the precessing and aligned spin components, keeping the masses fixed.  In addition, we construct the metric in the two-dimensional $\chi_{\mathrm{align}}$--$\chi_{\mathrm{p}}^{2}$ space.  The metric accurately reconstructs the posterior, whereas working in $\chi_{\mathrm{align}}$--$\chi_{\mathrm{p}}$ co-ordinates does not accurately capture the degeneracy.  From Figure \ref{fig:chip_chi_eff}, it is clear that the value of $\chi_{\mathrm{p}}$ is essentially undetermined from the phasing of the leading waveform component (the reason that the contour doesn't extend to $\chi_{\mathrm{p}} = 1$ is that the total spin $\sqrt{\chi_{\mathrm{p}}^{2} + \chi_{\mathrm{align}}^{2}}$ is required to be less than 1).  Nonetheless, we must include the degeneracy between $\chi_{\mathrm{p}}$ and the other mass and spin parameters when calculating posterior distributions for the parameters.

\subsection{Summary}
\label{ssec:pe_summ}

In the preceding sections, we have laid out how the gravitational-wave parameters are encoded into the gravitational wave signal and how, upon observing the signal and various specific features, we are able to extract estimates of the different parameters.  Here, we provide a brief summary of the above discussions.

The amplitude and phase evolution of the (leading harmonic of) the gravitational waveform enables us to measure:

\begin{enumerate}
    \item The redshifted chirp mass, $\mathcal{M}(1 + z)$ of the signal.
    \item The symmetric mass ratio $\eta$.
    \item A combination $\chi_{\mathrm{align}}$ of the aligned spins $\chi_{1z}$ and $\chi_{2z}$.  The individual spins are typically unmeasurable.
    \item The time of coalescence of the system, as measured at the detector.
\end{enumerate}
There is significant degeneracy between these parameters, most notably the mass ratio and aligned spins.

When the system is observed in a network of detectors
\begin{enumerate}
  \setcounter{enumi}{4}
    \item The right ascension of the source.
    \item The declination of the source.
\end{enumerate}
When the system is observed in at least three observatories, we typically localize to a region with an area of a few square degrees.  Signals observed in two detectors are only localized to (a fraction of) a ring in the sky with areas of hundreds of square degrees.

Using the detector antenna response, we are able to identify
\begin{enumerate}
  \setcounter{enumi}{6}
    \item The amplitude of the gravitational wave signal, which enables inference of a combination of distance to the source and its orientation, $(1 \pm \cos \theta_{JN})^{2}/d_{L}$.
    \item The phase of the dominant circular gravitational-wave polarization ($\phi \pm \psi$).
\end{enumerate}
If we are able to identify either the second polarization or power in higher multipoles, or both, this enables measurement of
\begin{enumerate}
  \setcounter{enumi}{8}
    \item The binary orientation, and consequently a more accurate distance measurement.
    \item A second phase measurement, which enables the separate inference of the coalescence phase $\phi_{o}$ and polarization $\psi$.    
\end{enumerate}
We note that for an aligned-spin system, this comprises all of the parameters when we simplify to a single, effective spin parameter: two masses, one spin parameter, four extrinsic parameters, sky location and time of arrival.  Thus, in principle, with a network of detectors, all of the parameters can be measured.  Those which are observed with the least accuracy tend to be the second combination of effective spin and mass ratio, and those which require an observation of the second polarization or higher multipoles.

With measurement of power in precession we can measure
\begin{enumerate}
  \setcounter{enumi}{10}
    \item The precession spin $\chi_{\mathrm{p}}$, provided the orientation is already constrained and, if not, then a combination of $\chi_{\mathrm{p}}$ and $\theta_{JN}$.
    \item The precession phase, $\alpha_{o}$ (also denoted $\phi_{JL}$).
\end{enumerate}

 \section{Simple PE Implementation}
\label{sec:simple_pe}

The intuitive understanding of how the binary parameters are encoded in the observed gravitational-wave signal, given in Section \ref{ssec:pe_summ}, can be used to develop a simple, computationally cheap parameter estimation routine.  Here, we introduce the \texttt{simple-pe}~\cite{simple_pe_docs, simple_pe_code} algorithm that has been developed for this purpose.  

The outline of the method is as follows.  First, we identify the peak of the likelihood in the mass and spin space in the network of detectors, maximizing the time of arrival, amplitude and phase of the signal independently in each detector.  The values of masses and spins at the peak are used as central values for those parameters.  The arrival times, relative amplitudes and phases of the signal in each observatory are used to obtain an estimate for the sky location of the source. Around this peak, we construct posterior distributions for the masses and spins of the binary, using the expected accuracies and known degeneracies presented in Section \ref{ssec:chirp}.  We matched filter the data at the peak of the likelihood to identify the \gls*{snr} in the second polarization, higher multipole and precession waveforms.  Then, based upon the expected \gls*{snr} in each of these features as a function of the masses, spins and binary orientation, we identify regions of parameter space that are consistent with the observed \glspl*{snr}.  In particular, the \gls*{snr} in the second polarization can be used to restrict the orientation, both mass ratio and orientation are constrained by the \gls*{snr} in higher multipoles and in-plane spins restricted using the \gls*{snr} in the leading precession correction to the waveform.  Finally, we infer the distance distribution based on the masses, network sensitivity and binary orientation.

\subsection{Find the maximum likelihood in mass-spin space}
\label{ssec:max_snr}

The first step is to identify the peak of the likelihood, or equivalently the maximum \gls*{snr}, across the mass and spin space.  In Section \ref{ssec:chirp}, we have shown that the masses and aligned spin can be  inferred from the phasing of the dominant waveform component.  In Section \ref{sec:waveform} we have argued that the higher multipoles and precession waveform components could contribute a non-negligible amount of \gls*{snr} to the overall signal.  Thus, when finding the peak of the \gls*{snr} or likelihood, we must consider whether it is necessary to incorporate the power in either of these features.

\begin{figure}
    \centering
    \includegraphics[width=\linewidth]{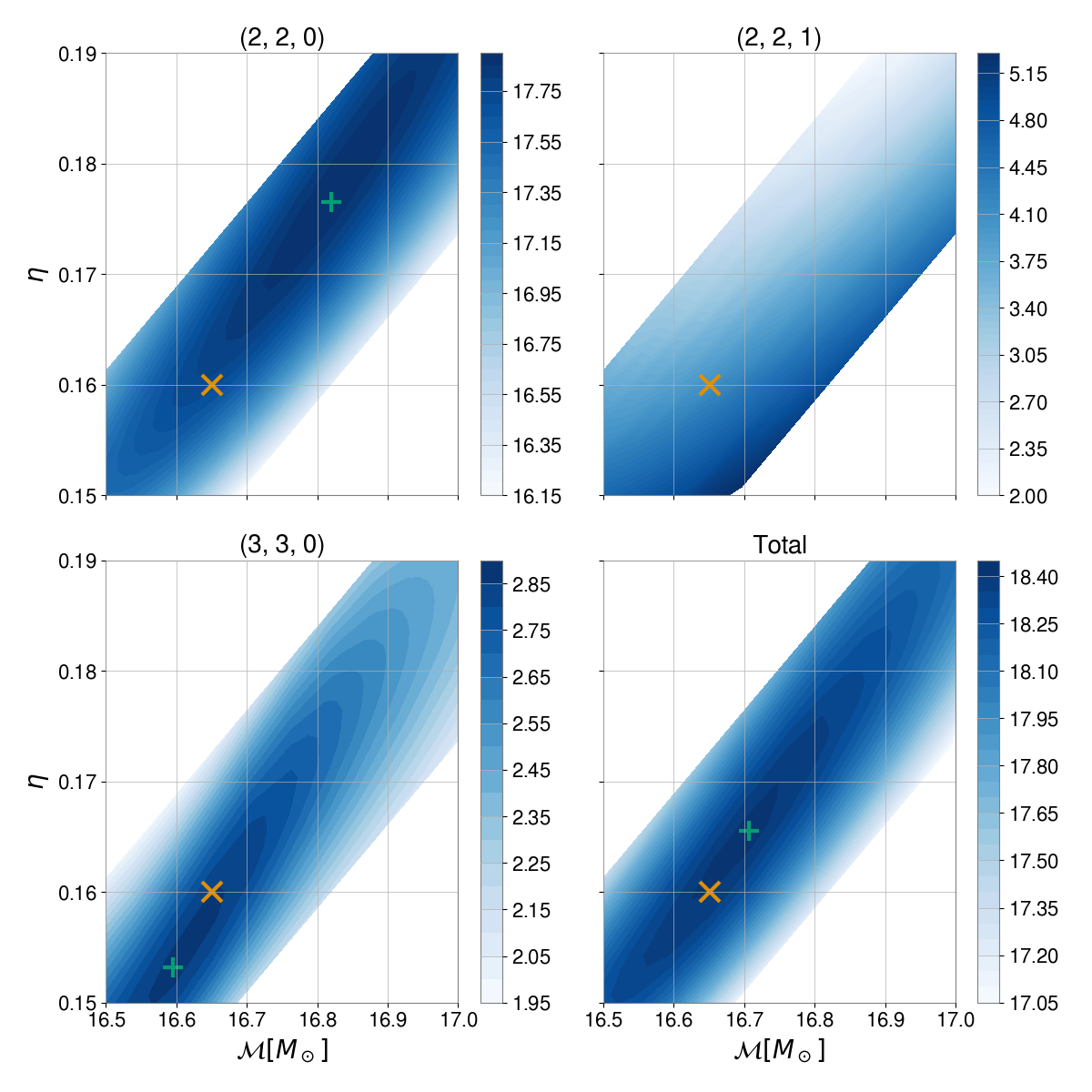}
    \caption{The \gls*{snr} distribution for across the mass space for the L1 detector for the modes (2, 2, 0), (2, 2, 1) and (3, 3, 0).  The system masses are marked with a $\times$.  The system has a precessing spin $\chi_{\mathrm{p}}$ of 0.4.  The peak of the matched-filter \gls*{snr} for each mode is marked with a + (the peak for the (2, 2, 1) component is off the plot).  The \gls*{snr} of the (2, 2, 0) waveform peaks away from the simulated mass values due to the presence of power in precession.  The combined \gls*{snr} of the (2, 2, 0), (2, 2, 1) and (3, 3, 0) components peaks closer to the simulated value.  The remaining discrepancy arises from using $\chi_{\mathrm{p}}$ and $\chi_{\mathrm{align}}$, rather than the full spin vectors, to describe the binary spins.
    }
    \label{fig:snr_vs_mass}
\end{figure}

Figure \ref{fig:snr_vs_mass} shows the \gls*{snr} of the  (2, 2, 0), (2, 2, 1) and (3, 3, 0) waveform components when matched-filtered against the same simulated signal which we have considered previously --- masses of $40$ and $10 M_{\odot}$ with aligned spin components of 0.5 and 0, respectively, and $\chi_{\mathrm{p}} = 0.5$.  We compute the \gls*{snr} for the L1 detector, across a range of masses, keeping the values of $\chi_{\mathrm{align}}$ and $\chi_{\mathrm{p}}$ fixed.  The simulated signal has an \gls*{snr} of $18.5$.  As expected, from Section \ref{ssec:final_waveform}, the \gls*{snr} in the (2, 2, 0) component is the largest and has a value of almost 18 by itself.  Interestingly, though, the peak of the \gls*{snr}, occurs at masses offset from the simulated values.  The offset in the peak \gls*{snr} is largely caused by precession.  For this signal, the overlap between the two precession harmonics (2, 2, 0) and (2, 2, 1) is $\approx 0.2$ and, therefore, one obtains a higher \gls*{snr} in the (2, 2, 0) harmonic at values of the masses where it picks up some of the power in the (2, 2, 1) harmonic. This can be clearly seen from the \gls*{snr} distribution for the (2, 2, 1) harmonic, which decreases significantly from $\rho_{p} = 4$ at the simulated values to $\rho_{p} = 3$ at the (2, 2, 0) peak.  The structure of the \gls*{snr} in the (3, 3, 0) waveform is very similar to the leading mode, varying between $\rho_{33} \approx 2.6$ to $2.9$ across the region of interest.  It will therefore have a limited contribution to the offset of the peak. When combining the \glspl*{snr} in the three waveform components, we find that the peak is approximately in the correct location. Interestingly, it is not in the exact location, and this arises because we are using $\chi_{\mathrm{align}}$ and $\chi_{\mathrm{p}}$ to describe the signal and these parameters do not perfectly describe the simulated waveform.  The simulated signal has spin on only the larger black hole while we assign the same (aligned)-spin value $\chi_{\mathrm{align}}$ to both components when identifying the signal.  While this has limited impact for non-precessing systems, the difference can be greater in precessing systems as the final black hole spin and spin orientation will be impacted by the component spins.

In performing parameter estimation on the signal, we identify the location of the peak \gls*{snr} as follows.  First, we find the peak \gls*{snr} of the (2, 2, 0) waveform across the mass and aligned spin space, using a fixed value of $\chi_{\mathrm{p}}$.  To do so, we filter the data from each detector independently and sum the maximum in quadrature (over time and phase) of the \gls*{snr} in each detector.  We use the \texttt{scipy.optimize} routine~\cite{2020SciPy-NMeth} with an initial guess offset from the peak --- in this example, we offset $\mathcal{M} = 16.6 M_{\odot}$, $\eta = 0.15$ and $\chi_{\mathrm{align}} = 0.4$, although we have varied the starting point to demonstrate that it has minimal impact on the optimization result.  
For our example signal, we obtain values of $\mathcal{M} = 16.87 M_{\mathrm{\odot}}$, $\eta = 0.180$ and $\chi_{\mathrm{align}} = 0.44$ for the peak of the (2, 2, 0) \gls*{snr}, which is consistent with the peak shown in Figure \ref{fig:mass_spin_measurement}. For a real signal, we would use the parameters returned by the search and, indeed, gravitational wave searches have now implemented a similar maximization procedure to obtain mass and spin measurements more accurately, see e.g.~\cite{DalCanton:2020vpm}. 

Based on the discussion above, and the plots in Figure \ref{fig:snr_vs_mass}, it is clear that we can obtain an improved estimate of the peak location using the two-harmonic \gls*{snr} \cite{Fairhurst:2019_2harm} which incorporates the power in the (2, 2, 1) harmonic.  To do so, we perform a second optimization step, again over the chirp mass, mass ratio and aligned spin space, to identify the peak of the two-harmonic waveform.  We use the previously identified peak to seed the second maximization.  Although we have included the precession correction, we \textit{do not} vary the precession spin $\chi_{\mathrm{p}}$ at this stage, as we find that it is not helpful in identifying the peak.  This is to be expected, given the significant degeneracy between the precession and aligned spins shown in in Figure \ref{fig:chip_chi_eff}.  The precession spin is better constrained by the amplitude of the (2, 2, 1) harmonic, which we consider later.  At present, the optimization routine is not able to accurately identify the peak of the two-harmonic \gls*{snr}.  We are continuing to investigate the reason for this.  Consequently to obtain an accurate peak, we currently construct a dense grid of points around the (2, 2, 0) peak and filter them against the two precession harmonics to find the peak \gls*{snr}.  We make use of the parameter-space metric to identify the eigen-directions and generate a grid of points which covers the $3\sigma$ uncertainty region around the (2, 2, 0) peak.  This method identifies the peak with good accuracy, but does slow down our analysis, as filtering the grid is computationally intensive.  The peak of two-harmonic \gls*{snr} is identified to occur at $\mathcal{M} = 16.69 M_{\mathrm{\odot}}$, $\eta = 0.171$ and $\chi_{\mathrm{align}} = 0.421$. 

\subsection{Obtain the sky location}
\label{ssec:sky_loc}

There are several rapid sky localization analyses, most notably BAYESTAR \cite{Singer:2015ema}, that can return the sky position of the signal quickly and accurately.  Our goal here is not to present a new localization method, as that problem is already well addressed \cite{Fairhurst:2009tc, Fairhurst:2010is}.  Nonetheless, we do require an estimate of the sky location of the source, and its uncertainty, for several purposes.  Most notably, knowing the sky position enables us to estimate the network sensitivity, and this is critical to obtaining a good estimate of the distance to the source, from Equation (\ref{eq:rho_o}).  In addition, the sky location is used to calculate sensitivity to the second polarization, $\alpha_{\mathrm{net}}$, and the observed power in the second polarization.

To estimate the sky location, we use a simple chi-squared minimization, as described in the Appendix of \cite{Fairhurst:2010is}, to identify the preferred location.  For the simulated signals discussed in this paper, the signal is observed in the LIGO-Virgo network, so that timing alone provides two locations (above and below the plane of the detectors).  We generate a localization for each, as described in Section \ref{ssec:loc} --- in this case they overlap, so we obtain a single sky patch.  Across the sky patch, we calculate the network sensitivity $|\boldsymbol{ F\sigma}|$ and the sensitivity to the second polarization $\alpha_{\mathrm{net}}$.  For the analysis presented here, we use the mean and variance of the network sensitivity and the mean value of $\alpha_{\mathrm{net}}$ in reconstructing the source parameters.

\subsection{Generate samples in the mass and spin space}
\label{ssec:mass_spin_pe}

Starting from the maximum likelihood point, we calculate the approximate uncertainties in the masses and spins using the parameter space metric, introduced in Section \ref{ssec:match}.  As has been discussed in detail \cite{Cutler:1994ys, Ohme:2013nsa, Baird:2012cu, PhysRevD.90.024018},  the choice of parameters used in the metric expansion can have a significant impact on the domain of validity of the quadratic approximation in Equation (\ref{eq:like_metric}),  
\begin{equation*}
    M(\delta\vec{\lambda}) \approx 1 - g_{ab} \delta\lambda^{a}\delta\lambda^{b} 
    \quad \mathrm{where} \quad
    \delta \vec{\lambda} = \hat{\lambda} - \vec{\lambda} \, .
\end{equation*}
We use the chirp mass $\mathcal{M}$, symmetric mass ratio $\eta$, aligned  $\chi_{\mathrm{align}}$ and in-plane spins, parameterized by $\chi_{\mathrm{p}}^{2}$ --- the reason for using $\chi_{\mathrm{p}}^{2}$ is discussed in Section \ref{ssec:prec}.  These provide a good basis for estimating the parameter accuracy.

In many cases, the metric is calculated by taking derivatives of the waveform in various parameters \cite{Owen:1995tm, Owen:1998dk}, to obtain the leading order variation.  Here, we instead choose to take finite differences when evaluating the metric.  This means that any higher order terms, which are of interest for the scale of variations being considered, will be appropriately included in the metric.  Since we are particularly interested in identifying the high-likelihood region, e.g. the 90\% confidence interval for the physical parameters, this provides a natural scale at which to evaluate the metric.  (A similar method has previously been introduced in \cite{PhysRevD.87.024004}).  For the four dimensional parameter space under consideration, the 90\% region is given by
\begin{equation}\label{eq:metric_norm}
    g_{ab} \delta \lambda^{a} \delta \lambda^{b} = \frac{3.9}{\rho_{h}^{2}}
\end{equation}

To generate the metric, we begin with the four basis vectors in the directions $(\mathcal{M}, \eta, \chi_{\mathrm{align}}, \chi_{\mathrm{p}}^{2})$.  We scale each basis vector to obtain the mismatch required for the observed \gls*{snr} of the system, as given in Equation (\ref{eq:metric_norm}).  In principle, the mismatch will be symmetric in steps $\pm \delta \lambda^{a}$ but, as can be seen in Figure \ref{fig:mass_spin_measurement} this is not exactly true for finite steps.  Therefore, we take the average mismatch between positive and negative variations.  These immediately provide the diagonal elements of the metric $g_{ab}$.  To obtain the off-diagonal elements, we calculate the mismatch for steps in directions $\pm \delta\lambda^{a} \pm \delta\lambda^{b}$, where $a$ and $b$ run over the four parameters.  We average over the mismatches to obtain an estimate of the off-diagonal terms in $g_{ab}$.  Thus, by calculating a small number of matches, we obtain an expression for the metric $g_{ab}$.

\begin{figure}
    \centering
    \includegraphics[width=\linewidth]{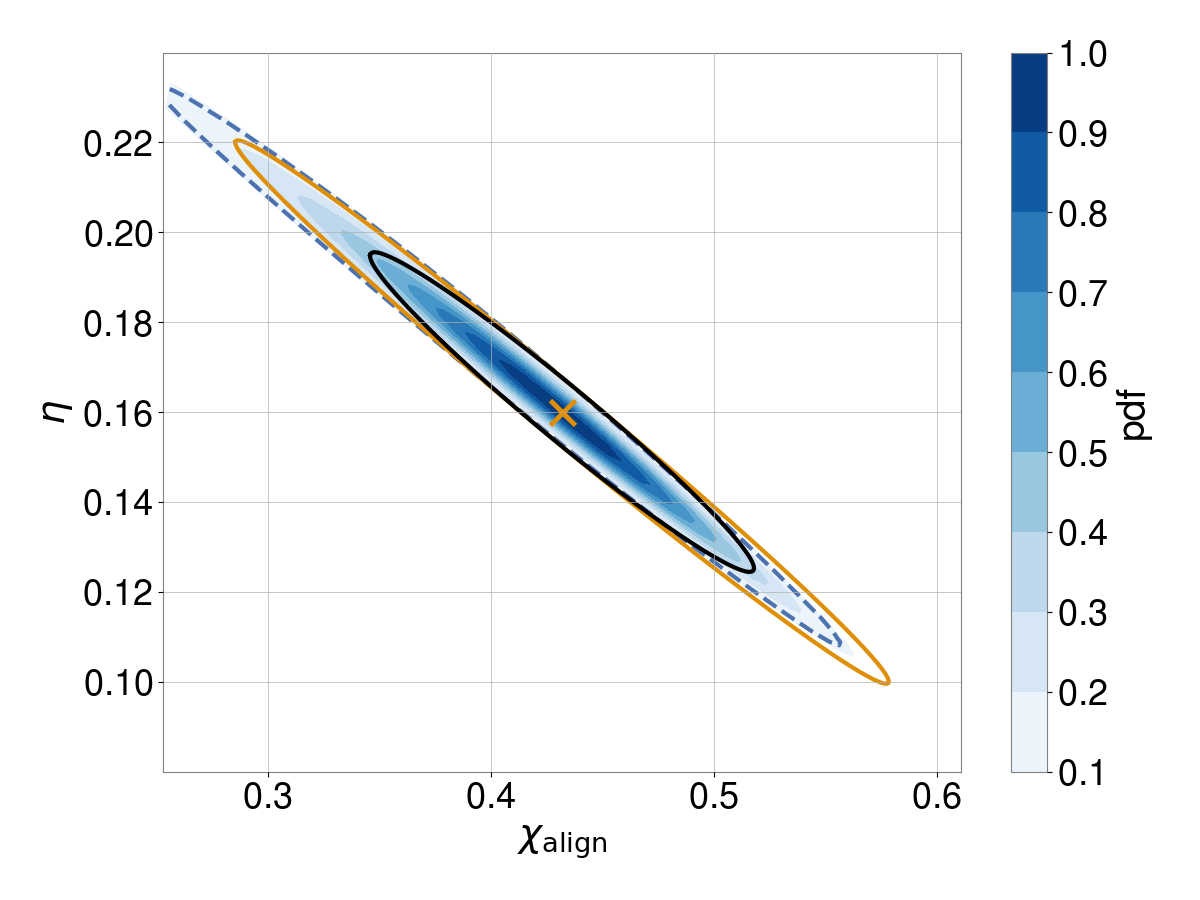}    \caption{Degeneracy between mass ratios and spins for a signal with \gls*{snr} of 10.  The black ellipse shows the predicted degeneracy obtained from a metric calculated with variations along $\eta$ and $\chi_{\mathrm{align}}$.  The orange ellipse shows the final result after with variations taken along the eigen-directions of the degeneracy.   The final metric matches the degeneracy well, as shown in the \gls*{pdf}, while the initial metric significantly underestimates it.
    }
    \label{fig:metric_iteration}
\end{figure}

Typically the co-ordinate directions are \textit{not} a good choice of basis for calculating the metric, since the degenerate directions do not lie along the physical parameters, as can be seen in Figure \ref{fig:mass_spin_measurement}.  Thus, while the uncertainties in individual parameters will be well approximated by calculating the metric along the physical parameters, the correlations will typically be less well estimated.  To improve the accuracy of the metric, we iteratively update it to use co-ordinates which lie along the eigen-directions in the parameter space.  Specifically, we first calculate the metric using variations $\delta \lambda^{a}$ along each physical parameter.  We then generate the eigen-directions of the initial metric and test whether they do, indeed, describe the principal axes of the parameter degeneracy ellipse.   If not, then we re-normalize them to have the desired mismatch (given in Equation (\ref{eq:metric_norm})), and re-calculate the metric in these new coordinates. We stop when the metric is no longer changing significantly, specifically, we test whether the eigen-directions of the old metric remain eigen-directions of the updated metric (within a given tolerance, i.e. that the vectors reproduce the desired mismatch and are orthogonal).  Figure \ref{fig:metric_iteration} shows the importance of this iterative process for a low \gls*{snr} system.  For higher \glspl*{snr}, the effect is less significant.

The metric provides an approximate likelihood over the mass and spin parameter space, using Equation (\ref{eq:like_metric}).  Since the likelihood is approximated as a multi-variate Gaussian, we can very quickly generate large numbers of samples across the mass and spin parameter space.  To do so, we generate the requested number of points drawn from four normal distributions, and then use the metric to project these to the physical parameter space.

\subsection{Generate samples in distance and orientation}
\label{ssec:dist_inc}

We assume that sources are distributed uniformly in volume and that their orientations are uniformly distributed.  While the latter assumption should be generically valid, at small or large distances sources will not be uniformly distributed -- they will either follow the local density of galaxies or cosmological effects, including a redshift dependence on the merger rate, become important.  Nonetheless, it is standard to generate parameter estimates using uniform volume distribution and then re-weight for astrophysical or cosmological distributions later \cite{LIGOScientific:2021djp, LIGOScientific:2021psn}.

A uniform in volume distribution of sources leads to a preference for the observation of face-on (or face-away) sources due to their greater gravitational wave amplitude (see, e.g. Equation (\ref{eq:rho_o})) \cite{Schutz:2011tw}.  In Section \ref{ssec:2nd_pol}, we have argued that the majority of observed sources will have significantly greater \gls*{snr} in one of the two circular polarizations.  Therefore, we wish to obtain the distribution for $\cos \theta_{JN}$ for a signal observed with a fixed \gls*{snr} in the right or left circular polarization.  This is given as
\begin{align}
    p(\cos \theta_{JN}) &\propto 
    \int d_{L}^{2} d d_{L} \, 
    d \cos \theta_{JN} \, 
    \nonumber \\
    &
    \delta\left(\rho_{o} -  
    |\boldsymbol{\sigma} \mathbf{F} | 
    \frac{d_{o}}{d_{L}} 
    \left(\frac{1 \pm \cos \theta_{JN}}{2}\right)^{2} 
    \right)
\end{align}     
where $\delta$ denotes the Dirac delta function.  It is straightforward to marginalize over the distance distribution and identify that the (originally flat in $\cos\theta_{JN}$) distribution becomes
\begin{equation}\label{eq:theta_distrib}
    p(\cos \theta_{JN}) \propto  (1 \pm \cos\theta_{JN} )^{6}
\end{equation}
where the positive/negative corresponds to right/left polarization respectively.  We use this distribution to generate samples in $\theta_{JN}$ corresponding to left and right circularly polarized signals.

Given the sky location of the source and the (complex) \gls*{snr} observed in each detector, it is straightforward to obtain the \gls*{snr} in the right and left circular polarizations, $\rho_{R, L}$. As described in \cite{Fairhurst:2017mvj}, we achieve this by first rotating to the dominant polarization frame, and calculating the network's sensitivity to the $+$ and $\times$ polarizations, $w_{+}$ and $w_{\times}$ respectively. We then use the given sky location and project the (complex) \gls*{snr} observed in each detector onto the space of circularly polarized signals using,
\begin{equation}
    P^{ij}_{\mathrm{circ}} = \left[\frac{(w_{+}^{i} \pm iw_{\times}^{i})(w^{j}_{+}\pm iw_{\times}^{j})}{|w_{+}|^{2} + |w_{\times}|^{2}}\right],
\end{equation}
where $i$ and $j$ run over the detectors in the network and the $+/-$ gives the \gls*{snr} in the left and right polarizations respectively.  The relative \gls*{snr} in each of the circular polarizations can be used to appropriately weight the probability that the signal is predominantly either left or right circularly polarized.  Since the likelihood is proportional to $\exp[\rho_{R, L}^{2}/2]$, this provides the appropriate normalization factor to weight the number of samples drawn from the left and right circular polarization distributions for $\theta_{JN}$.  In many cases, the signal will be preferentially right (or left) circularly polarized, in which case the majority of samples will correspond to face-on (or face-away) orientation.  If $\alpha_{\mathrm{net}}$ is small, it is often impossible to distinguish between left and right circular polarizations.  In this case, there will be large numbers of samples for both face-on and face-away signals, with a minimum at edge on since these systems emit the weakest gravitational wave signal.

Given the binary orientation $\theta_{JN}$, we can obtain the distance from Equation (\ref{eq:rho_o}), which we repeat below,
\begin{equation*}
     \rho_{o} = \frac{d_{o}}{d_{L}}
    \frac{ |\boldsymbol{\sigma} \mathbf{F}| }
     {(1 + \tau^{2})^{2}} \, .
\end{equation*}
The inferred distance depends upon the network response $|\boldsymbol{\sigma F}|$ through the detector response encoded in $\boldsymbol{F}$ and the masses and spins through the detector sensitivity to the signal encoded in $\sigma$.  We incorporate both of these effects when estimating the distance.  For the network sensitivity, we simply draw samples from a Gaussian, based upon the previously measured mean and variance.  To incorporate variations in the mass and spin space, we must re-calculate $\sigma$ for each sample, given the values of mass and spin.  Since $\sigma$ is a slowly varying function across mass and spin (it will typically vary by at most tens of percent over the mass-spin posterior), we first interpolate across the space and then use the interpolation function to evaluate $\sigma$ at each of the samples.  Finally, the observed \gls*{snr} has measurement uncertainty, which is well modelled by a non-central $\chi^{2}$ distribution with two degrees of freedom and a non-centrality parameter $\rho_{R, L}$.  Thus, given a value of $\theta_{JN}$ and $\sigma$ for each sample, we randomly sample $\rho_{o}$ and $|\boldsymbol{F}|$ from the appropriate distributions and use Equation (\ref{eq:rho_o}) to calculate the distance.

\subsection{Restrict the parameters using additional waveform components}
\label{ssec:snr_pe}

In the previous subsections, we have described a method to obtain samples in masses, spins, distance and binary orientation.  
This provides a good initial estimate of the binary parameters but we have additional information which can still be used to improve the parameter estimates.  In particular, we have not yet used the measured \glspl*{snr} in the second polarization, higher multipoles or precession.

The measured \gls*{snr} in the second polarization, precession and higher multipoles can be used to improve measurement of the binary parameters, as discussed in detail in Sections \ref{ssec:2nd_pol}, \ref{ssec:higher_harm} and \ref{ssec:prec}.  For each of the samples, we calculate the \textit{expected} \gls*{snr} in the each of these features, using the given masses, spins, distance and orientation.  In particular, the \gls*{snr} in the second polarization is given by Equation (\ref{eq:rho_2pol}) and depends upon the orientation $\theta_{JN}$ and $\alpha_{\mathrm{net}}$.  The expected \gls*{snr} in higher multipoles is given in Equation (\ref{eq:rho_33}) and depends upon the orientation and relative significance of the higher multipoles, encoded in $\alpha_{\ell m}$, which is primarily determined by the mass ratio \cite{Mills:2020thr}.  The expected \gls*{snr} in precession is given in Equation (\ref{eq:rho_prec}) and depends upon the binary orientation and the opening angle between the orbital and total angular momenta.  The opening angle is largely determined by the in-plane spins, $\chi_{\mathrm{p}}$, but also varies with mass ratio and aligned spin.   Those samples where the expected \gls*{snr} in these waveform features matches the observed \gls*{snr} are given a higher weighting than those where the expected and observed \glspl*{snr} differ significantly.

We calculate the observed \gls*{snr} in the second polarization, precession and higher multipoles by matched filtering the waveform components, evaluated at the maximum likelihood point identified in Section \ref{ssec:max_snr}, against the data.  In detail, the precession and higher multipole \glspl*{snr} are obtained by matched filtering the $h_{22,1}$ and $h_{33,0}$ waveforms against the data from the network of detectors and evaluating the \gls*{snr} in each detector at the time where the \gls*{snr} for $h_{22,0}$ is maximum.  To obtain the \gls*{snr} in the second polarization, we project the \gls*{snr} in the (2, 2, 0) component into the space orthogonal to the right/left circular polarization, as described in Section \ref{ssec:wf_network}.
For each sample, we calculate the likelihood of obtaining the observed \gls*{snr} in these waveform components --- it is given by a non-central chi-squared distribution with two degrees of freedom, where the non-centrality parameter is the expected \gls*{snr} at the parameter values of the sample and the distribution is evaluated at the observed \gls*{snr}.  Thus, points in the parameter space which accurately predict the observed \gls*{snr} in the second polarization, higher multipoles and precession are preferred to those which predict either too much or too little \gls*{snr} in these features.  We assign weights to each of the samples, based upon the product of the probabilities for obtaining the given \gls*{snr} in each of the three features, and then use these weights to importance sample the points to produce our final result.

In the above, we have used only the \glspl*{snr} calculated at the maximum likelihood point.  This has the benefit of significantly reducing computational cost, as we do not need to re-compute the likelihood at every sample.
From Figure \ref{fig:snr_vs_mass}, we see that the distribution of the higher multipole \gls*{snr} has a similar structure to the leading waveform component, although in the example show the peak is somewhat offset.  Furthermore, in the region of mass and spin space consistent with the (2, 2, 0) waveform (shown, e.g., in Figure \ref{fig:mass_spin_measurement}) the \gls*{snr} in the (3, 3, 0) waveform component varies by around 10\%.  Therefore, it is a reasonable approximation to take this to be constant.  For the second polarization, the variation of \gls*{snr} with an incorrect estimate of $\theta_{JN}$ (which will lead to an over/under-estimation of the \gls*{snr} in the second polarization) should be independent of the masses and spins.  For precession, the \gls*{snr} distribution in mass and spin space is significantly different than the (2, 2, 0) waveform.  Nonetheless, the variation in precession \gls*{snr} over the mass and spin space of interest is $\approx 30\%$ and so we can reasonably treat this as constant.  Finally, for speed, we pre-compute slowly varying quantities such as $\bar{b}$ and $\alpha_{33}$ coarsely over the parameter space and then interpolate them when calculating the expected \gls*{snr} in each waveform component.

\subsection{Summary}
\label{ssec:simple_pe_summary}

The \texttt{simple-pe} algorithm generates a set of discrete samples that approximate the mass, spin, distance and orientation posterior distribution for the observed gravitational-wave signal.  The mass and spin distributions are approximated as multi-dimensional Gaussian distributions centred around the point with the maximum \gls*{snr} in the (2,2) multipole (incorporating two precession harmonics).  While the sky location is inferred from the \glspl*{snr} in the network, the information is only used in estimating the overall network sensitivity to the two gravitational wave polarizations, and their variation over the localization region.  The distance and orientation are generated assuming uniformly distributed events in both volume and binary orientation, and weighting for observability.  Correlations between the inferred distance/orientation and mass/spin parameters arise through the requirement that the binary parameters accurately predict the observed power in the second polarization, higher multipoles and precession.

Finally, we must discuss priors.  So far, we have been evaluating the likelihood, either through the metric expansion or via observed \glspl*{snr} in different waveform components.  We have explicitly considered priors on distance and orientation, where we chose a distribution uniform in volume and in $\cos\theta_{JN}$. This means that, over the mass and spin space, we have effectively imposed flat priors in the parameters that we are using to generate samples, namely $\mathcal{M}$,$\eta$, $\chi_{\mathrm{align}}$ and $\chi_{\mathrm{p}}^{2}$.  In the majority of analyses, priors are chosen to be uniform in component masses, spin magnitudes, and orientations, see e.g. Refs.~\cite{abbott2019gwtc,Abbott:2020niy,ligo_scientific_collaboration_and_virgo_2021_5117703,LIGOScientific:2021djp}.  While, for many systems, the masses are well constrained and the differing prior causes little impact, this is not the case for equal mass systems.  To transform to a prior which is uniform in component masses requires 
\begin{equation}
    p(\mathcal{M}, \eta) \propto 
    \frac{\mathcal{M}}{\sqrt{1 - 4 \eta}} 
\end{equation}
which blows up at equal mass ($\eta = \tfrac{1}{4}$).  
To address this, we limit the denominator to 50, effectively truncating the prior at $\eta = 0.2499$.  This leads to a minor under-sampling of near-equal mass systems (with a mass ratio between $1$ and $1.04$).  Similarly, we transform from uniform priors on $\chi_{\mathrm{align}}$ and $\chi_{p}^2$ to uniform priors on component spin magnitudes and orientations using the transformations given in \cite{Callister:2021gxf}.  These priors are implemented by drawing a large number of samples from the underlying distributions and then using rejection sampling to retain samples with a distribution matching the desired priors, see, e.g. Appendix C in Ref.~\cite{Romero-Shaw:2020owr}. 

The entire analysis presented above runs in a few minutes on a single CPU.\footnote{In the current implementation, filtering the two-harmonic \gls*{snr} over a grid to find the peak takes around 15 minutes and dominates the computational time. Other techniques, such as relative binning~\cite{Zackay:2018qdy}, may reduce this computational cost, and we leave an investigation future work.}
It can be neatly separated into two parts: in the first, we matched-filter the data to identify the maximum \gls*{snr} point, the sky location and power in each of the waveform components: second polarization, higher multipoles and precession.  For the estimation of the binary parameters, we make use of these values and then calculate distributions in masses, spins, orientation and distance.  This second step requires the \gls*{psd} of the detector data to obtain the metric, but not the data themselves.  \section{Simple PE Results}
\label{sec:pe_results}

\begin{table}[t]
    \centering
    \begin{tabular}{|l|c|}
    \hline
        Source frame primary mass, [$M_{\odot}$] & 40.0 \\
        Source frame secondary mass, [$M_{\odot}$] & 10.0 \\
        Source frame chirp mass, $\mathcal{M}$ [$M_{\odot}$] & 16.65 \\
        symmetric mass ratio, $\eta$ & 0.16 \\
    \hline
$y$th component of the primary spin & 0 \\
        $z$th component of the primary spin & 0.5 \\
        Secondary spin magnitude & 0 \\
        Effective spin, $\chi_{\mathrm{eff}}$ & 0.4 \\
        Aligned spin, $\chi_\mathrm{align}$ & 0.42 \\
\hline
Right ascension, $\alpha$ [rad] & 1.2 \\
        Declination, $\delta$ [rad] & 0.3 \\
        Polarization, $\psi$ [rad] & 0.5 \\
        Network alignment factor, $\alpha_{\mathrm{net}}$ & 0.364 \\
        GPS merger time, [s] & 1677672000.0 \\
\hline
    \end{tabular}
    \caption{Common parameters of the simulated signals}
    \label{tab:common_params}
\end{table}

We use four simulated signals to demonstrate an application of the \texttt{simple-pe} algorithm described in Section \ref{sec:simple_pe}.  The goal of the examples is to demonstrate the accuracy of the parameter estimation across the mass, spin, distance and orientation space.  In particular, we are interested in seeing that the quadratic approximation from the metric is appropriate for generating samples in the mass and spin space.  Furthermore, we would like to investigate how the observed \gls*{snr} in the additional waveform components can impact parameter recovery.  To do so, we choose signals with many parameters in common, and vary only those which will impact the \gls*{snr} in the second polarization, higher multipoles and precession.  The common parameters of the systems are provided in Table \ref{tab:common_params}.  In all cases, the signals have masses of $40 M_{\odot}$ and $10 M_{\odot}$, with aligned spin components of $0.5$ on the more massive black hole and zero on the smaller system.  The binaries are located at a fixed sky location (the same as shown in Figure \ref{fig:localization}).  

\begin{table}[t]
    \centering
    \begin{tabular}{|l|c|c|c|c|}
     \hline
        Signal & (1) & (2) & (3) & (4) \\
               &  -  & HM  & prec & HM-prec \\
    \hline
Effective precession spin, $\chi_{\mathrm{p}}$ & 0.01 & 0.1 & 0.7 & 0.4 \\
        Opening angle, $\beta$ & 0.003 & 0.06 & 0.4 & 0.2 \\
    \hline
        Angle between $L$ and $N$, $\iota$ & 0.005 & 0.6 & 0.1 & 0.6 \\
        Angle between $J$ and $N$, $\theta_{JN}$ & 0.006 & 0.6 & 0.4 & 0.6 \\
        Luminosity distance, $d_{L} [\mathrm{Mpc}]$ & 982 & 825 & 859 & 814 \\
    \hline
        SNR in $(3, 3)$ multipole, $\rho_{33}$ & <0.1 & 4.3 & 2.8 & 4.2 \\
        SNR in precession, $\rho_{p}$ & <0.1 & 1.02 & 4.4 & 4.4\\
        SNR in second polarization & <0.1 & 0.1 & <0.1 & 0.2 \\
    \hline
    \end{tabular}
    \caption{Parameters that vary for the simulated signals
    }
    \label{tab:varying_params}
\end{table}

The binary orientation and in-plane spins are varied to give examples with negligible or significant power in both the higher multipoles (HM) and precession (prec).  The power in the second polarization also varies but, in none of the cases we have considered is it sufficient to enable confident observation of the second polarization.  We fix the distance to the signals to ensure a \gls*{snr} of 25 in the LIGO-Virgo network with expected O4 sensitivity.  The set of varying parameters for each signal are given in Table \ref{tab:varying_params}, along with the \gls*{snr} in each of the waveform components.

\begin{figure*}
    \centering
    \includegraphics[width=\linewidth]{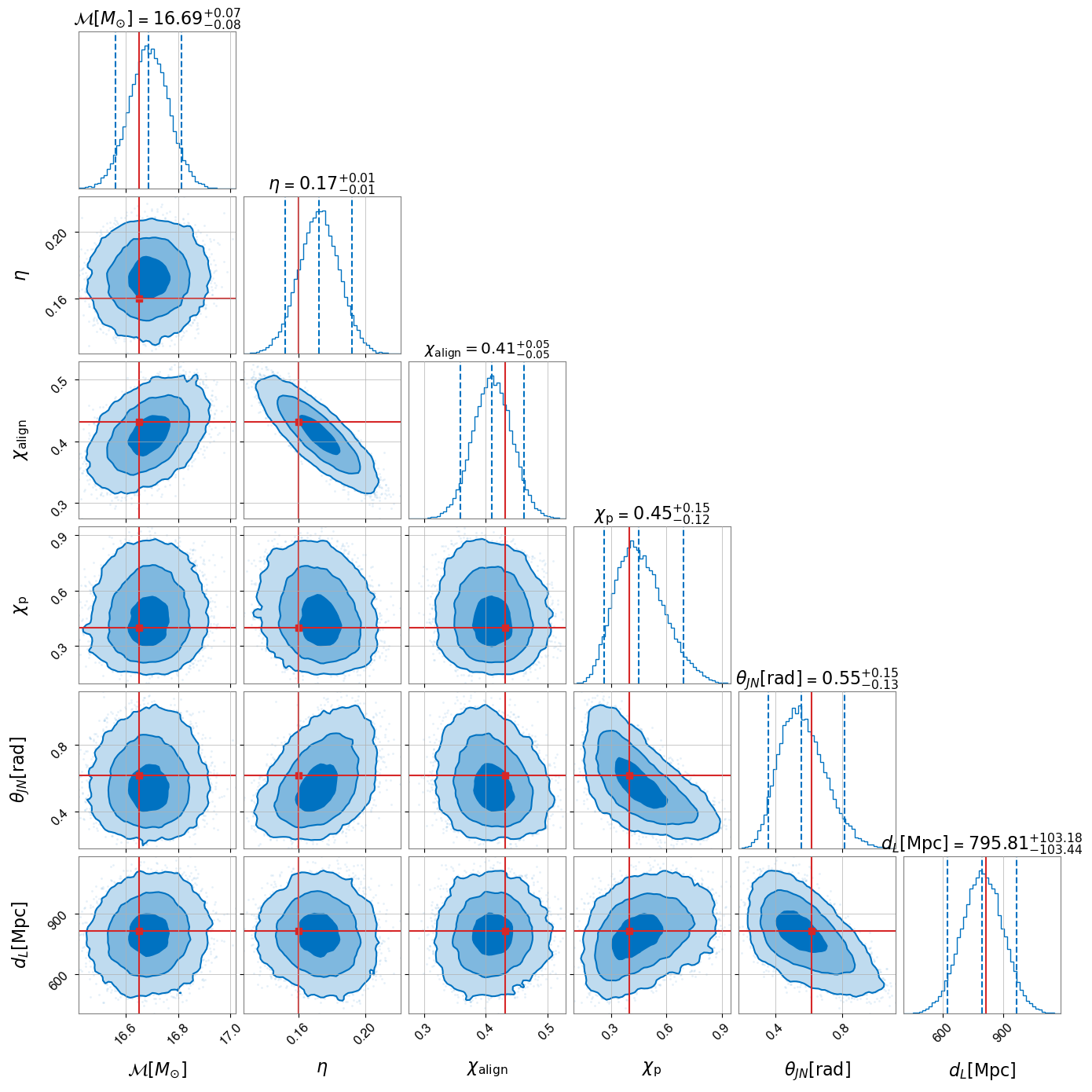}
    \caption{The posterior distribution for masses (chirp mass $\mathcal{M}$ and symmetric mass ratio $\eta$) spins (aligned $\chi_{\mathrm{align}}$ and precessing $\chi_{\mathrm{p}}$), distance $d_{L}$ and binary orientation $\theta_{JN}$. The shaded regions in the two-dimensional figures show the 1, 2 and 3 $\sigma$ regions for the parameters and the one-dimensional distributions show the median and 90\% symmetric confidence interval.
    The simulated values are shown as red lines.  For all of the parameters, the inferred values are in good agreement with the simulation.  The binary was deliberately generated to have negligible power in the second polarization, and significant power in both higher multipoles and precession, signal (4) in Table.~\ref{tab:varying_params}.
    }
    \label{fig:high_p_high_33_posterior}
\end{figure*}

\begin{figure*}[t!]
    \includegraphics[width=0.32\textwidth]{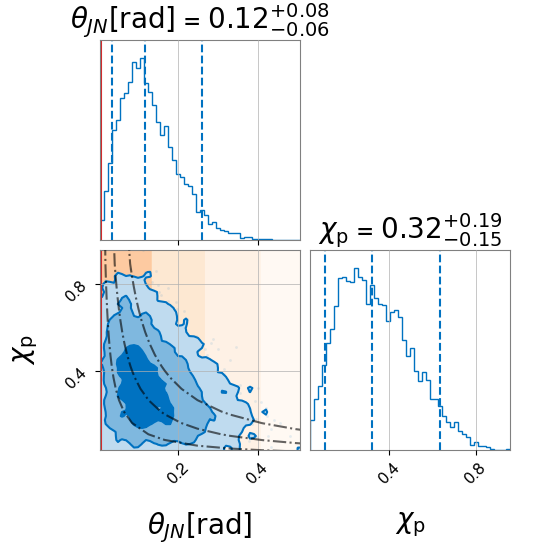}
    \includegraphics[width=0.32\textwidth]{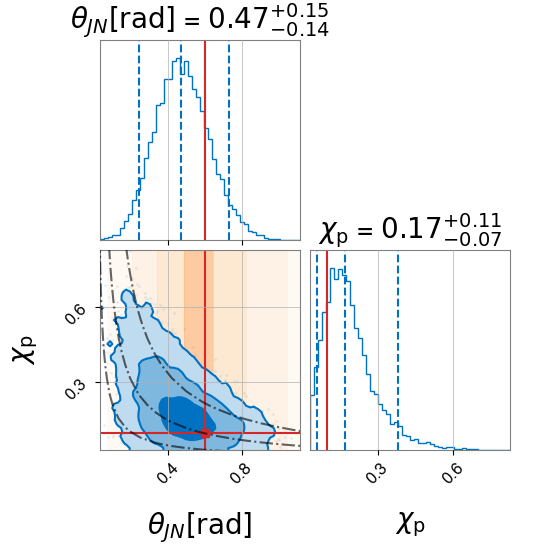}
    \includegraphics[width=0.32\textwidth]{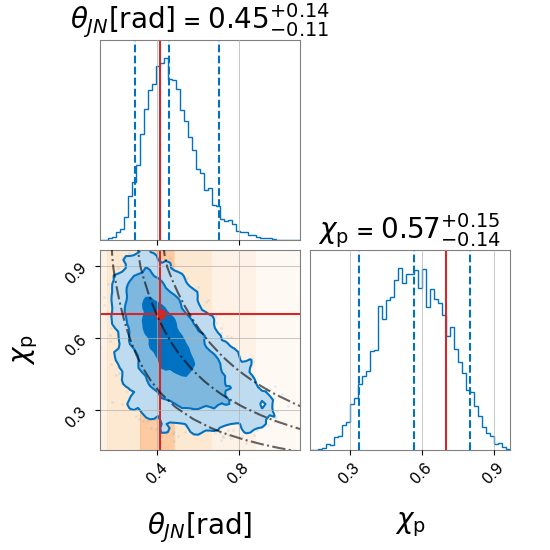}
    \caption{The posterior distribution for the binary's orientation $\theta_{JN}$ and precessing spin $\chi_{\mathrm{p}}$ for 3 simulated signals; \emph{Left}: a binary with insignificant power in precession and higher-order multipoles, signal (1).
    \emph{Middle}: a binary with significant power in higher-order multipoles, signal (2), and
    \emph{Right}: a binary with significant power in precession and some power in higher-order multipoles, signal (3). The blue shaded regions in the two-dimensional figures show the 1, 2 and 3 $\sigma$ regions for the parameters. The black dash-dotted lines show lines of constant $\rho_{p}$, and the orange shaded regions show lines of constant $\rho_{33}$: the darker region encompasses the injected value. The one-dimensional distributions show the median and 90\% symmetric confidence interval.
    The simulated values are shown as red lines. We note that the priors on $\theta_{JN}$ and $\chi_{\mathrm{p}}$ vanish at zero leading to zero posterior support there, and consequently posteriors peaked away from the injected values, as shown in the left hand panel.
    }
    \label{fig:injected_theta_jn_chi_p}
\end{figure*}

In Figure \ref{fig:high_p_high_33_posterior}, we show the \texttt{simple-pe} posterior obtained for signal (4), with negligible power in the second polarization, but significant power in both higher multipoles and precession.  The posteriors obtained are in good agreement with the simulated values, with both the $\theta_{JN}$ and $\chi_{\mathrm{p}}$ distributions clearly peaked away from zero.  In addition, we see clear correlations between several pairs of parameters.  As expected, the mass ratio and aligned spin are (anti-)correlated due to their impact on the phase evolution of the waveform.  The distance and binary orientation are also anti-correlated based upon the measured \gls*{snr} in the (2, 2, 0) waveform component --- to achieve a given \gls*{snr} an inclined signal must be at a smaller distance.  Finally, the binary orientation and precessing spin are (anti-)correlated as larger values of either leads to a larger \gls*{snr} in precession.  We also observe a slight correlation between orientation and mass ratio.  This arises from the necessity to obtain the correct \gls*{snr} in higher multipoles --- the \gls*{snr} increases with $\theta_{JN}$ but decreases for increasing $\eta$ (more equal mass binaries).

Next, we consider how the parameter recovery varies for the three other systems listed in Table \ref{tab:varying_params}.  While the distributions of all parameters do change somewhat, we focus on the binary orientation, $\theta_{JN}$, and precessing spin, $\chi_{\mathrm{p}}$.  These parameters have the greatest impact on the observed \gls*{snr} in precession and higher multipoles and we therefore expect them to be most impacted by the changes.  Of course, the distance will also vary (as is clear from Table \ref{tab:varying_params}), but this is primarily due to our requirement of a fixed \gls*{snr} in the signal.   Although the mass ratio remains fixed, one might expect its recovery to vary when higher multipoles and precession are observed (as was the case for GW190412, \cite{LIGOScientific:2020stg}).  However, since the signal is already identified as unequal mass from the (2, 2, 0) waveform, they have little impact on the mass ratio distribution. For all 3 signals: one with negligible power in both higher multipoles and precession, one with significant power in higher multipoles, and another with significant power in precession, we see that the recovered $\theta_{JN}$ and $\chi_{\mathrm{p}}$ are in good agreement with the simulated values. For all 3 cases, the simulated values lie within the $1\sigma$ confidence interval. Furthermore, the inferred ranges of $\theta_{JN}$ and $\chi_{\mathrm{p}}$ can be explained by the \gls*{snr} in higher-order multipoles and precession.  In each case, the binary orientation is restricted by the measured (3, 3, 0) \gls*{snr}. The precession \gls*{snr} increases with both $\chi_{p}$ and $\theta_{JN}$ as, to a good approximation is scales with $\chi_{p} \cdot \theta_{JN}$.  Thus, the measured precession \gls*{snr} limits the permitted range in the two-dimensional space.  As expected a higher observed (3, 3) multipole \gls*{snr} leads to a larger inferred $\theta_{JN}$ while a larger precession \gls*{snr} leads to larger values of both $\theta_{JN}$ and $\chi_{p}$.

The events we have investigated were deliberately chosen to lie in a region of the parameter space where higher-order multipoles, precession and power in the second polarization are all likely to be observed.  In particular, we selected a high \gls{snr} signal with unequal masses, observed well away from face-on.  For this example, even when there is no power in higher multipoles or precession, the lack of power can be used to restrict the physical parameters, as shown in \ref{fig:injected_theta_jn_chi_p}.  The majority of \gls{gw} observations are expected to be low \gls{snr} and close to equal mass.  For these systems, it is very unlikely that additional waveform features will be identifiable.  In this case, lack of power in precession, higher multipoles and second polarization have limited impact on the inferred parameters.  In Appendix \ref{app:equal_mass}. \subsection{Comparison with other samplers}
\label{sec:pe_comp}

\begin{figure*}
    \centering
    \includegraphics[width=\linewidth]{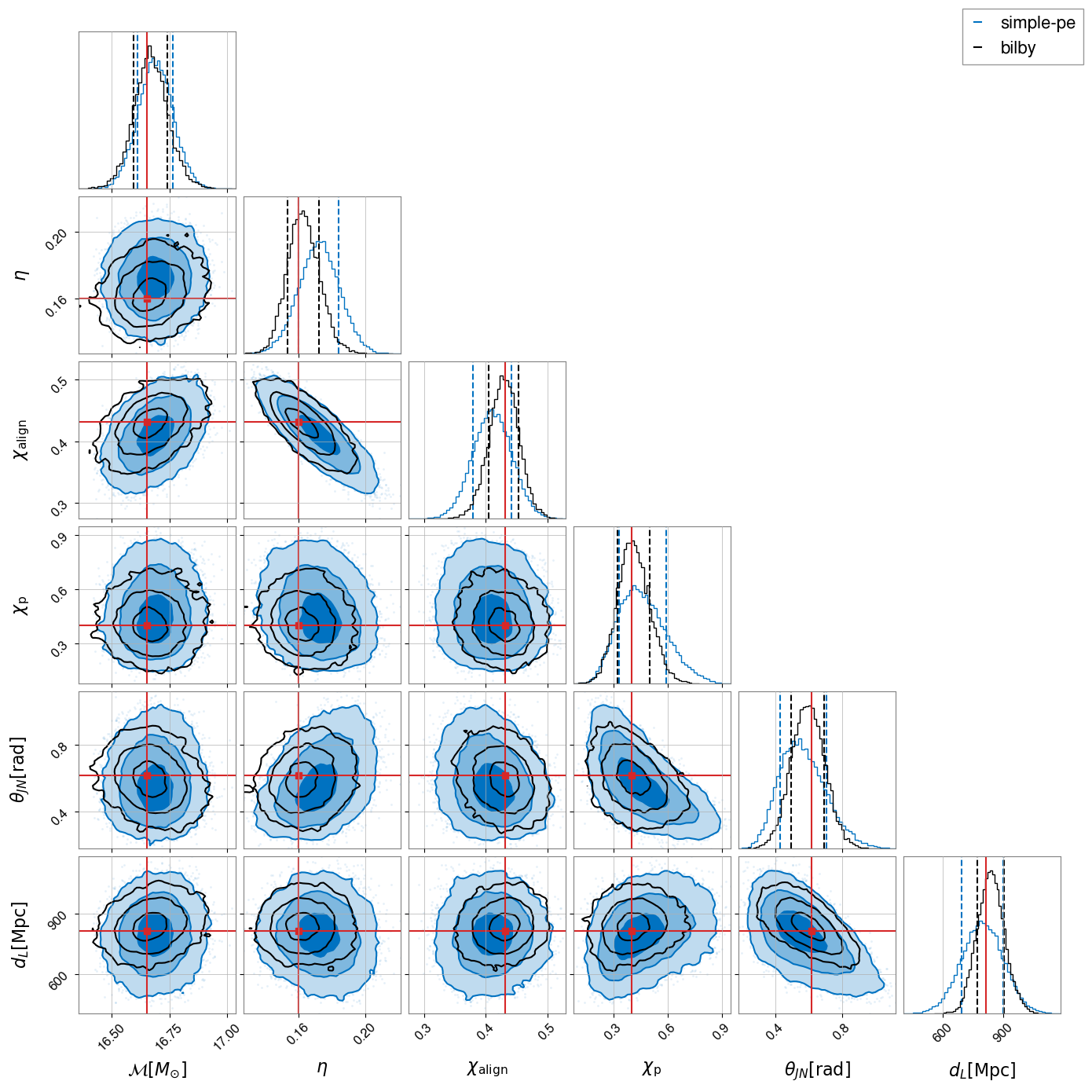}
    \caption{Same as Fig.~\ref{fig:high_p_high_33_posterior} but with the posteriors obtained with {\texttt{dynesty}} (generated through the {\texttt{bilby}} infrastructure~\cite{Ashton:2018jfp,Romero-Shaw:2020owr,Smith:2019ucc}) overlaid in black. {\texttt{dynesty}} took $\sim 20$ hours to complete when parallelised over 400 CPUs while {\texttt{simple-pe}} took $\sim 20$ minutes on a single CPU.
    }
    \label{fig:comparison_with_bilby}
\end{figure*}

In Fig.~\ref{fig:comparison_with_bilby}, we show the posterior distributions for a signal with power in the second polarization, and significant power in higher multipoles and precession obtained using \texttt{simple-pe} (as shown in Fig.~\ref{fig:high_p_high_33_posterior}) with that obtained using {\texttt{dynesty}}~\cite{Speagle:2019ivv} (generated through the {\texttt{bilby}} infrastructure~\cite{Ashton:2018jfp,Romero-Shaw:2020owr,Smith:2019ucc}). {\texttt{dynesty}} was chosen as it is commonly used for gravitational-wave follow-up analyses (see e.g. Refs.~\cite{LIGOScientific:2021djp,Nitz:2021zwj}). We see that in general, there is good agreement between the two sampling techniques, with the \texttt{simple-pe} results algorithm presented here consistent with {\texttt{dynesty}} for all parameters.  The \texttt{simple-pe} posteriors are somewhat broader than those obtained with the full Bayesian parameter estimation routine.  Most notably, the distance and precession spin posteriors are significantly broader.  The distance uncertainty arises from three effects: a degeneracy between distance and orientation, a degeneracy between distance and mass, and a variation of distance over the sky region.  For this signal, it is the variation of detector sensitivity over the sky patch that gives the greatest contribution to the distance uncertainty.  While our localization routine returns a reasonable approximation to the sky patch, it does give a significantly larger region than {\texttt{dynesty}} and this explains the difference in distance inference.  The precessing spin is also less well measured by the \texttt{simple-pe} result.  Similar results are obtained for the three other systems discussed in Section \ref{sec:pe_results}, namely that the \texttt{simple-pe} posteriors are comparable with, although typically broader than thos obtained with \texttt{dynesty}.  Importantly, {\texttt{dynesty}} took $\sim 20$ hours to complete when parallelised over 400 CPUs while the work presented here took $\sim 20$ minutes on a single CPU. \section{Discussion} 
\label{sec:discussion}

In this paper, we have presented a detailed discussion of the primary way in which different physical parameters impact the observed gravitational-wave signal emitted during a black hole binary coalescence.  Using these insights, we have developed a parameter estimation method, \texttt{simple-pe}, that uses this information to provide estimates of the physical parameters of the system.  By restricting focus to the \textit{observable} impact on the gravitational waveform, we are able to obtain rapid parameter estimates which naturally incorporate degeneracies between physical parameters.
We have presented parameter estimation results for a small set of signals and, in Section \ref{sec:pe_comp}, shown that our results are in good agreement with those obtained by the bilby \cite{Ashton:2018jfp} sampler which is used in interpreting gravitational wave observations \cite{LIGOScientific:2021djp}.  The \texttt{simple-pe} results are obtained in a fraction of the time and recover broadly the same results as the more intensive samplers.  There have been numerous other analyses developed which provide rapid parameter estimation for gravitational wave observations \cite{Singer:2015ema,Lange:2018pyp,Smith:2019ucc,Gabbard:2019rde,Green:2020hst,Dax:2021tsq,Dax:2022pxd,Shen:2019vep,Green:2020dnx,Tiwari:2023mzf, Pathak:2022ktt, Islam:2022afg}.  However, many of these use waveforms that exclude either precession or higher multipoles or both.  Thus, while they are able to provide reasonable estimates of masses, aligned spins and sky locations, they are unable to probe some of the interesting astrophysics that will only be uncovered through higher multipoles (which are one of the best ways to clearly identify unequal mass systems) or precession (which is the only way to probe in-plane spins).  Furthermore, the \texttt{simple-pe} analysis has the advantage of providing clear interpretation as to \textit{why} and \textit{how} the various parameters are measured and the accuracy achieved.  For example, in Figure \ref{fig:injected_theta_jn_chi_p}, we can understand the inferred orientation and in-plane spins based upon the higher multipole and precession \glspl*{snr}.  

Here, we have restricted attention to a limited number of simulated signals, primarily to describe the methods employed and the properties of the observed signal which enable measurement of various parameters.  In the main paper, we have focused on an example where additional waveform features are observable and can therefore be used to improve parameter recovery.  In Appendix \ref{app:equal_mass}, we present an example for a low \gls*{snr} signal with close to equal masses for which higher multipoles and precession have minimal impact.  This does not constitute a comprehensive test of the analysis.  In a future paper, we plan to present an in depth investigation of \texttt{simple-pe} results using both a large set of simulations and also existing \gls*{gw} observations \cite{LIGOScientific:2021djp}.

In future observing runs, the rate of observed binary mergers will increase as instrumental sensitivity improves.  For example, in the O4 run \cite{KAGRA:2013rdx}, several events per week are expected and in O5 and beyond we could regularly be observing multiple events per day.  Thus, it becomes ever more important to obtain detailed estimates of the binary properties in a short time-frame --- comparable to the time between events.  The \texttt{simple-pe} analysis provides a tool that can become part of the process for quickly understanding the observations.  We are able to provide good parameter estimates in a matter of minutes.  The analysis, inevitably, introduces some approximations, for example by considering only the most significant higher multipole and precession harmonic.  In addition, in the current implementation, we do not extract the phase, polarization or precession phase angles, although we have briefly discussed how this could be done.  Finally, the \texttt{simple-pe} analysis does not, currently, include calibration uncertainties which can broaden the recovered parameter widths \cite{Vitale:2020gvb}.  Thus, in its current implementation, \texttt{simple-pe} \textit{will not} provide the most accurate estimate of the binary parameters.  Nonetheless, a fast and reasonably accurate estimate of the binary parameters is useful to rapidly identify interesting events for prioritized analysis and provide parameter estimates to guide follow-up electromagnetic observations, where knowledge of masses, mass ratio and binary orientation can impact the expected signal \cite{Metzger:2016pju}.  Furthermore, there may be applications, for example, studies of the black hole population \cite{LIGOScientific:2021psn} or searches for lensed events \cite{LIGOScientific:2021izm}, where the accuracy provided by \texttt{simple-pe} is sufficient to obtain the results.  Finally, where full parameter estimation is required, the initial parameter estimates provided by \texttt{simple-pe} can be used to inform the detailed parameter estimation routines.  For example, the posteriors obtained by \texttt{simple-pe} could be importance sampled by bilby to produce a final set of posterior samples, as has been done previously to obtain faster results for higher multipoles and eccentricity  \cite{Payne:2019wmy, Romero-Shaw:2020thy}.

There are a number of additional features which can lead to an observable effect in the emitted gravitational waveform, including matter effects, eccentricity and mode asymmetries.  For binaries containing one or two neutron stars, the waveform will carry information about the structure of the star, this arises as additional post-Newtonian corrections during the inspiral and also a difference in the merger and post-merger signal from that of a black hole binary \cite{De:2018uhw, LIGOScientific:2018cki}. For binaries on non-circular orbits, the emitted gravitational-wave signal carries an imprint of the orbital eccentricity \cite{Peters:1964zz}. For several of the binaries observed in O3, there is tantalising evidence of eccentricity\cite{Romero-Shaw:2020thy, Gayathri:2020coq}. Finally, for binaries where the orbit precesses, there are asymmetries between the gravitational wave signal emitted above and below the orbital plane.  These correspond to the ``bobbing'' of the binary and the final kicks given to the system following merger~\cite{boyle2014gravitational,Ramos-Buades:2020noq}.  Detailed studies of these additional features, and their impact on parameter measurements, have been pursued by several groups.  Studies of neutron star structure include \cite{Pratten:2021pro, Finstad:2022oni}; eccentricity had been studied in \cite{Romero-Shaw:2020thy, Knee:2022hth} and mode asymmetries are discussed in Refs.~\cite{Ramos-Buades:2020noq,Kolitsidou:2023aaa}.

All three of these features are amenable to the approach presented in this paper. The presence of neutron-star structure will impact the overall frequency and phase evolution of the system, and is likely to be primarily measured from the leading waveform component. Mode asymmetries lead to a difference between, e.g. the (2, 2) and (2, -2) multipoles.  By separating into the leading (symmetric) and sub-dominant (anti-symmetric) components, we obtain an additional waveform component whose amplitude will help constrain the in-plane spins.
Eccentricity impacts the overall evolution of the waveform, and it is well known that the eccentricity is degenerate with mass (both higher eccentricity and higher mass lead to a faster merger). However, eccentricity also leads to additional harmonics in the waveform, at multiples of the orbital frequency \cite{PhysRevD.70.064028}. Therefore, it seems likely that a combination of techniques will be required to incorporate eccentricity. Finally, we note that most waveform models include only a subset of the physical effects we have discussed, as each additional effect expands the dimensionality of the parameter space to be simulated. Our approach of identifying the leading effect on the waveform of each new (astro-)physical phenomenon has the potential to help guide development of future waveform models. Additionally, even if complete models are not available, we could incorporate all of their leading order effects into the \texttt{simple-pe} analysis to obtain a more complete picture of each observed binary.

 \section*{Acknowledgements}

We thank Tom Dent, Ilya Mandel, Frank Ohme, Javier Roulet, Rory Smith, Eric Thrane, Veronica Villa  and Barak Zackay for useful discussions and suggestions. We thank Ben Farr and Javier Roulet for detailed comments on an earlier draft of the paper.  We thank Mark Hannam and Vivien Raymond for continued discussions and insights throughout this project.
SF, RG, CH, CM and SU thank STFC for support through the grants ST/V005618/1 and ST/N005430/1. CH thanks the UKRI Future Leaders Fellowship for support through the grant MR/T01881X/1.  S. F. was supported by a Leverhulme Trust International Fellowship. 
The authors are also grateful for computational resources provided by LIGO Laboratory and supported by National Science Foundation Grants PHY-0757058 and PHY-0823459 and provided by Cardiff University, and
funded by STFC grant ST/I006285/1.

{\texttt{simple-pe}} is programmed in Python and implements modules from \texttt{numpy}~\cite{harris2020array}, \texttt{scipy}~\cite{2020SciPy-NMeth}, \texttt{pycbc}~\cite{pycbc-software} and \texttt{pesummary}~\cite{Hoy:2020vys}. Plots were prepared with \texttt{matplotlib}~\cite{2007CSE.....9...90H}, and {\texttt{pesummary}}~\cite{Hoy:2020vys}.
As part of this paper, the {\texttt{parallel-bilby}} parameter estimation software~\cite{Smith:2019ucc}, which made use of the {\texttt{dynesty}} nesting sampling package~\cite{Speagle:2019ivv}, was used for comparisons.
 
\appendix
\section{Multipole decomposition of the gravitational waveform}
\label{app:waveform}

In the body of the text, we derived the expression for the gravitational waveform as a sum over multipole moments $(\ell, m)$ with an additional ``splitting'' of the harmonics due to precession, given in Eq.~(\ref{eq:prec_modes}), which we repeat here:
\begin{equation}\label{ap_eq:prec_modes}
h = \sum_{\ell, m, n} {}^{-2}Y_{\ell, m}(\theta, \phi) D^{\ell}_{n, m}(\alpha, \beta, \epsilon) h^{\mathrm{NP}}_{\ell, n}(t, \vec{\lambda}) \, .
\end{equation}
Our aim is to explicitly extract the waveform dependence upon the opening angle $\beta$ and to show that we can decompose the waveform into components with a natural hierarchy in the parameter $b = \tan(\beta/2)$.  

We begin by expanding the spherical harmonics using
\begin{equation}\label{ap_eq:y_lm}
{}^{-2}Y_{\ell, m}(\theta_{JN}, -\alpha_o) =
\sqrt{\frac{2 \ell + 1}{4\pi}} d^{\ell}_{m, 2}(\theta_{JN}) e^{- im \alpha_{o}}
\end{equation}
and 
\begin{equation}\label{ap_eq:wigner_D}
D^{\ell}_{n, m}(\alpha, \beta, \epsilon) = e^{i m \alpha} d^{\ell}_{n, m}(-\beta) e^{-i n \epsilon} \, ,
\end{equation}
to obtain
\begin{align}
h &= 
\sum_{\ell} \sum_{m, n =-\ell}^{\ell}
\sqrt{\frac{2\ell + 1}{4\pi}} 
\\
& \quad
 h^{\mathrm{NP}}_{ln} 
d^{\ell}_{m, 2} (\theta_{JN})
d^{\ell}_{n, m}(- \beta) 
e^{- i n \epsilon}
e^{i m  (\alpha -\alpha_{o})} \, .
 \nonumber
\end{align}
Next, we impose the restriction that the gravitational wave emission is symmetric above and below the plane of the binary so that $h_{\ell, n} = (-1)^{\ell} h^{\ast}_{\ell, -n}$.\footnote{This is a reasonable approximation but, particularly for large in-plane spins, there is an asymmetry in the emitted radiation \cite{Varma:2019csw}.  The formalism presented here can be extended to that more general case, but we leave that to future work.}
This allows us to express the waveform in terms of multipoles with $n \ge 0$ as
\begin{align}\label{ap_eq:h_prec_dmat}
h &= \sum_{\ell} 
\sum_{n =0}^{\ell} 
\sum_{m =-\ell}^{\ell} 
\sqrt{\frac{2\ell + 1}{4\pi}}
d^{\ell}_{m, 2} (\theta_{JN})
e^{i m  (\alpha -\alpha_{o})}
\\
&
\quad
\left[
h^{\mathrm{NP}}_{ln} d^{\ell}_{n, m}(- \beta) e^{- i n \epsilon}
+
(-1)^{\ell} (h^{\mathrm{NP}}_{ln})^{\ast} d^{\ell}_{-n, m}(- \beta) e^{i n \epsilon}
\right]
\nonumber
\end{align}
Next, we collect together the $\beta$ terms as we want to obtain an expansion in terms of the parameter $b = \tan(\beta/2)$.  To do so, we use the discrete symmetries of the Wigner d-matrices, $d^{\ell}_{n, m}(- \beta)$, 
\begin{equation}
	d^{\ell}_{n, m} = 
	(-1)^{m-n} d^{\ell}_{m, n} = 
	(-1)^{m - n} d^{\ell}_{-n, -m}
\end{equation}
and a re-labelling of $m \rightarrow -m$ in the second term of Eq.~(\ref{ap_eq:h_prec_dmat}) to extract the factor $d^{\ell}_{n,m}(-\beta)$ from both terms.  In addition, for consistency with previous work, we relabel the indices by switching $n$ and $m$.  This enables us to use $(\ell, m)$ to label the multipole moments of the waveform in the
frame aligned with the orbital angular momentum and $n$ ($\in [-\ell, \ell]$)
to denote the precession harmonics.\footnote{Following \cite{Fairhurst:2019_2harm}, we will
later introduce $k \in [0, 2\ell +1]$ as the harmonic labelling precession, and show that the harmonics
have a power-law structure with index $k$. } 
Combining these steps leads to a final expression for the waveform in terms of non-precessing multipoles as
\begin{align} \label{ap_eq:spherical_decomp}
h &= \sum_{\ell} \sum_{m =0}^{\ell} \sum_{n =-\ell}^{\ell} \sqrt{\frac{2\ell + 1}{4\pi}}
d^{\ell}_{m, n}(- \beta)
\times
 \nonumber
\\
&
\quad
\left[
d^{\ell}_{n, 2} (\theta_{JN})
\left(
h^{\mathrm{NP}}_{lm}  e^{- i m \epsilon}
e^{i n  (\alpha -\alpha_{o})}
\right)
+
\nonumber
\right.
\\ &
\qquad
\left.
(-1)^{\ell - m}
d^{\ell}_{n, -2} (\theta_{JN})
\left(
h^{\mathrm{NP}}_{lm}  e^{- i m \epsilon}
e^{i n  (\alpha -\alpha_{o})}
\right)^{\ast}
\right] \, .
\end{align}
In the following sub-sections, we examine the structure of the waveform in equation \ref{ap_eq:spherical_decomp} by considering the (2, 2) and (3, 3) multipoles explicitly, as well as the general case of the $(\ell, \ell)$ and $(\ell, \ell-1)$ multipoles.

\subsection{The (2, 2) two-harmonic waveform}
\label{app_sec:wf_22}

The two-harmonic waveform introduced in \cite{Fairhurst:2019_2harm} can be recovered by restricting to the $(2, 2)$ multipole in Eq.~(\ref{ap_eq:spherical_decomp})
and then further restricting to the two leading order terms in $b$.  We briefly summarize the calculation below.

The Wigner d-matrices for $\ell = m = 2$ are given by
\begin{align}\label{ap_eq:wigner_22}
	d^{2}_{2,  n}(-\beta) 
	&:= C_{2,n} \cos^{2 + n}(-\beta/2) \sin^{2 -n}(\beta/2) \nonumber \\
	&= C_{2, n} \frac{(-b)^{2 - n}}{( 1 + b^{2})^{2}} 
\end{align}
where $C_{2,\pm2} = 1$, $C_{2,\pm 1} = 2$, $C_{2,0} = \sqrt{6}$ and 
$b = \tan(\beta/2)$.
This enables us to express the waveform amplitudes as a power series in $b$.  Similarly, it is convenient to express the Wigner d-matrices for $\theta_{JN}$ appearing in Eq.~(\ref{ap_eq:spherical_decomp}) in terms of
\begin{equation}\label{ap_eq:tau}
    \tau := \tan(\theta_{JN}/2) \, .
\end{equation}

Restricting to $\ell = m = 2$ in Equation (\ref{ap_eq:spherical_decomp}) and substituting the explicit form of the Wigner d-matrices for both $-\beta$ and $\theta_{JN}$ gives
\begin{align}
    h_{22} &= 
    \sum_{n = -2}^{2} 
    \sqrt{\frac{5}{4\pi}} 
    \frac{(C_{2, n})^{2} b^{2 - n}}{( 1 + b^{2})^{2}} \times 
    \\
    & 
    \quad
    \left[ \frac{\tau^{2 - n}}{(1 + \tau^{2})^{2}} 
    \left(h^{\mathrm{NP}}_{22}(t) e^{- 2 i \epsilon} 
    e^{i n (\alpha - \alpha_{o})}\right) + \right. \nonumber 
    \\
    &
    \quad \left. \;  \frac{(-\tau)^{2 + n}}{(1 + \tau^{2})^{2}} 
    \left(h^{\mathrm{NP}}_{22}(t) e^{- 2 i \epsilon} 
    e^{i n (\alpha - \alpha_{o})} \right)^{\ast} 
    \right] \, .
    \nonumber 
\end{align}
This shows the desired structure, with each of the precession harmonics appearing with a factor of $b^{2 - n}$.  Using the fact that $b < 1$, we can restrict to the two leading precession harmonics ($b^{0}$ and $b^{1}$).  Specifically, we introduce
\begin{align}
    h_{22, 0} &= \sqrt{\frac{5}{4\pi}} 
    \frac{1}{(1 + b^{2})^{2}}
    h^{\mathrm{NP}}_{22}(t) e^{- 2 i \epsilon} e^{2i (\alpha - \alpha_{o})} \, ,
    \nonumber \\
    \label{ap_eq:h22}
    h_{22, 1} &= \sqrt{\frac{5}{4\pi}} 
    \frac{b}{(1 + b^{2})^{2}}
    h^{\mathrm{NP}}_{22}(t) e^{- 2i \epsilon} e^{i (\alpha - \alpha_{o})} \, .
\end{align}
Both waveforms accumulate a secular phase of $2i \epsilon$ relative to the non-precessing waveform, as first noted in \cite{Apostolatos:1994mx}.  The amplitude of the second precession harmonic is reduced by a factor of $b$ relative to the first and the frequency is reduced by the precession frequency $\Omega_{p}$, where $\dot{\Omega}_{p} = \alpha$.

The two-harmonic waveform is given as
\begin{equation}
    h_{22} \approx  
    \frac{(h_{22,0} + \tau^{4} h_{22,0}^{\ast})}{(1 + \tau^{2})^{2}} 
    + \frac{4\tau (h_{22,1} - \tau^{2} h_{22,1}^{\ast})}{(1 + \tau^{2})^{2}} 
    \, .
\end{equation}
reproducing the expression provided in \cite{Fairhurst:2019_2harm}.

\subsection{The (3, 3) waveform}
\label{app_sec:wf_33}

Let us now obtain a similar expression for the (3, 3) multipole, following a similar procedure as in the previous section.  As before, we wish to evaluate the multipole amplitude explicitly in terms of $b$ and $\tau$.  The required Wigner d-matrices are
\begin{align}\label{ap_eq:wigner_33}
	d^{3}_{3,  n}(-\beta)
	&:= C_{3,n} \cos^{3+n}(\beta/2) \sin^{3-n}(-\beta/2) \nonumber \\
	&= C_{3, n} \frac{(-b)^{3-n}}{( 1 + b^{2})^{3}}
\end{align}
where $C_{3,\pm 3} = 1$ and $C_{3,\pm 2} = \sqrt{6}$ and, as before, $b = \tan(\beta/2)$.
This gives
\begin{align} \label{ap_eq:33_decomp}
    h_{3 3} &= 
    \sum_{n =-3}^{3} \sqrt{\frac{7}{4\pi}}
    C_{3, n} \frac{(-b)^{3-n}}{( 1 + b^{2})^{3}}
    \times
    \nonumber \\
    &
    \quad
    \left[
    d^{3}_{n, 2} (\theta_{JN})
    \left(
    h^{\mathrm{NP}}_{3 3}  e^{- 3i \epsilon}
    e^{i n  (\alpha -\alpha_{o})}
    \right) + 
    \right.
    \nonumber \\ 
    & 
    \qquad
    \left.
    d^{3}_{n, -2} (\theta_{JN})
    \left(
    h^{\mathrm{NP}}_{3 3}  e^{- 3i \epsilon}
    e^{i n  (\alpha -\alpha_{o})}
    \right)^{\ast}
    \right] \, .
\end{align}
Thus, we see that the terms appear with an amplitude factor $b^{3-n}$.  For the (2, 2) multipole, we restricted to the two most significant precession contributions ($b^{0}$ and $b^{1})$.  We will do the same for the (3, 3) multipole.  For the leading-order precession contribution $b^{0}$, we can evaluate the $d^{3}_{3, \pm 2}  (\theta_{JN})$ terms  Eq.~(\ref{ap_eq:wigner_33}).  For the first order precession correction, we need to evaluate $d^{3}_{2, \pm 2}$ which are given by
\begin{align}\label{ap_eq:wigner322}
    d^{3}_{2, 2}(\theta) &=   \frac{1 - 5\tau^{2}}{(1 + \tau^{2})^{3}} 
    \nonumber \\
    d^{3}_{2, -2}(\theta) &=   \frac{5\tau^{4} - \tau^{6}}{(1 + \tau^{2})^{3}} 
\end{align}
where $\tau = \tan(\theta/2)$.

We introduce the precession contributions to the (3, 3) multipole as\footnote{We have chosen the normalization of the (3, 3) waveform to match the one used in \cite{Mills:2020thr}.  This differs from the more ``natural'' normalization where the pre-factor would be $\sqrt{7/4\pi}$.  This choice of normalization has no impact on the results in the paper.}
\begin{equation}\label{ap_eq:h_33k}
    h_{33, k}  =  \frac{1}{4} \sqrt{\frac{21}{2\pi}} 
    \frac{b^{k}}{(1 + b^{2})^{3}}
    \left(h^{\mathrm{NP}}_{33} e^{- 3 i \epsilon} e^{(3-k) i (\alpha - \alpha_{o})}\right) \, .
\end{equation}
We can then express the (3, 3) multipole, to first order in $b$ as
\begin{align}\label{ap_eq:h33}
h_{3 3} &\approx
     \frac{4\tau (h_{33,0} + \tau^{4} h_{33,0}^{\ast})}{(1 + \tau^{2})^{3}} + 
     \\
     & \quad\frac{-4 \tau (1 - 5 \tau^{2}) h_{33,1} + 4\tau (\tau^{2} - 5) h_{33, 1}^{\ast}}{(1 + \tau^{2})^{3}} \nonumber 
\end{align}

We are also interested in isolating the $h_{33,0}$ and $h_{33,1}$ waveform harmonics.  To do so, we follow \cite{Fairhurst:2019_2harm} and generate a waveform at different orientations and combine them to produce a waveform containing only the leading precession harmonic for the (3, 3) multipole.  For the (2, 2) multipole, only the 0 and 4 precession harmonics are non-zero for face-on systems so the simplest way to generate the leading precession harmonic is simply to generate a face-on system --- this will be correct up to $O(b^{4})$, and even that contribution can be easily removed by subtracting two waveforms with different phase and polarization (see \cite{Fairhurst:2019_2harm} for details).  
For the (3, 3) multipole, the leading precession harmonic vanishes for a face-on system.  Indeed, it is often stated that the (3, 3) multipole vanishes for face-on systems, actually this is not true if the system is precessing as the $b^{1}$ and $b^{5}$ precession harmonics are non-zero.  Thus, to $O(b^{4})$ the face on (3, 3) multipole gives the $b^{1}$ precession harmonic.  Specifically:
\begin{equation}
    h_{33,1} \approx h_{33}(\theta_{JN} = 0,  \alpha_o = 0, \phi = 0, \psi=0) \, .
\end{equation}
This will be correct to $O(b^{4})$.  The contribution from the $k=5$ harmonic can be removed by taking the average of the above waveform and one generated with $\alpha_{o} = \tfrac{\pi}{4}$, $\phi = - \tfrac{\pi}{6}$.

To generate the leading precession harmonic for the (3, 3) multipole, we require the waveform for $\theta_{JN} = \tfrac{\pi}{2}$ ($\tau = 1$).  In this case, the amplitude of the left and right circular polarizations, given by the $\tau$ and $\tau^{5}$ terms in Equation (\ref{ap_eq:h33}), are equal and the signal is linearly polarized, with power only in the $+$ polarization.  The same is true for the $b^2$, $b^{4}$ and $b^{6}$ harmonics, while the $b^{1}$ and $b^{5}$ harmonics have power only in the $\times$ polarization and the $b^{3}$ harmonic vanishes.  Thus, the $+$ polarized waveform at $\theta_{JN} = \tfrac{\pi}{2}$ provides the leading precession harmonic, accurate to $b^{2}$.  We can further improve the accuracy by noting that the precession phase $\alpha_{o}$ appears with a different factor for the different modes.  In particular, generating the waveform at $\alpha_0 = 0, \tfrac{\pi}{2}$ leads to the a phase shift of $\pi$ in the leading harmonic and $2\pi$ in the $b^{2}$ harmonic.  Therefore, to accuracy of $b^{4}$ we can write
\begin{align}
    h_{33, 0} &\approx \frac{1}{2} \left[
    h_{33+}(\theta_{JN} = \tfrac{\pi}{2}, \alpha_o = 0, \phi = 0)   \right. 
     \\
    & \left. \qquad  +
    h_{33+}(\theta_{JN} = \tfrac{\pi}{2}, \alpha_o = \tfrac{\pi}{2},
    \phi = \tfrac{\pi}{6}) \right] \, .
    \nonumber
\end{align}
Again, it is not difficult to obtain the waveform accurate to all powers in $b$ by generating waveforms with $\alpha_{o} = \pm\tfrac{\pi}{4}$ and appropriately combining them to remove the $b^{4}$ contribution.

\subsection{The waveform for $\ell = m$ multipoles}

\label{app_sec:wf_l=m}

Here, we briefly sketch the derivation for a generic $(\ell, \ell)$ multipole.  The calculation is essentially identical to the one above for the $(3, 3)$ multipole.  In particular, we first evaluate the $d^{\ell}_{\ell,  n}(-\beta)$ term using the explicit form of the Wigner d-matrices
\begin{align}\label{ap_eq:wigner_ll}
	d^{\ell}_{\ell,  n}(-\beta)
	&:= C_{\ell,n} \cos^{\ell+n}(\beta/2) \sin^{\ell-n}(-\beta/2) \nonumber \\
	&= C_{\ell, n} \frac{b^{\ell-n}}{( 1 + b^{2})^{\ell}}
\end{align}
where $C_{\ell,n}$ is a real number (we give the value where necessary).  Thus we can write 
\begin{align} \label{ap_eq:ll_decomp}
    h_{\ell \ell} &= 
    \sum_{n =-\ell}^{\ell} 
    \sqrt{\frac{2\ell + 1}{4\pi}}
    C_{\ell, n} 
    \frac{(-b)^{\ell-n}}{( 1 + b^{2})^{\ell}}    
    \times
    \\
    &
    \quad
    \left[
    d^{\ell}_{n, 2} (\theta_{JN})
    \left(
    h^{\mathrm{NP}}_{\ell \ell}  e^{- i \ell \epsilon}
    e^{i n  (\alpha -\alpha_{o})}
    \right)
    \right.
    \nonumber \\ 
    & 
    \quad
    \left. \;
    + 
    d^{\ell}_{n, -2} (\theta_{JN})
    \left(
    h^{\mathrm{NP}}_{\ell \ell}  e^{- i \ell \epsilon}
    e^{i n  (\alpha -\alpha_{o})}
    \right)^{\ast}
    \right] \, . \nonumber
\end{align}
As before, we immediately see that the $(\ell, \ell)$ multipole has the same decomposition in terms of $b$, with a clearly identified leading order term when $n = \ell$.  Restricting to this leading precession contribution we note that $C_{\ell, \pm \ell} = 1$.  In addition, we can evaluate the $d^{\ell}_{2, \pm \ell}  (\theta_{JN}) = d^{\ell}_{\ell, \pm 2}  (\theta_{JN})$ term using Eq.~(\ref{ap_eq:wigner_ll}), where we note the values $C_{22} = 1$, $C_{32} = \sqrt{6}$, $C_{42} = 2\sqrt{7}$.

Then, to leading order in the precession parameter $b$,
\begin{equation}
h_{\ell \ell} \approx
     \frac{2 (2 \tau)^{\ell - 2} (h_{\ell \ell ,0} + \tau^{4} h_{\ell \ell,0}^{\ast})}{(1 + \tau^{2})^{\ell}} 
\end{equation}
with the normalization again chosen to match that introduced in \cite{Mills:2020thr} and
\begin{equation}\label{ap_eq:h_ll0}
    h_{\ell \ell, 0}  \propto  
    \frac{1}{(1 + b^{2})^{\ell}}
    \left(h^{\mathrm{NP}}_{\ell \ell} 
    e^{- i \ell \epsilon} 
    e^{i \ell (\alpha - \alpha_{o})}\right) \, .
\end{equation}
The proportionality constant is determined through the desired normalization for $h_{\ell \ell}$.

As with the (2, 2) multipole, if we were to keep sub-dominant precession harmonics, their amplitudes would be suppressed by powers of $b$ while their frequencies would be reduced by multiples of the precession frequency.

\subsection{The waveform for $m = \ell - 1$ multipoles}
\label{app_sec:wf_l=m+1}

The other multipoles which can contribute significantly to the observed gravitational wave signal are the (2,1) and (3,2) multipoles \cite{Khan:2019kot}, so we also look briefly at the multipoles with $m = \ell -1$.  Now, the Wigner d-matrix
can be written as
\begin{align*}
    d^{\ell}_{\ell -1, n}(-\beta) 
    & = \frac{A_{\ell, n} (-b)^{\ell - n + 1}
    + 
    B_{\ell, n} (-b)^{\ell - n - 1}}{(1 + b^{2})^{\ell}}
\end{align*}
where $A_{\ell n}$ and $B_{\ell n}$ are real numbers and, since the power of $b$ must be non-negative, $A_{\ell, -\ell} = B_{\ell, \ell} = 0$.  

As for the other multipoles, we see that the form of the $h_{\ell, (\ell -1)}$ waveform can be factorized into terms with a prefactor of $b^{k}$. Since the (3, 2) and (2, 1) multipoles are sub-dominant \cite{Mills:2020thr}, we restrict to the leading order, $b^{0}$, contributions.  These arise from  the $|n| = m = \ell -1$ terms, for which $A_{\ell, -(\ell - 1)} = B_{\ell, (\ell-1)} = 1$.  Therefore, we obtain
\begin{align} \label{ap_eq:h_ll-1}
    h_{\ell (\ell - 1)} &= 
    \sqrt{\frac{2\ell + 1}{4\pi}}
    \frac{1}{(1 + b^2)^{\ell}}
    \\
    & \qquad \left[
    d^{\ell}_{\ell - 1, 2} (\theta_{JN})
    \left(
    h^{\mathrm{NP}}_{\ell (\ell - 1)}
    e^{i (\ell - 1)  (\alpha -\alpha_{o} - \epsilon)}
    \right)
    \nonumber
    \right.
    \\ &
    \qquad
    \left. -
    d^{\ell}_{\ell - 1, -2} (\theta_{JN})
    \left(
    h^{\mathrm{NP}}_{l (\ell - 1)} 
    e^{i (\ell - 1)  (\alpha -\alpha_{o} - \epsilon)}
    \right)^{\ast}
    \right] \, .
\end{align}
Finally, we note that for the $(\ell, \ell-1)$ multipoles the sub-dominant precession term would include two additional harmonics --- one at a frequency $\Omega_{p}$ above the dominant harmonic and one a frequency $\Omega_{p}$ below.
 \section{simple-pe example for low SNR event}
\label{app:equal_mass}

\begin{figure*}
    \centering
    \includegraphics[width=\linewidth]{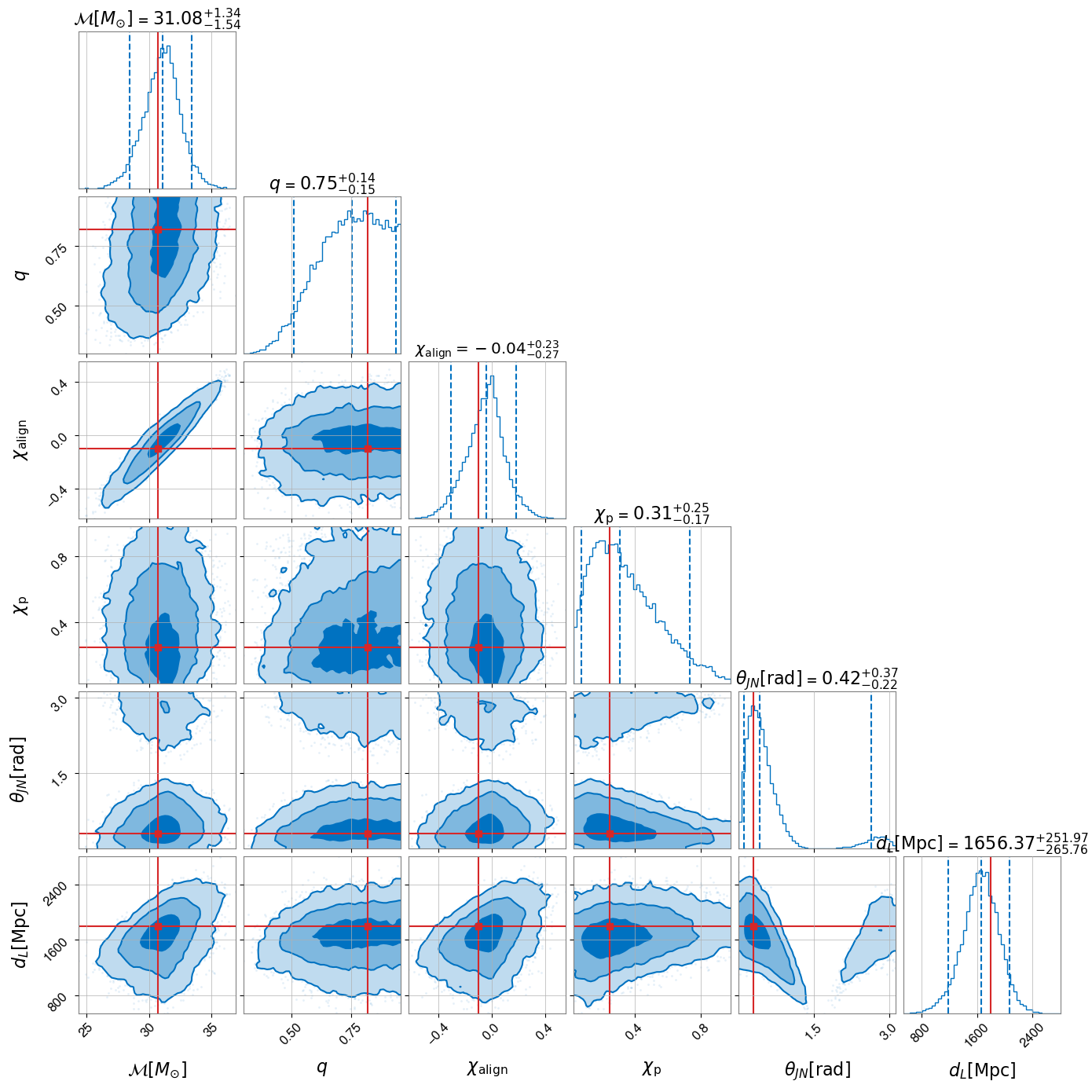}
    \caption{The posterior distribution for masses (chirp mass $\mathcal{M}$ and  mass ratio $q = m_{2}/m_{1}$) spins (aligned $\chi_{\mathrm{align}}$ and precessing $\chi_{\mathrm{p}}$), distance $d_{L}$ and binary orientation $\theta_{JN}$ for a binary with masses $39 M_{\odot}$ and $32 M_{\odot}$ at a \gls*{snr} of 12. The shaded regions in the two-dimensional figures show the 1, 2 and 3 $\sigma$ regions for the parameters and the one-dimensional distributions show the median and 90\% symmetric confidence interval.
    The simulated values are shown as red lines.  At this low \gls*{snr}, we observe broad posteriors for many parameters, in particular orientation and precessing spin.
    }
    \label{fig:low_snr_example}
\end{figure*}

In this appendix, we show the results for a \texttt{simple-pe} analysis of a low \gls*{snr} event with close to equal masses.  As discussed in Section \ref{sec:pe_results}, such a system will have limited \gls*{snr} in higher-order multipoles, precession and the second \gls*{gw} polarization for the majority of possible orientations and in-plane spin configurations.  Therefore, we expect that the lack of \gls*{snr} in these features will do little to restrict the inferred properties of the binary, most notably the orientation and in-plane spins.  This is borne out in the results shown in Figure \ref{fig:low_snr_example}.  The binary is simulated with masses $39 M_{\odot}$ and $32 M_{\odot}$ (in the detector frame), giving a redshift of $z=0.33$ and a network \gls*{snr} of 12.  There is minimal power in precession, higher-order modes and the second polarization with all features having \gls*{snr} $< 0.5$.  This leads to broad distributions in $\chi_p$, mass ratio $q$ and orientation $\theta_{\mathrm{JN}}$.  Furthermore, there is only peak preference for a left (rather than right) circular polarization, which leads to the observed bimodality in orientation.  While a detailed investigation of \texttt{simple-pe} results for \gls*{gw} observations is beyond the scope of this paper, we note that the widths of parameter posteriors are broadly consistent with observed signals \cite{LIGOScientific:2021djp}.

\bibliographystyle{utphys}
\bibliography{references}

\end{document}